\documentclass[
 prd,aps,amsmath,superscriptaddress,nofootinbib]{revtex4}

\usepackage{amsmath}
\usepackage{graphicx}
\usepackage{bm}
\usepackage{xspace}
\usepackage{epsfig}
\usepackage{color}
\usepackage{xcolor}
\usepackage[hypertex]{hyperref}

  \newcount\hour \newcount\minute
  \hour=\time \divide \hour by 60
  \minute=\time
  \newcommand{\mydate}{\ \today \ - \number\hour :\ifnum \minute<10 0\fi 
\number\minute}

\def\OMIT#1{}

\newcommand{\nn}{\nonumber} 

\newcommand{\bn}{{\bar n}}

\newcommand{\mcdot}{\!\cdot\!}

\newcommand{\plus}{\ensuremath{\! + \!}}
\newcommand{\minus}{\ensuremath{\! - \!}}

\newcommand{\sh}{\hat s}

\newcommand{\half}{\frac{1}{2}}

\newcommand{\tbreve}{\tilde}

\newcommand{\im}{\mbox{Im}}  
  
\newcommand{\ugu}{\!\!\!&=&\!\!\!}  
  
\newcommand{\fr}{\frac}

\newcommand{\cB}{{\cal B}}  

\newcommand{\cO}{{\cal O}}  
\newcommand{\cL}{{\cal L}}

\newcommand{\al}{\alpha} 
 \newcommand{\bt}{\beta} 
 \newcommand{\ga}{\gamma} 
 \newcommand{\de}{\delta} 
    
\newcommand{\s}{\sigma}

 \newcommand{\Ga}{\Gamma}

\def\tb{{\bar{t}}}
\def\sht{\hat{s}_t}
\def\shtb{\hat{s}_\tb}

\begin{document}



\title{\boldmath
  Two-loop Jet-Function and Jet-Mass for Top Quarks
\vspace{0.4cm}
}

\author{Ambar Jain}
\affiliation{Center for Theoretical Physics, Massachusetts Institute of
  Technology, Cambridge, MA 02139\vspace{0.1cm}}

\author{Ignazio Scimemi}
\affiliation{Center for Theoretical Physics, Massachusetts Institute of
  Technology, Cambridge, MA 02139\vspace{0.1cm}}
\affiliation{Departamento de Fisica Teorica II,
Universidad Complutense de Madrid,
28040 Madrid, Spain\vspace{0.4cm}}

\author{Iain W.~Stewart\vspace{0.5cm}}
\affiliation{Center for Theoretical Physics, Massachusetts Institute of
  Technology, Cambridge, MA 02139\vspace{0.1cm}}
\begin{abstract}
\vspace{0.3cm}  

We compute the two-loop heavy quark jet-function in the heavy quark limit. This
is one of the key ingredients in next-to-next-to-leading order (NNLO) and
next-to-next-to-leading-log order (NNLL) computations of the invariant mass
distribution of top-jets at a future $e^+e^-$ collider. The shape of the top
invariant mass distribution is affected by large logs which we compute at NNLL
order. Exploiting the non-abelian exponentiation theorem, a definition of the
top jet-mass is given which is transitive and whose renormalization group
evolution is determined by the cusp-anomalous dimension to all orders in
perturbation theory. Relations of the jet-mass to the pole, $\overline {\rm
  MS}$, and 1S masses are presented at two-loop order.
\\[25pt]
 \hbox{MIT-CTP 3916}\\
 \hbox{0801.0743 [hep-ph]}

\end{abstract}

\maketitle

\newpage 

\section{Introduction}

We are about to step into the LHC era, opening up a new energy regime for the
discovery of physics beyond the standard model. In this era precise measurements
of standard model parameters will still be important for disentangling new
physics scenarios. A prime example of this is the effect of the mass of the
top quark on precision electroweak constraints. The latest Tevatron analyses
give $m_t = 170.9 \pm 1.8\,{\rm GeV}$~\cite{unknown:2007bxa}, a measurement at
the 1\% level. However for a minimal standard model Higgs sector the indirect
determination of the Higgs mass $m_H = 76{}^{+33}_{-24} \,{\rm GeV}$ has such a
strong sensitivity to the top-mass, that a $2\,{\rm GeV}$ upward shift in $m_t$
causes this $m_H$ central value to shift upward by 15\% (with the same upward
shift for the 95\% CL bound $m_H < 182\,{\rm GeV}$)~\cite{private1}.
Such strong sensitivities to $m_t$ are also a feature of many new physics
scenarios, such as supersymmetric extensions of the standard model.

In addition to uncertainties related to the experimental analysis, the
top-mass also suffers from a theoretical uncertainty related to the
specification of the mass-scheme in which the measurement is being made. Many
observables used in measurements of the top quark mass incorporate only lowest
order theory results, which does not allow one to distinguish between quark mass
schemes.  In general quark mass schemes are connected by a perturbative series in
the strong coupling, with relations of the form
\begin{align} \label{mschemeAB}
 m_t^{\rm scheme A} = m_t^{\rm scheme B} (1 +\alpha_s + \alpha_s^2 +\ldots) \,,
 \qquad\qquad
 m_t^{\rm scheme A} = m_t^{\rm scheme B} + R (\alpha_s + \alpha_s^2 + \ldots) \,,
\end{align}
where $R$ is a scheme parameter (examples of schemes with relations of both of
these types are discussed in the body of the paper). For high precision
measurements of the $b$-quark mass~\cite{Yao:2006px,HFAG} a useful class of
schemes are the so-called threshold mass-schemes, with examples being the
kinetic, 1S, and shape-function
schemes~\cite{Bigi:1996si,Hoang:1998hm,Hoang:1998ng,Hoang:1999zc,Bosch:2004th}.
For the $b$-quark these schemes are optimized to avoid $\Lambda_{\rm QCD}$
sensitivities, while still maintaining a power counting in $\Lambda_{\rm
  QCD}/m_b$ to handle non-perturbative corrections in observables.  The
pole-mass is not used for precision analyses because of its infrared
sensitivity, which introduces an ambiguity $\delta m_b^{\rm pole}\sim
\Lambda_{\rm QCD}$ from infrared renormalons~\cite{Beneke:1998ui}.  For the
top quark the infrared physics is cut off by it's width $\Gamma_t = 1.43\,{\rm
  GeV}$, since tops decay before they hadronize.  However the top pole-mass
still suffers from a $\Lambda_{\rm QCD}$ infrared renormalon
ambiguity~\cite{Smith:1996xz}.

Although we do not know the precise scheme for the top-mass measurement of
Ref.~\cite{unknown:2007bxa}, we do know that it falls in a category of ``top
resonance mass schemes'', which differ from $m_t^{\rm pole}$ by an amount
$\lesssim \Gamma_t$. This follows from the fact that top-mass measurements rely
on an underlying Breit-Wigner to incorporate the top-width, and only top
resonance mass schemes are compatible with a Breit-Wigner
line-shape~\cite{Fleming:2007qr}. For these observables using a short distance
resonance mass-scheme avoids infrared sensitivity while maintaining a
$\Gamma_t/m_t$ expansion.  Examples of top resonance mass schemes include the
jet-mass~\cite{Fleming:2007qr} and kinetic-mass~\cite{Bigi:1996si}. For the
top-mass this scheme dependence issue could easily add an additional theoretical
uncertainty of $\lesssim 2\,{\rm GeV}$ when the measured top-mass is used in
practical applications. To see this, lets say that the mass measurement
corresponds to $m_t^{\rm scheme B}$, and that we want to use a result for a
resonance mass $m_t^{\rm scheme A}$ in an analysis. Since scheme B is not
precisely known, the perturbative series introduces an additional uncertainty
that we estimate to be of ${\cal O}(\alpha_s^2 m_t)$ or ${\cal O}(\alpha_s R)$
from Eq.~(\ref{mschemeAB}), leading to the $\lesssim 2\,{\rm GeV}$ theoretical
error estimate.

In order to obtain higher precision measurements of the top-mass one needs
accurate theoretical predictions for a realistic experimental observable in a
definite mass scheme. This has been achieved for $e^+e^-\to t\bar t$ at
threshold~\cite{Hoang:2000yr,Hoang:2001mm,Pineda:2006ri,Martinez:2002st,Juste:2006sv,Hoang:2006pd},
where state of the art computations incorporate next-to-next-to-leading-log
(NNLL) and next-to-next-to-leading order (NNLO) QCD corrections for the
cross-section, as well as subleading lifetime effects.  Theoretically the
necessary setup is also clear for $e^+e^- \to t\bar t$ far from threshold, where
the center-of-mass energy $Q^2\gg m_t^2$~\cite{Fleming:2007qr,Fleming:2007xt}.
Here the top quark decay products form well separated collinear jets together
with soft-radiation between the jets. A suitable observable is the event-shape
cross-section $d^2\sigma/dM_t^2 dM_{\bar t}^2$. Here $M_t^2 = (\sum_{i\in a}
p_i^\mu)^2$ and $M_{\bar t}^2 = (\sum_{i\in b} p_i^\mu )^2$ are hemisphere invariant
masses, and the hemispheres a and b are separated by a plane perpendicular to
the thrust axis for each event. The different physics components of
$d^2\sigma/dM_t^2 dM_{\bar t}^2$ can be separated by a factorization theorem
derived in Ref.~\cite{Fleming:2007qr}
\begin{align} \label{Fthm}
  \fr{d\s}{d M_t^2d M_\tb^2} &= \sigma_0
  H_Q(Q,\mu_m)H_m\Big(m,\fr{Q}{m},\mu_m,\mu\Big)
 \int\! d\ell^+ d\ell^-\
 B_+\Big(\sht-\fr{Q\ell^+}{m},\Ga_t,\mu\Big) B_-\Big(\shtb-\fr{Q \ell^-}{m},\Ga_t,\mu\Big)
 \: S(\ell^+,\ell^-,\mu)
\nn\\[4pt]
&\quad
 + \cO\Big(\fr{m\al_s(m)}{Q},\frac{m^2}{Q^2},\fr{\Ga_t}{m},\fr{s_t}
{m^2},\fr{s_\tb}{m^2}\Big)
 \ .
\end{align}
In Eq.~(\ref{Fthm}) $\sigma_0$ is the tree level Born cross section, $H_Q$ and
$H_m$ are hard-functions which encode the perturbative corrections at the scales
$Q$ and $m$, where from now on we use $m$ for the mass of the top quark. The
invariant mass variables $\hat s_t$ and $\hat s_{\bar t}$ are defined as
\begin{align} \label{shat}
  \sht &=\fr{s_t}{m}=\fr{M_t^2-m^2}{m} \,,
  & \shtb& =\fr{s_\tb}{m}=\fr{M_\tb^2-m^2}{m}\ ,
\end{align}
and the most sensitive region for mass measurements is the peak region where
$\hat s_{t,\bar t} \lesssim \Gamma_t + Q\Lambda_{\rm QCD}/m$. Finally, $B_{\pm}$
in Eq.~(\ref{Fthm}) are heavy-quark jet functions for the top quark/antiquark,
and $S$ is the soft function describing soft radiation between the jets. Our
main focus in this article will be on the functions $B_{\pm}$, which are defined
in the heavy-quark limit $m_t \gg \Gamma_t$ using heavy quark effective theory
 (HQET)~\cite{Manohar:2000dt,Neubert:1993mb}. The soft function $S$ is universal to
massless and massive jets and a suitable model can be found in
Ref.~\cite{Hoang:2007vb}, extending earlier work in
Ref.~\cite{Korchemsky:2000kp}.  The factorization theorem in Eq.~(\ref{Fthm})
was derived using soft collinear effective theory
(SCET)~\cite{Bauer:2000ew,Bauer:2000yr,Bauer:2001yt,Bauer:2001ct} and effective
theory methods for unstable
particles~\cite{Fadin:1987wz,Beneke:2003xh,Beneke:2004km,Hoang:2004tg}.  A
similar factorization theorem with the soft-function and different jet functions
is known to apply for jets initiated by massless
quarks~\cite{Korchemsky:1994is,Korchemsky:1999kt,Bauer:2002ie,Bauer:2003di}.

In this paper we carry out the first step towards NNLO and NNLL predictions for
the invariant mass spectrum, $d^2\sigma/dM_t^2 dM_{\bar t}^2$, by computing the
top quark jet function at two-loop order.  We also carry out the resummation of
large logs for this jet function at NNLL order. This translates into a
resummation of all the large logs in the cross-section that can modify the
invariant mass distribution~\cite{Fleming:2007qr}. On the conceptual side we
introduce a definition of the top jet-mass scheme that has a well defined mass
anomalous dimension at any order in perturbation theory (unlike definitions
based on cutoff moments or peak locations). In this jet-mass scheme we prove
that the quark-mass anomalous dimension is completely determined by the cusp
anomalous dimension at any order in perturbation theory. As an intermediate step
to demonstrating this we show that in position space the heavy quark jet
function exponentiates. This follows from the fact that this jet function
satisfies the criteria for the non-abelian exponentiation
theorem~\cite{Gatheral:1983cz,Frenkel:1984pz}.

Because of the simplifying nature of HQET our two-loop computation of the jet
function is significantly simpler than a direct two-loop computation of the
cross-section. In particular, as we discuss below in sections~\ref{sect:Bdef}
and~\ref{sect:Wloop}, even for a finite width and an arbitrary mass scheme the
jet function computation can be reduced to the perturbative evaluation of a
vacuum matrix element of Wilson lines. For heavy quarks two loop computations
are already available for the partonic heavy-quark shape
function~\cite{Korchemsky:1992xv,Gardi:2005yi,Becher:2005pd} and heavy-quark
fragmentation function~\cite{Korchemsky:1992xv,Gardi:2005yi,Neubert:2007je}. The
hadronic versions of these functions that appear in observables are
non-perturbative. The hadronic shape function describes the light-cone momentum
distribution of b-quarks in a heavy B-meson~\cite{Neubert:1993ch,Bigi:1994ex},
while the hadronic fragmentation function describes the probability that a
b-quark fragments to a B-meson with a particular light-cone momentum
fraction~\cite{Jaffe:1993ie}. The jet function is fundamentally different since
it is defined by a matrix element evaluated between vacuum states, and due to
the smearing from the finite top-width can be reliably computed in perturbation
theory. We elaborate on similarities and differences below in
sections~\ref{sect:Bdef} and~\ref{sect:Wloop}.

Our outline is as follows. In section~\ref{sect:Bdef} we discuss the basic
formalism for the top quark jet function, including its renormalization and
anomalous dimension. We then give a summary of our two-loop results for the
jet function and for the solution of its renormalization group equation, with
details relegated to appendices. In section~\ref{sect:Wloop} we determine the
Wilson line representation of the jet function and compare it with the shape
function and fragmentation function. Then in section~\ref{sect:NAExp} we work
out the implications of the non-abelian exponentiation theorem for the
heavy-quark jet function and for the partonic shape-function, including the
combined implications of this theorem and the all-orders solution of the
renormalization group equation.  In section~\ref{sect:jetmass} we discuss
possible jet-mass scheme definitions, and present a scheme based on the position
space jet function that remains transitive to all orders in perturbation theory.
We also give two loop relations of the jet-mass to the pole-mass, $\overline
{\rm MS}$-mass, and 1S-mass schemes. Finally, in section~\ref{sect:results} we
present results for the NNLO jet function with NNLL resummation, including
numerical analysis. We conclude in section~\ref{sect:conc}.


\section{Heavy Quark jet function} \label{sect:Bdef}
    
In this section we describe the basic properties of the heavy-quark jet
functions $B_\pm$.  Up to a change of variable $B_+$ for the top quark and $B_-$
for the antitop quark are identical by charge conjugation, so we will only refer
to the computation of $B_+$. To simplify the notation we also drop the
subscript, so that $B=B_+$. These subscripts $\pm$ are restored when we
simultaneously consider the top and antitop system in the final factorization
theorem.  We start by reviewing definitions and results for the HQET jet
function from Refs.~\cite{Fleming:2007qr,Fleming:2007xt}. $B$ is given by the
imaginary part of a forward scattering matrix element,
\begin{eqnarray}
B( {\hat{s}},\delta m,\Ga_t,\mu  )\ugu\im \big[\cB({\hat{s}},\delta m,\Ga_t,\mu) \big]\ ,
\label{eq:BCB}
\end{eqnarray}    
where $\cB$ are vacuum matrix elements of a time-ordered product of fields and
Wilson lines
\begin{eqnarray} \label{eq:CB}
\cB(2 v_+\cdot r,\delta m,\Ga_t,\mu)\ugu\fr{-i}{4\pi N_c m}
 \int\! d^4x\, e^{i r\cdot x}\left\langle\,
   0\left|  T\{\bar{h}_{v_+}(0) W_{n}(0)
   W_{n}^\dagger(x)h_{v_+}(x)\}\right|0\right\rangle 
 \,.
\end{eqnarray}
Here $v_+^\mu$ is the velocity of the heavy top quark, and we introduce
null-vectors $n^\mu$ and $\bn^\mu$ so that we can decompose momenta as $p^\mu=
n^\mu \bn\mcdot p/2 + \bn^\mu n\mcdot p/2 + p_\perp^\mu$. The vectors satisfy
$v_+^2=1$ and $n^2=\bn^2=0$.  The dot-products of these vectors encode the boost
of the top quarks relative to the center-of-mass frame of the $e^+e^-$
collision, $n\cdot v_+=m/Q$, and $\bn\cdot v_+=Q/m$.  In Eq.~(\ref{eq:CB}) the
Wilson lines are
\begin{align} \label{WW}
 W^\dagger_{n}(x) &=  {\rm P}\exp\bigg(ig\int_0^\infty ds\, \bar{n}\cdot A_n(\bar
 n s+x)\bigg),
 & W_{n}(x) &=\overline {\rm P}\exp\bigg(-ig\int_0^\infty ds\, 
  \bar{n}\cdot A_n(\bar n s+x)\bigg) \,.
\end{align}
These Wilson lines make ${\cal B}$ gauge-invariant and encode the residual
interactions from the antitop jet. Both the HQET fields $h_{v_+}$ and the gluon
fields in $W_{n}$ (which we call $A_n^\mu$) are only sensitive to fluctuations
with $p^2\ll m^2$.  In the rest-frame of the top quark these are
soft-fluctuations, while in the $e^+e^-$ center-of-mass frame they are
``ultra-collinear'' along the direction of the energetic top quark. The gluon
fields $A_n^\mu$ have zero-bin subtractions~\cite{Manohar:2006nz} for the region
of the soft function $S$ in Eq.~(\ref{Fthm}) as explained in Appendix~B of
Ref.~\cite{Fleming:2007xt}.

The HQET fields $h_{v_+}$ have the leading order Lagrangian
\begin{eqnarray} \label{eq:LHQET}
 \cL_h &= \bar{h}_{v_+} \Big(iv_+\cdot D-\de
 m+\fr{i}{2}\Ga_t\Big)h_{v_+} \,.
\end{eqnarray}
Here $\Ga_t$ is the top quark total width, obtained from matching the top-decay
amplitudes in the standard model (or a new physics model) onto HQET at leading
order in the electroweak interactions, and at any order in $\al_s$. This gives
the correct description of finite lifetime effects for cross-section in
Eq.~(\ref{Fthm}) to $\cO(m^2/Q^2,\Ga/m)$ in the power counting for separation of
the jets from the decay products~\cite{Fleming:2007qr}. The residual mass term
$\de m$ in Eq.~(\ref{eq:LHQET}) fixes the definition of the top mass $m$ for the
HQET computations~\cite{Falk:1992fm}, where
\begin{eqnarray} \label{dm}
 \de m\ugu m_{pole}-m \,.
\end{eqnarray} 
For predictions in the peak region consistency with the power counting requires
$\de m\sim\Ga_t\sim \sht\sim \shtb$~\cite{Fleming:2007qr}, a condition which is
true of the jet-mass scheme that we discuss below in section~\ref{sect:jetmass}.

From the definitions in Eqs.~(\ref{eq:BCB}-\ref{eq:CB}) and the Lagrangian in
Eq.~(\ref{eq:LHQET}) one can deduce a series of properties of the jet function.
As a first, instead of computing $B(\hat{s},\delta m, \Ga_t,\mu)$ and ${\cal
  B}(\hat s,\delta m,\Gamma_t,\mu)$, one can consider computing these functions
for a (fictitious) top quark having zero width. Furthermore, due to
Eq.~(\ref{eq:LHQET}) the $\hat s$ and $\delta m$ dependence occurs in the
combination $(\hat s-2\delta m)$, so it is useful to also have a notation for
computations done with a zero residual mass term in the Lagrangian. Thus we define
\begin{align}
 B(\hat{s},\delta m,\mu) &\equiv B(\hat{s},\delta m, 0,\mu) \,,
 & {\cal B}(\hat s,\delta m,\mu) &\equiv {\cal B}(\hat s,\delta m,0,\mu)
  \,,
  \nn\\
 B(\hat s,\mu) &\equiv B(\hat s,0,0,\mu) \,,
 & {\cal B}(\hat s,\mu) & \equiv {\cal B}(\hat s,0,0,\mu) \,.
\end{align}
These jet functions and vacuum matrix elements are related by
\begin{align} \label{eq:BcB}
  B(\hat s,\delta m,\mu) &= {\rm Im} \big[ {\cal B}(\hat s,\delta m,\mu) \big]
  \,,
  & B(\hat s,\mu) &= {\rm Im} \big[ {\cal B}(\hat s,\mu) \big]
  \,,
\end{align}
and $B(\hat s,\mu)$ has support for $\hat s\ge 0$.  The form of the Lagrangian
in Eq.~(\ref{eq:LHQET}) implies that having calculated ${\cal B}(\hat s,\mu)$ we
can include the width and $\delta m$ terms by simple shifts,
\begin{eqnarray}  \label{eq:bbbb}
 \cB(\hat{s},\delta m, \Ga_t,\mu) 
   = \cB(\hat{s}+i \Ga_t,\delta m,\mu)
   = \cB(\hat{s}-2\delta m +i \Ga_t,\mu)  \,.
\end{eqnarray}
As discussed in Ref.~\cite{Fleming:2007xt} the stable and unstable HQET jet
functions can also be related with a dispersion relation,
\begin{align} \label{eq:dr}
B(\hat{s},\delta m, \Ga_t,\mu) 
  &= \int_{-\infty}^{\infty}\!\! d\hat{s}'\
 B(\hat s-\hat{s}',\delta m,\mu)\, \fr{\Ga_t}{\pi (\hat{s}'^{\,2}+\Ga_t^2)}
= \int_{-\infty}^{\infty}\!\! d\hat{s}'\
 B(\hat s-\hat{s}'-2\delta m,\mu)\, \fr{\Ga_t}{\pi (\hat{s}'^{\,2}+\Ga_t^2)}
 \,.
\end{align}
The width of the top quark acts as an infrared cutoff through this smearing with
the Breit-Wigner. Finally we remark that the $\mu$-dependence indicated by the
last argument of $B(\hat s,\delta m,\Gamma_t,\mu)$ and ${\cal B}(\hat s,\delta
m,\Gamma_t,\mu)$ is independent of $\Ga_t$ and $\delta m$.  Additional scale
dependence may be induced by the choice of mass-scheme, ie. by a parameter
$\delta m=\delta m(\mu)$.  When we consider $B(\hat s_t,\delta m,\Gamma_t,\mu)$
as a function of $M_t$ this additional $\mu$-dependence from $\delta m$ cancels against that in
the mass $m(\mu)$ in Eq.~(\ref{shat}). This cancellation occurs at leading order
in the HQET power counting.

We will also find it useful to consider the Fourier transformed jet functions
\begin{align} \label{By}
 \tilde B(y,\delta m,\Gamma_t,\mu) &= \int_{-\infty}^{+\infty} \!\!\! d\hat s \ e^{-iy\, \hat s
   } \ B(\hat s,\delta m,\Gamma_t,\mu) \,,
 & \tilde B(y,\delta m,\mu) & = \int_{-\infty}^{+\infty} \!\!\! d\hat s \ e^{-iy\, \hat s
   } \ B(\hat s,\delta m,\mu) \,,
\end{align}
where $y=y-i0$ to ensure convergence as $\hat s\to \infty$. In Fourier space the
connection between the jet functions computed with zero and non-zero width and
residual mass terms becomes particularly simple,
\begin{align}
  \tilde B(y,\delta m,\Gamma_t,\mu) 
    &=  \tilde B(y,\delta m,\mu)\ e^{- |y|\, \Gamma_t }
    =  \tilde B(y,\mu)\ e^{-2i y \delta m}\ e^{- |y|\, \Gamma_t }
    \,.
\end{align}
This formula is quite interesting, since as we discuss in
section~\ref{sect:NAExp} below, the result for $\tilde B(y,\mu)$ also
exponentiates to all orders in perturbation theory.

\subsection{Renormalization and Anomalous Dimension} \label{subsec:ren}

We use dimensional regularization with $d=4-2\epsilon$ and the $\overline {\rm
  MS}$ scheme to renormalize the jet function. The renormalization properties of
$B(\hat{s},\mu)$ and $\cB(\sh,\mu)$ are the same, so in the following we work
with $\cB(\hat s,\mu)$ for simplicity.  The divergences of loop calculations are
removed with $Z$-factors, so one can pass from bare to renormalized matrix
elements by
\begin{align} \label{eq:bren}
\cB(\hat{s},\mu) &=
  \int d\hat{s}' \; Z_{B}^{-1}(\hat{s}-\hat{s}',\mu)\ \cB^{\rm{bare}}(\hat{s}')
   \,.
\end{align}
This equation can be thought of as the generalization of a $Z$ matrix which
renormalizes a set of operators indexed by $\hat s$, to the case where $\hat s$
is continuous~\cite{Bauer:2000ew}.  Here $Z_B$ and its inverse satisfy
\begin{eqnarray} \label{ZZinv}
\int d\hat{s}' \;
Z^{-1}_{B}(\hat{s}''-\hat{s}',\mu)Z_{B}(\hat{s}'-\hat{s},\mu)=\de(\hat{s}''-\hat{s})
\ . 
\end{eqnarray}
From the $\mu$ independence of ${\cal B}^{\rm bare}$ one obtains the
renormalization group equation
\begin{eqnarray} \label{Brge}
\mu\fr{d}{d\mu}\cB(\hat{s},\mu) \ugu 
  \int d\hat{s}' \;\ga_B(\hat{s}-\hat{s}',\mu)\: \cB(\hat{s}',\mu)\ ,
\end{eqnarray}
where the anomalous dimension is
\begin{align}
\ga_B(\hat{s},\mu) 
 &= -\int d\hat{s}' \; Z^{-1}_{B}(\hat{s}-\hat{s}',\mu)\mu\fr{d}{d\mu} Z_{B}(\hat{s}',\mu)
= \int d\hat{s}' \; Z_{B}(\hat{s}-\hat{s}',\mu)\mu\fr{d}{d\mu} Z^{-1}_{B}(\hat{s}',\mu)
\,.
\end{align}
Since $\gamma_B(\hat s,\mu)$ is real we can also simply take the imaginary part
of Eq.~(\ref{Brge}) to obtain the renormalization group equation for $B(\hat
s,\mu)$.  In the $\overline {\rm MS}$ scheme $Z_B$ and $Z_B^{-1}$ have the
$\epsilon$ dependence
\begin{align}
  Z_B(\hat s,\mu) &= \delta(\hat s)
       +\sum_{k=1}^\infty \fr{1}{\epsilon^k} Z^{(k)}(\hat s,\mu) \,,
&  Z^{-1}_B(\hat s,\mu) &= \delta(\hat s)
       +\sum_{k=1}^\infty \fr{1}{\epsilon^k} \bar {Z}^{(k)}(\hat s,\mu) \,,
\end{align}
where $Z^{(k)}$ and $\bar Z^{(k)}$ are $\epsilon$ independent. Eq.~(\ref{ZZinv})
implies that $\bar Z^{(1)} = - Z^{(1)}$, and $\bar Z^{(k)} = - Z^{(k)} -
\sum_{j=1}^{k-1} \bar Z^{(j)} \otimes Z^{(k-j)}$ for $k\ge 2$. Demanding that
$\gamma_B(\hat s,\mu)$ is finite as $\epsilon\to 0$ and using the
$\beta$-function equation
\begin{eqnarray}
  \mu\fr{d}{d\mu}\al_s(\mu)\ugu-2\epsilon \al_s(\mu)+\bt[\al_s] \,,
\end{eqnarray}
gives the standard dimensional regularization result that the anomalous
dimension is determined by the residue of the $1/\epsilon$ term at any order in
perturbation theory,
\begin{align} \label{eq:gamma}
   \ga_B(\hat s,\mu) 
     &= - 2 \al_s\fr{\partial}{\partial\al_s}{\bar Z}^{(1)}(\hat s,\mu)
      =  2 \al_s\fr{\partial}{\partial\al_s} {Z}^{(1)}(\hat s,\mu)
     \,.
\end{align}
We find that the higher $1/\epsilon$ poles lead to the consistency equations
[$\ell \ge 1$]
\begin{align} \label{eq:consistency}
   2\al_s\fr{\partial}{\partial\al_s} Z_B^{(\ell +1)}(\hat s,\mu) 
  &= \Big(\mu\fr{\partial}{\partial\mu}+ \bt\fr{\partial}{\partial \al_s}\Big)
 Z_B^{(\ell)}(\hat s,\mu)
  \ +\ \sum_{k=1}^\ell \int\! d\hat s'\
   {\bar Z}_B^{(k)}(\hat s\minus \hat s',\mu)\bigg[-2\al_s\fr{\partial}{\partial \al_s} 
   Z_B^{(\ell-k+1)}(\hat s',\mu)
   \nn\\ 
  & \quad +
   \Big(\mu\fr{\partial}{\partial\mu}+ \bt\fr{\partial}{\partial \al_s}\Big)
     Z_B^{(\ell-k)}(\hat s',\mu)\bigg] \,,
\end{align}
where for convenience we let $Z_B^{(0)}(\hat s,\mu)=\delta(\hat s)$.  The result
in Eq.~(\ref{eq:consistency}) agrees with the form of the counterterm
consistency condition derived in Ref.~\cite{Becher:2005pd} for the heavy-quark
shape function.

At any order in perturbation theory the anomalous dimension in
Eq.~(\ref{eq:gamma}) has the form
\begin{eqnarray} \label{gBsform}
 \ga_B(\hat s,\mu)\ugu 
  -2 \Ga^{\rm c}[\alpha_s]\: \frac{1}{\mu} \left[\fr{\mu\, \theta(\hat s)}{\hat s}\right]_+
 +  \ga[\alpha_s]\, \delta(\hat s) \ ,
\end{eqnarray}
where our definition of this plus-function is given below in Eq.~(\ref{cLk}).
Here $\Gamma^{\rm c}[\alpha_s]$ and $\gamma[\alpha_s]$ have an infinite power
series expansions in $\alpha_s$ that starts at linear order.  $\Gamma^{\rm
  c}[\alpha_s]$ is the cusp-anomalous
dimension~\cite{Polyakov:1979gp,Dotsenko:1979wb,Brandt:1981kf,Korchemsky:1987wg},
while $\gamma[\alpha_s]$ is the part of the anomalous dimension that is
unrelated to the cusp.  In position space the renormalization group equation and
anomalous dimension are simpler,
\begin{align} \label{gByform}
  \mu \frac{d}{d\mu} \tilde B(y,\mu)  &= \tilde \gamma_B(y,\mu)\, \tilde {
   B}(y,\mu) \,,
  & \tilde\gamma_B(y,\mu) & = 2 \Gamma^{\rm c}[\alpha_s]\: \ln\big(i
  e^{\gamma_E} y \, \mu\big) + \gamma[\alpha_s] \,.
\end{align}
The form of the anomalous dimensions given in Eqs.~(\ref{gBsform}) and
(\ref{gByform}) is guaranteed to all orders in perturbation theory by a theorem
regarding the renormalization of Wilson-line operators with cusps proven in
Ref.~\cite{Brandt:1981kf,Korchemsky:1987wg}, which ensures it can not have
dependence on the position space variable other than the $\ln(y\mu)$. To solve
Eq.~(\ref{gByform}) one first writes $\ln(i e^{\gamma_E} y\mu)=
\ln(ie^{\gamma_E} y\mu_0) + \ln(\mu/\mu_0)$, then rewrites $\ln(\mu/\mu_0) =
\int_{\alpha_s{\mu_0}}^{\alpha_s(\mu)} d\alpha'/\beta[\alpha']$, and finally
integrates with a change of variables $d\ln\mu = d\alpha/\beta[\alpha]$. This
gives a solution that connects the result at the scale $\mu_0$ to that at the
scale $\mu$,
\begin{align} \label{Byrun}
  \tilde {B}(y,\mu) & = e^{K(\mu,\mu_0)}\,
    \big(ie^{\gamma_E} y\,\mu_0 \big)^{\omega(\mu,\mu_0)}\
    \tilde {B}(y,\mu_0) \,,
\end{align}
where the two evolution functions are
\begin{align} \label{wLfull}
  \omega(\mu,\mu_0) &= 2
  \int_{\alpha_s(\mu_0)}^{\alpha_s(\mu)} \frac{d\alpha}{\beta[\alpha]}\,
  \Gamma^c[\alpha] \,,
 &  K(\mu,\mu_0) &= 
  \int_{\alpha_s(\mu_0)}^{\alpha_s(\mu)}
  \frac{d\alpha}{\beta[\alpha]}\: \gamma[\alpha]
  + 2 \int_{\alpha_s(\mu_0)}^{\alpha_s(\mu)}
  \frac{d\alpha}{\beta[\alpha]}\, \Gamma^{\rm c}[\alpha]
  \int_{\alpha_s(\mu_0)}^\alpha \frac{d\alpha'}{\beta[\alpha']}  \,.
\end{align}
Taking the Fourier transform of Eq.~(\ref{Byrun}) then gives the solution to the
momentum space renormalization group evolution (RGE) equation
\begin{align} \label{Brgesol}
  {B}(\hat s,\mu) &= \int_{-\infty}^{+\infty} \!\! d\hat s'\
   U_B(\hat s-\hat s',\mu,\mu_0)\ {B}(\hat s',\mu_0)\,,
  &  U_B(\hat s-\hat s',\mu,\mu_0) &= \frac{e^{K}
      \big(e^{\gamma_E}\big)^{\omega}}{\mu_0\, \Gamma(-\omega)}\: 
     \bigg[ \frac{ \mu_0^{1+\omega}\, \theta(\hat s-\hat s')}
     {(\hat s-\hat s')^{1+\omega}}\bigg]_+ \,,
\end{align}
where $K=K(\mu,\mu_0)$ and $\omega=\omega(\mu,\mu_0)$. All results in this
subsection are valid to all orders in the $\alpha_s$ expansion, and can thus be
used to sum logs in $B$ at leading log (LL), next-to-leading log (NLL), NNLL, and beyond. To our knowledge, the
results in Eq.~(\ref{wLfull}) and (\ref{Brgesol}) were first derived for the
$B$-meson shape function, first at one-loop in Ref.~\cite{Balzereit:1998yf} and
then to all-orders in Ref.~\cite{Neubert:2004dd}.

\subsection{NNLO Result for ${\cal B}(\hat s,\mu)$}

To obtain results at NNLO we consider the $\alpha_s$ expansion of quantities
defined in subsection~\ref{subsec:ren}. The bare and renormalized jet functions
can be written as
\begin{align} \label{BB}
 \cB^{\rm{bare}}(\hat{s})
    &=\sum_{j=0}^{\infty} \Big[ \frac{\al_s^{\rm bare}}{\pi} \Big]^j \
    \cB^{\rm{bare}}_{j}(\hat{s})
  \,, 
 &  {\cal B}(\sh,\mu )&= \sum_{j=0}^\infty
   \Big[ \frac{\al_s(\mu)}{\pi} \Big]^j\ {\cal B}_{j}(\sh,\mu )
  \,.
\end{align}
We also expand the anomalous dimensions and $\beta$-function as
\begin{align} \label{Gammas}
   \Ga^{\rm c}[\alpha_s] &= 
      \sum_{j=0}^\infty \Ga^{\rm c}_j \Big[\fr{\al_s(\mu)}{4\pi}\Big]^{j+1} 
     \,, 
  & \ga[\alpha_s] &=
    \sum_{j=0}^\infty \ga_j \Big[\fr{\al_s(\mu)}{4\pi}\Big]^{j+1} 
   \,,
 & \beta[\alpha_s] &=
  - 2 \alpha_s(\mu) \sum_{n=0}^\infty 
  \beta_n\left[\fr{\al_s(\mu)}{4\pi}\right]^{n+1} \,,
\end{align}
where up to three-loop order~\cite{Gross:1973id,Politzer:1973fx,Jones:1974mm,Caswell:1974gg,Tarasov:1980au,Larin:1993tp}
\begin{align}
  \beta_0 &= \frac{11 C_A}{3} -\frac{2n_f}{3} \,,
    \hspace{4cm}
   \beta_1 = \frac{34 C_A^2}{3}  - \frac{10C_A n_f}{3}  - 2 C_F n_f \,, \nn\\
  \beta_2 &= \frac{2857 C_A^3}{54} +\Big( C_F^2 -\frac{205 C_F C_A}{18} -
  \frac{1415 C_A^2}{54}\Big) n_f + \Big(\frac{11 C_F}{9}+ \frac{79 C_A}{54}\Big)
  n_f^2 \,.
\end{align}
To incorporate the $\delta m$ term from the Lagrangian in Eq.~(\ref{eq:LHQET})
we evaluate ${\cal B}_j^{\rm bare}(\hat s-2 \delta m)$ and then expand in
$\alpha_s(\mu)$ with 
\begin{align}
  \delta m = \sum_{j=1}^\infty \Big[\frac{\alpha_s(\mu)}{\pi} \Big]^j\: \delta m_j(\mu)
    = \frac{\alpha_s(\mu)}{\pi}\: \delta m_1(\mu)
   + \frac{\alpha_s^2(\mu)}{\pi^2}\:  \delta m_2(\mu) +\ldots \,.
\end{align}
This is simpler than treating $\delta m$ as a Feynman rule insertion, and
equivalent.  The bare and renormalized couplings are related by
\begin{align} \label{asbare}
 \alpha_s^{\rm bare} &= \iota^\epsilon \mu^{2\epsilon}\alpha_s(\mu) Z_g^2 \,,
  \qquad\qquad\quad
  \iota \,\equiv\, \exp(\gamma_E)/(4\pi) \,,
\end{align}
where $Z_g$ is the $Z$-factor for the strong coupling and the iota dependence,
$\iota^\epsilon$, ensures we are in the $\overline {\rm MS}$ scheme rather than
the ${\rm MS}$ scheme. To determine the renormalized jet function we expand the
counterterms as
\begin{align}
  Z^{-1}_B(\hat s,\mu) &=
  \de(\hat s)+\sum_{j=1}^\infty \Big[\frac{\al_s(\mu)}{\pi}\Big]^j
  \: \bar {Z}_j(\hat s,\mu)=
  \de(\hat s)+\sum_{k=1}^\infty\sum_{j=1}^\infty \frac{1}{\epsilon^k}\,
   \Big[\frac{\al_s(\mu)}{\pi} \Big]^j\,
  {\bar Z}^{(k)}_j(\hat s,\mu)\ , \\
 Z_g &= 1 + \sum_{j=1}^\infty \Big[\frac{\alpha_s(\mu)}{\pi}\Big]^j\: z_{g j}  
 \,.\nn
\end{align}
Using this notation, converting $\alpha_s^{\rm bare}$ to $\alpha_s(\mu)$ with
Eq.~(\ref{asbare}), and then equating powers of $\alpha_s(\mu)$ in
Eq.~(\ref{eq:bren}) these expansions determine the renormalized ${\cal B}_j(\hat
s,\delta m, \mu)$. The tree, one-loop, and two-loop coefficients are
respectively,
\begin{align} \label{twoloopren}
{\cal B}_{0}(\sh,\delta m,\mu) &= {\cal B}^{\rm{bare}}_{0}(\sh)
  \,, \\
{\cal B}_{1}(\sh,\delta m,\mu) &= 
  \iota^\epsilon \mu^{2\epsilon} {\cal B}^{\rm{bare}}_{1}(\sh)  
  +   \int d\sh' \, \bar Z_1(\sh\minus \sh',\mu)\ {\cal B}^{\rm{bare}}_{0}(\sh') 
  - 2\, \delta m_1\, \frac{d{\cal B}_0(\hat s,\mu)}{d\hat s} 
  \,,\nn \\
{\cal B}_{2}(\sh,\delta m,\mu) &= 
  \iota^{2\epsilon} \mu^{4\epsilon} {\cal B}^{\rm{bare}}_{2}(\sh)
   + 2 z_{g1}\, \iota^\epsilon \mu^{2\epsilon} {\cal B}^{\rm bare}_1(\sh) 
   + \int d\sh' \, \bar Z_1(\sh\minus \sh',\mu)\ 
      \iota^\epsilon \mu^{2\epsilon} {\cal B}^{\rm{bare}}_{1}(\sh')  
   + \int d\sh' \, \bar Z_2(\sh\minus \sh',\mu)\ {\cal B}^{\rm{bare}}_{0}(\sh') 
  \nn\\
  & - 2\, \delta m_2\, \frac{d{\cal B}_0(\hat s,\mu)}{d\hat s} 
    + 2\, (\delta m_1)^2\, \frac{d^2{\cal B}_0(\hat s,\mu)}{d^2\hat s} 
    - 2\, \delta m_1\, \frac{d{\cal B}_1(\hat s,\mu)}{d\hat s} 
  \,, \nn
\end{align}
where we used a subscript notation for the $[\alpha_s(\mu)/\pi]^j$ expansion
coefficients as in Eq.~(\ref{BB}). The one and two-loop $Z$-factors have
terms
\begin{align} \label{ZZ12}
 \bar Z_1 &= \frac{1}{\epsilon} \bar Z_1^{(1)} + \frac{1}{\epsilon^2} \bar Z_1^{(2)}
   \,,
& \bar Z_2 &= \frac{1}{\epsilon} \bar Z_2^{(1)} + \frac{1}{\epsilon^2} \bar
 Z_2^{(2)} + \frac{1}{\epsilon^3} \bar Z_2^{(3)} + \frac{1}{\epsilon^4} \bar Z_2^{(4)}
 \,,
\end{align}
where the coefficients $\bar Z_j^{(k)}$ are defined so that the ${\cal B}_j(\hat
s,\delta m,\mu)$ are finite as $\epsilon\to 0$.

\begin{figure*}[t!]
  \centerline{
   \mbox{\epsfxsize=4truecm \hbox{\epsfbox{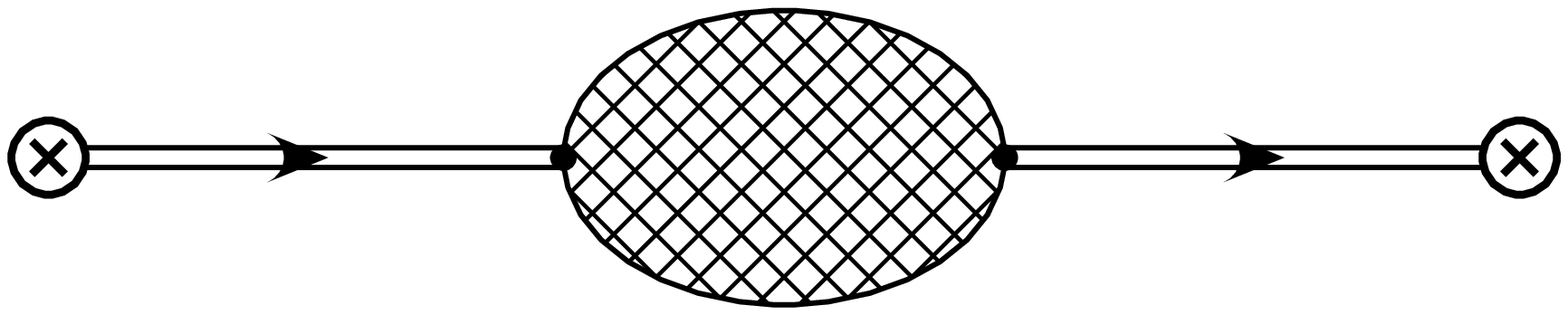}} } \quad
   \mbox{\epsfxsize=4truecm \hbox{\epsfbox{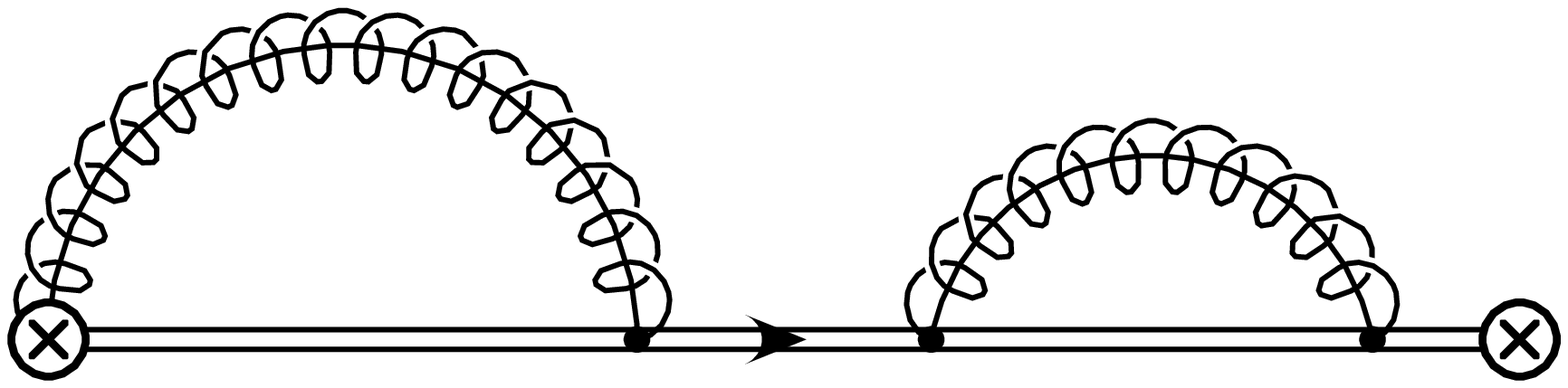}} } \quad
   \mbox{\epsfxsize=4truecm \hbox{\epsfbox{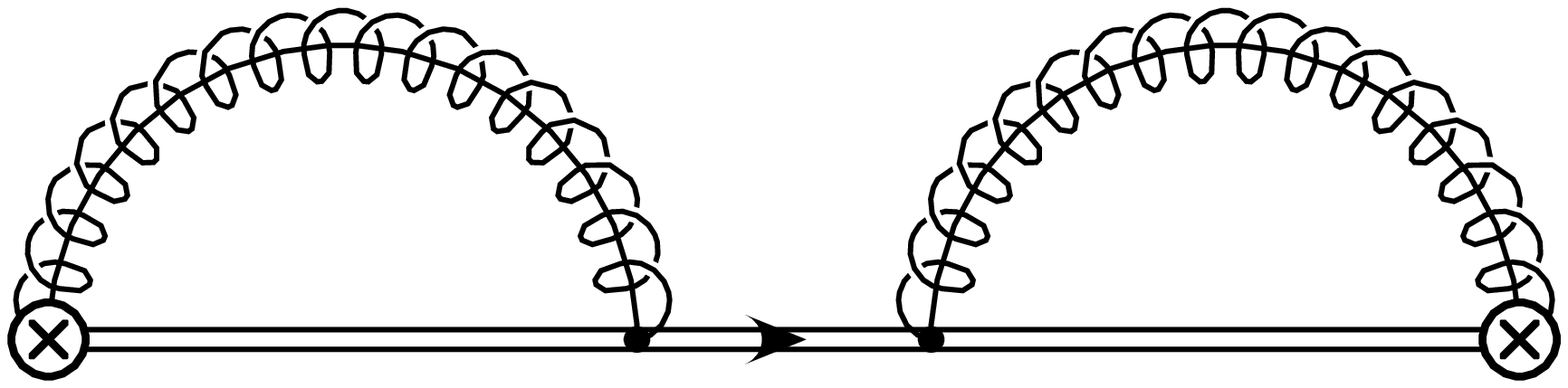}} } \quad
   \mbox{\epsfxsize=4truecm \hbox{\epsfbox{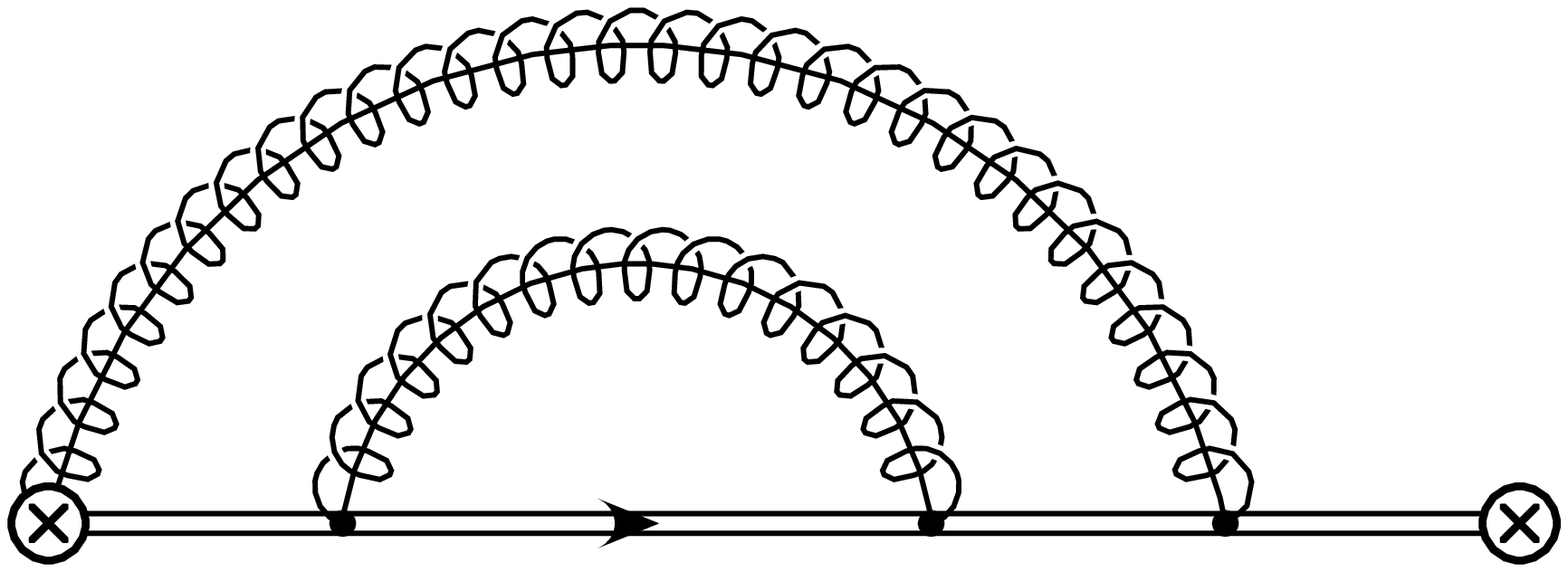}} }
  }
  \vspace{0.3cm}
  \centerline{
     \mbox{\epsfxsize=4truecm \hbox{\epsfbox{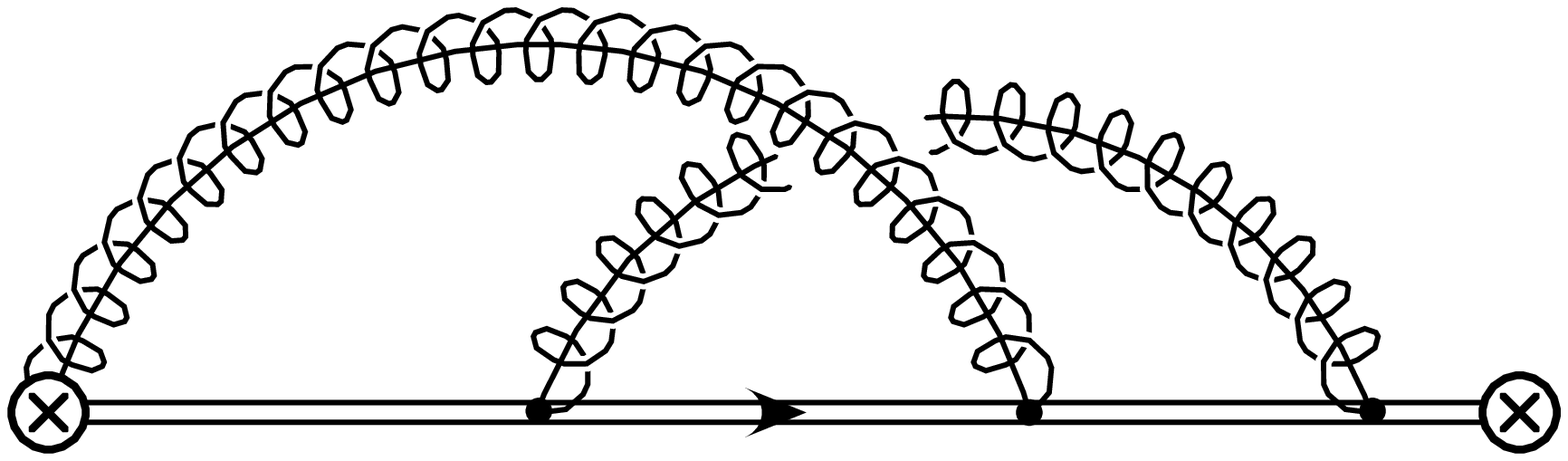}} } \quad
     \mbox{\epsfxsize=4truecm \hbox{\epsfbox{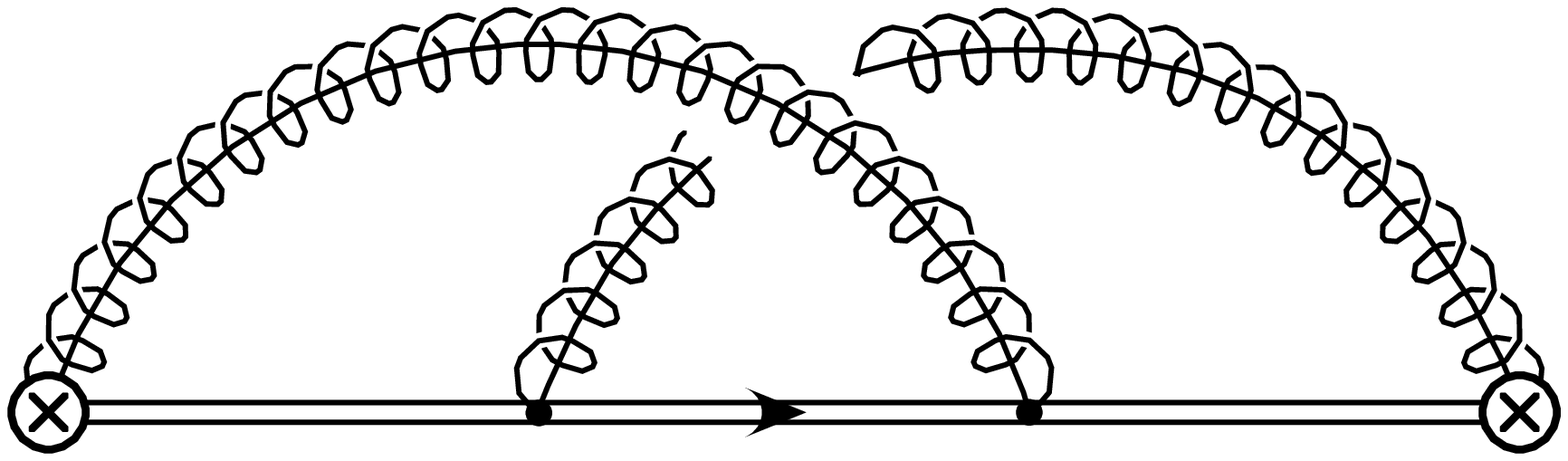}} } \quad
     \mbox{\epsfxsize=4truecm \hbox{\epsfbox{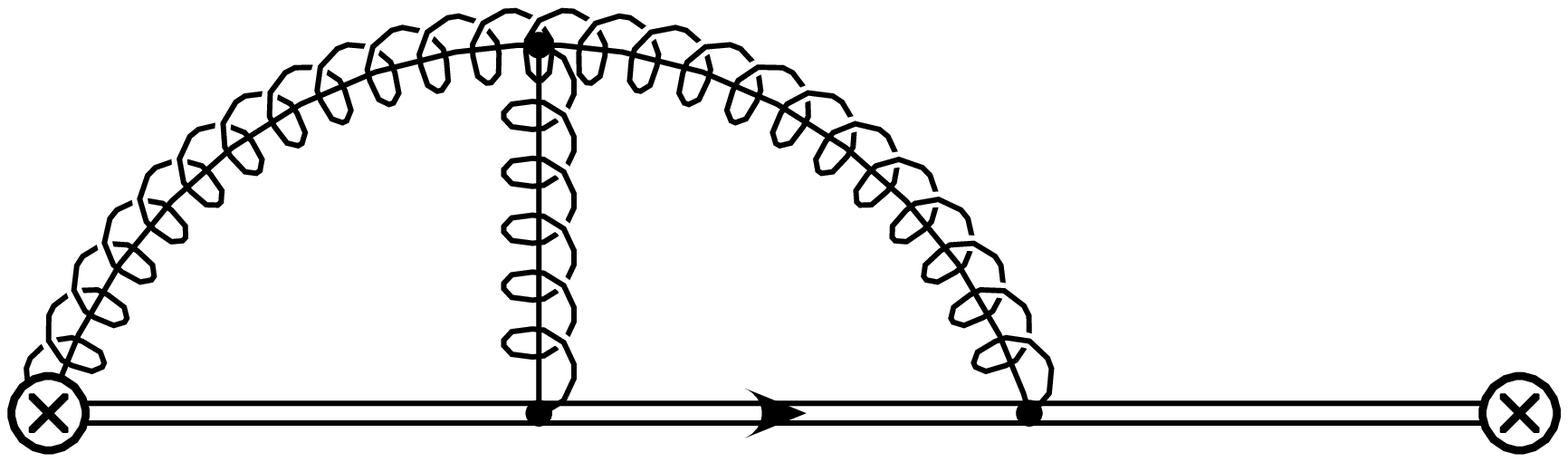}} } \quad
     \mbox{\epsfxsize=4truecm \hbox{\epsfbox{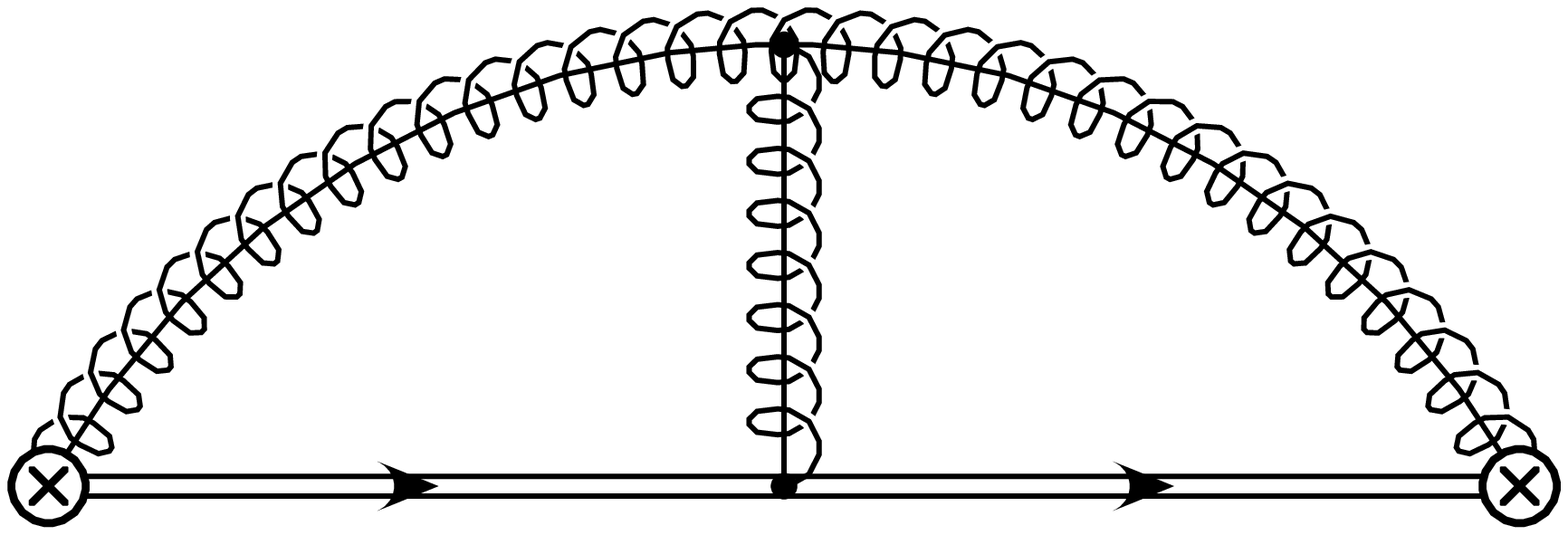}} } 
   }
  \vspace{0.3cm}
  \centerline{
    \mbox{\epsfxsize=4truecm \hbox{\epsfbox{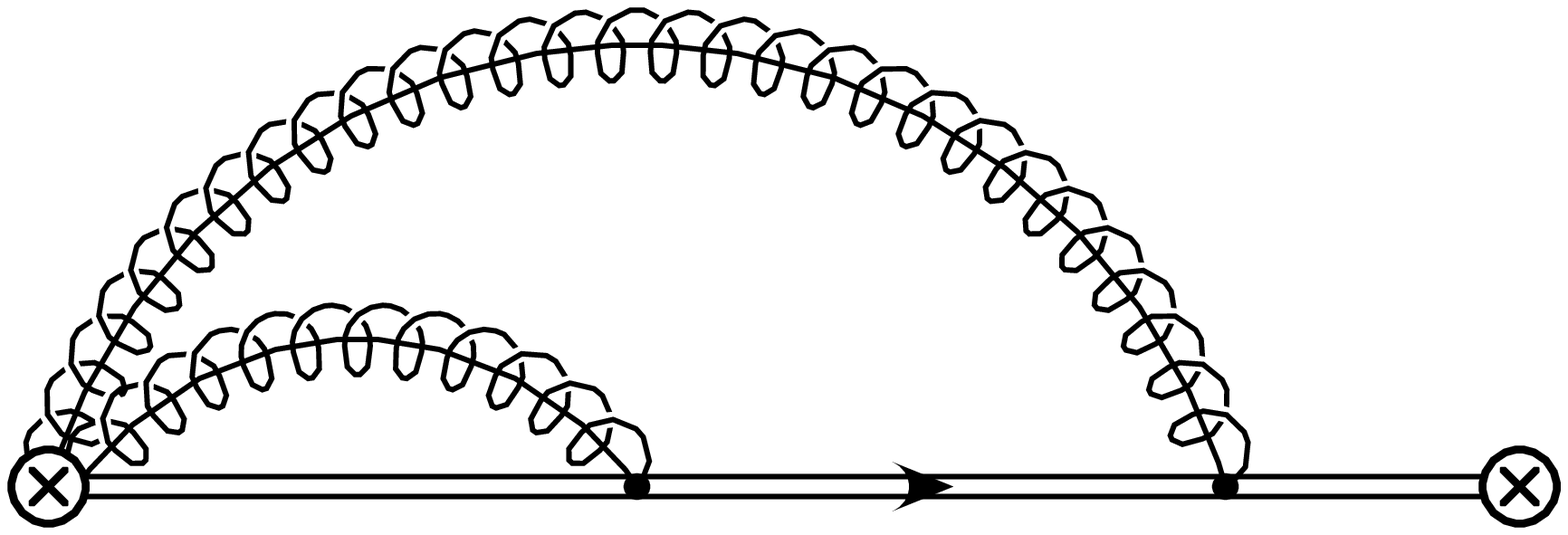}} } \quad
    \mbox{\epsfxsize=4truecm \hbox{\epsfbox{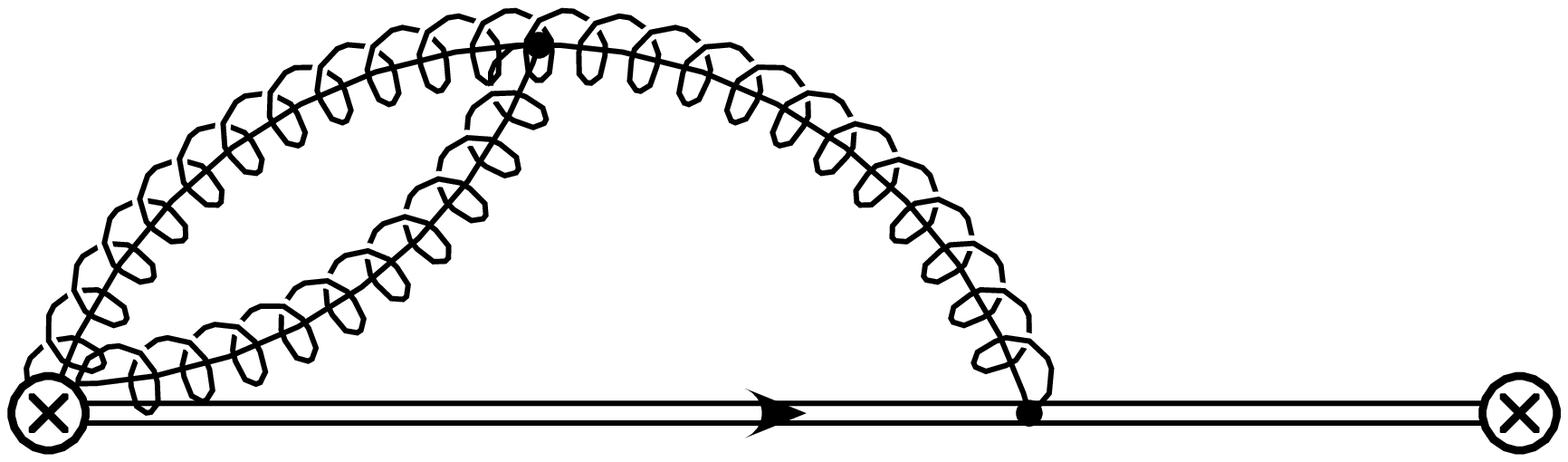}} } \quad
     \mbox{\epsfxsize=4truecm \hbox{\epsfbox{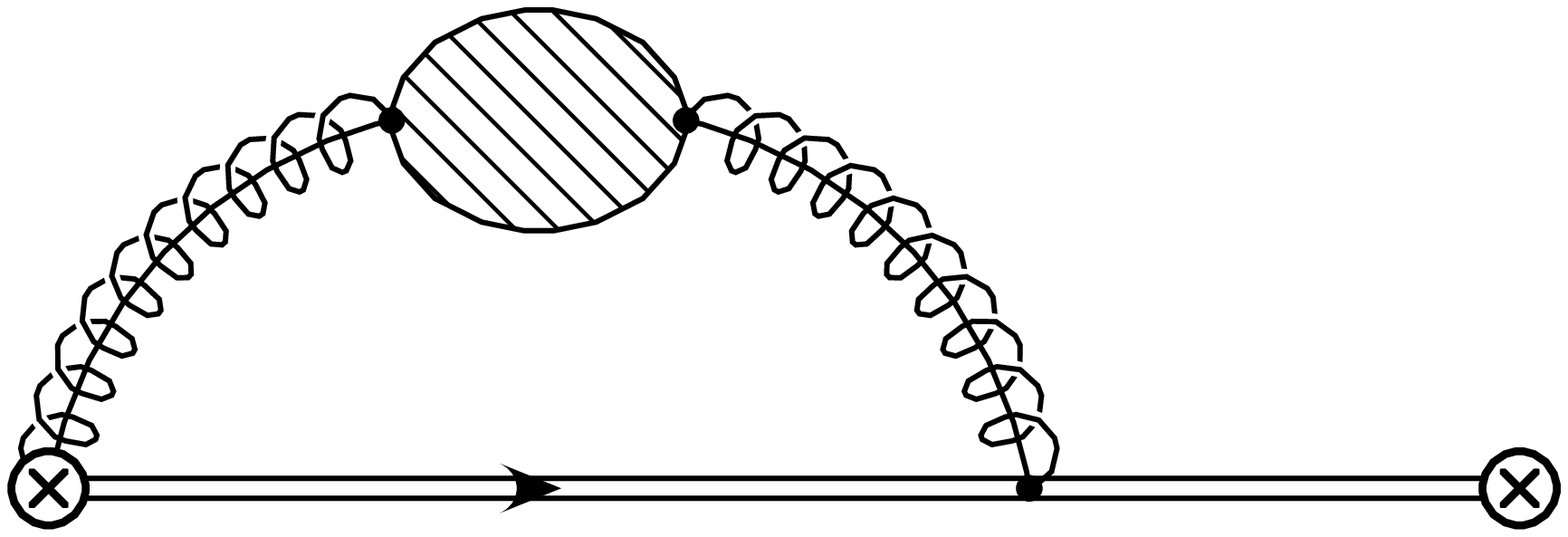}} } \quad
     \mbox{\epsfxsize=4truecm \hbox{\epsfbox{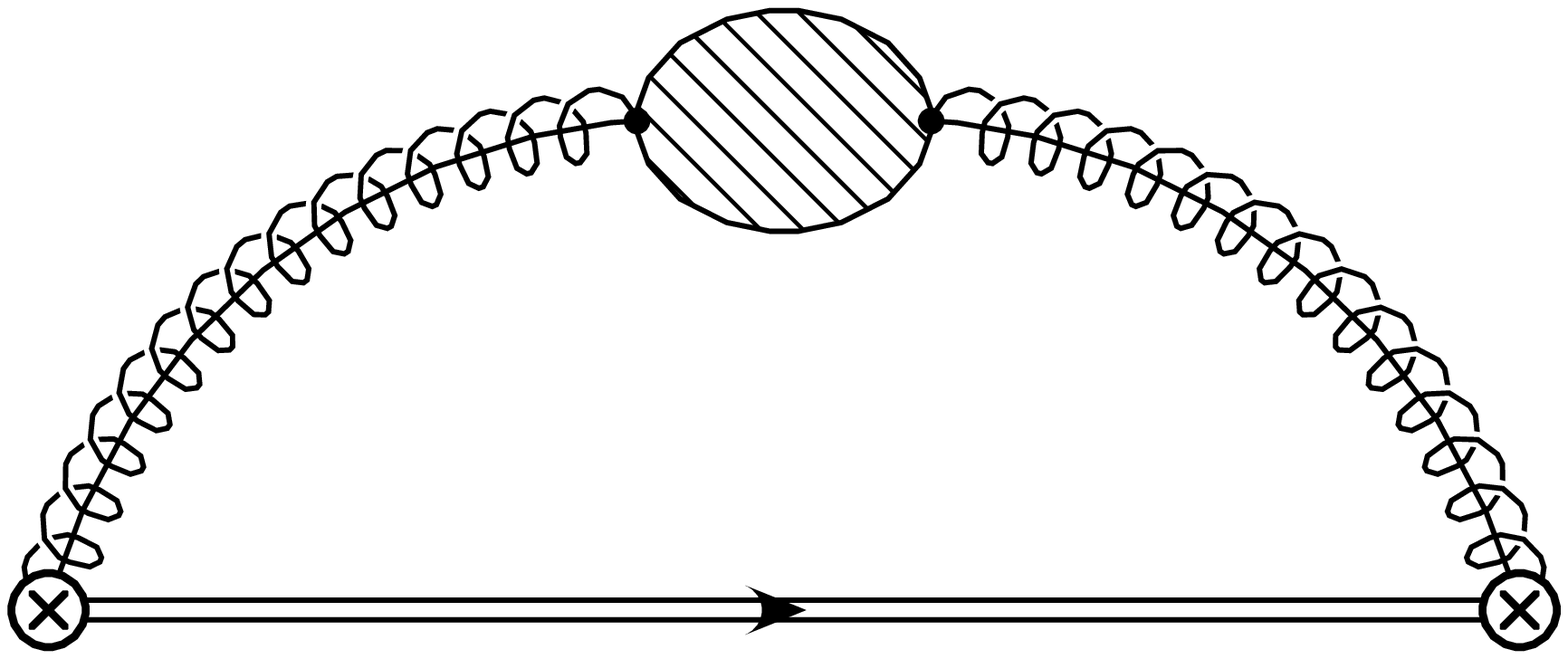}} }
    }
  \vspace{0.4cm}
  \centerline{
    \mbox{\epsfxsize=4truecm \hbox{\epsfbox{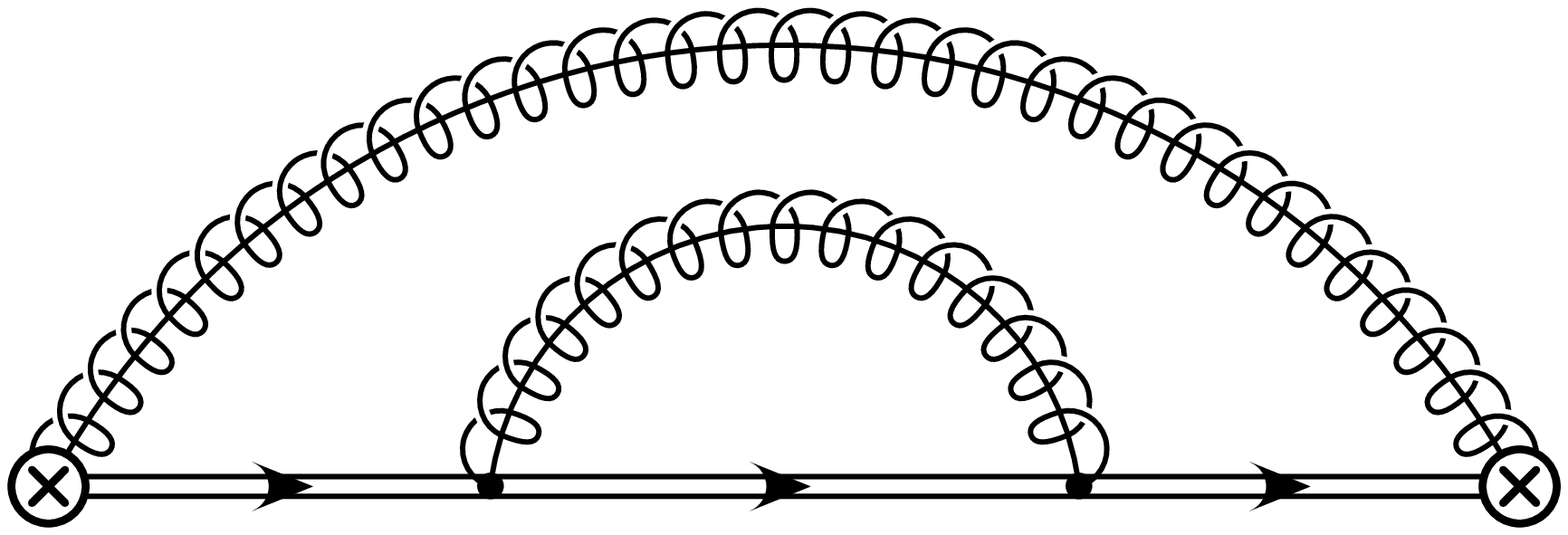}} } \quad
    \mbox{\epsfxsize=4truecm \hbox{\epsfbox{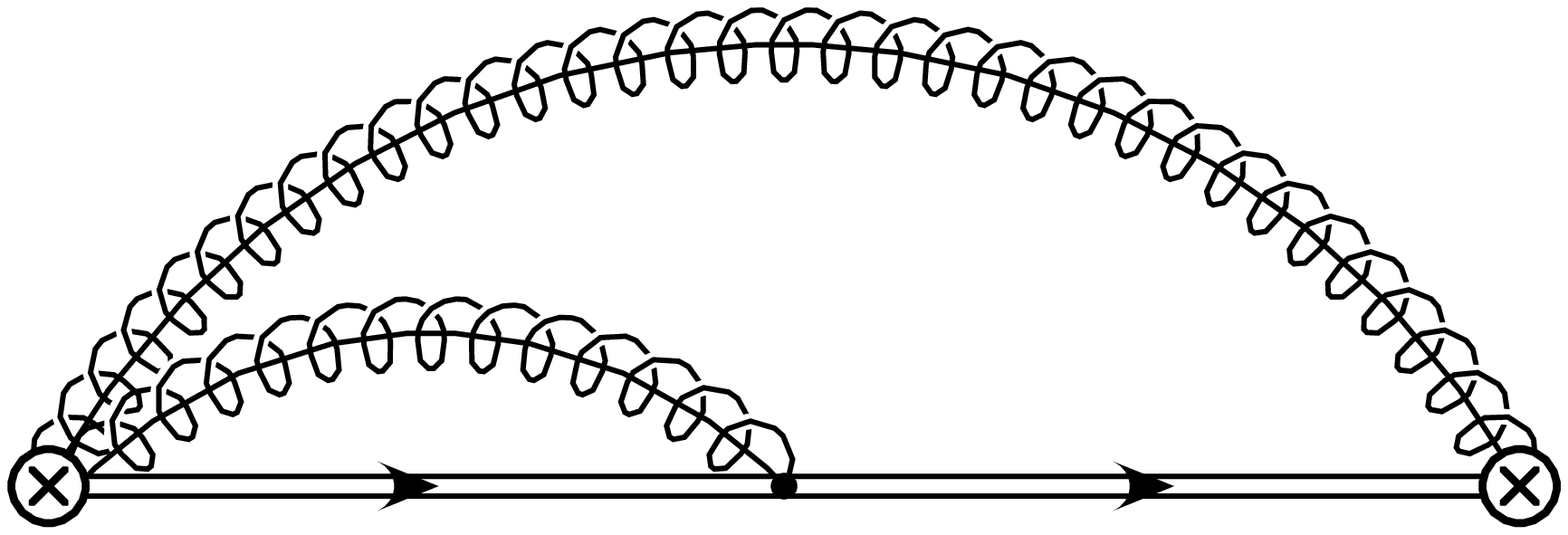}} } \quad
     \mbox{\epsfxsize=4truecm \hbox{\epsfbox{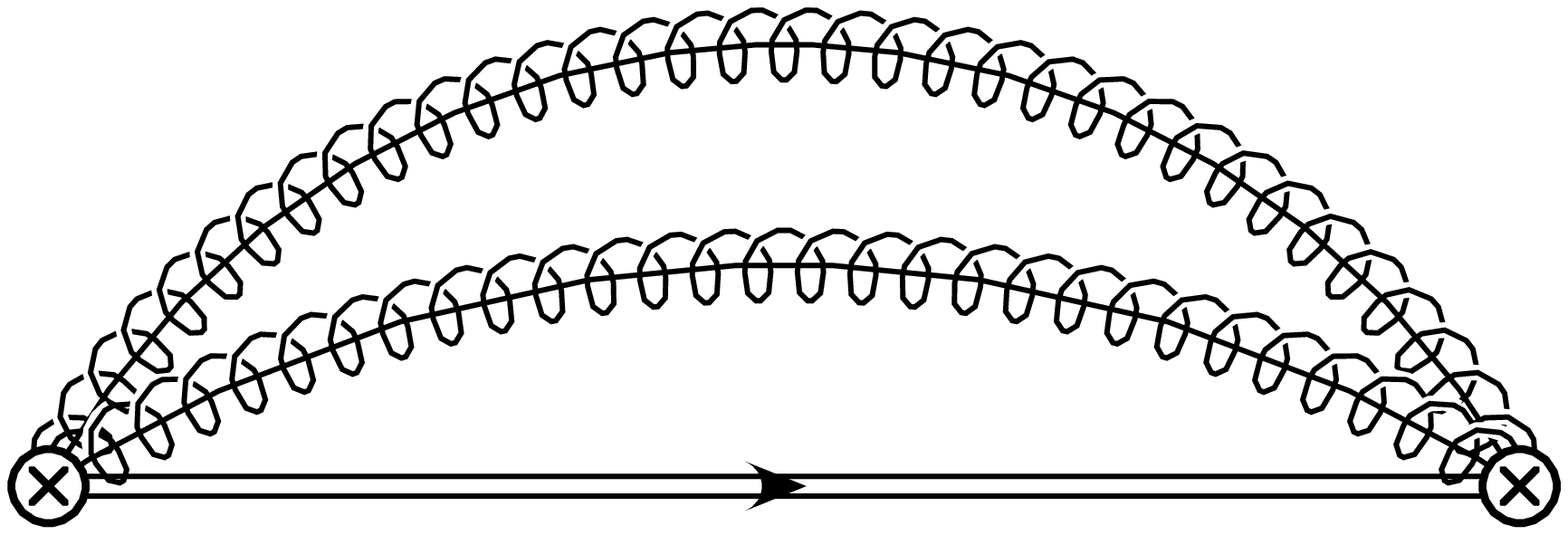}} } \quad
     \mbox{\epsfxsize=4truecm \hbox{\epsfbox{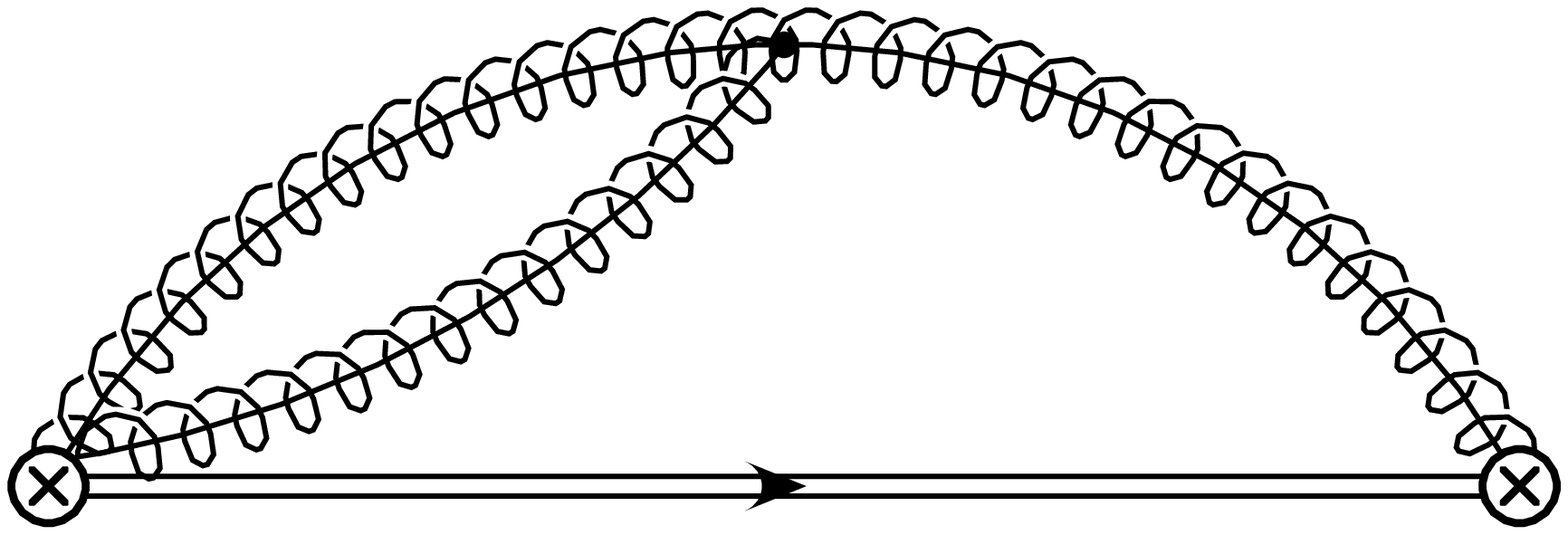}} }
    }
  \vskip0.2cm
  \caption{Graphs for ${\cal B}^{\rm bare}_2(\hat s)$. Gluons from an
    $\otimes$ are from the Wilson lines, the hatched blob is the two-loop vacuum
    polarization of the heavy quark, and the blobs with diagonal lines include
    all one-loop vacuum polarization graphs for the gluon. Numbering the graphs
    from 1 to 16  from left-to-right and top-to-bottom, we note that graphs 2, 4,
    5, 7, 9, 10, 11, 14, and 16 implicitly also stand  for their left-right
    symmetric counterpart.}
   \label{fig:2loop}
\end{figure*}

The results for $\cB_0$ and $\cB_1$ were obtained in
Ref.~\cite{Fleming:2007qr,Fleming:2007xt}. In an arbitrary mass-scheme we have
\begin{align} \label{cB01ren}
 m\, \cB_0(\sh,\delta m,\mu) &=  L^0
\,,
& m \cB_1(\sh,\delta m, \mu) &= 
   C_F \Big\{ L^2+L^1+\Big(1+\fr{5\pi^2}{24}\Big)L^0 \Big\}
   - {2\, \delta m_1} (L^0)^\prime
  \,,
\end{align}
where the prime denotes a derivative with respect to $\hat s$, and for
convenience we have defined
\begin{eqnarray}  \label{Lk}
 L^k= \frac{1}{\pi(-\sh-i0)}\: \ln^k\!\Big(\fr{\mu}{-\sh-i 0}\Big)\, .
\end{eqnarray}
The corresponding two-loop result, $\cB_2$, is one of the main results of this
paper and involves the Feynman diagrams shown in Fig.~\ref{fig:2loop}.  Details
of the computation of ${\cal B}_2$ using Eq.~(\ref{twoloopren}) in Feynman gauge
are given in Appendix~\ref{app:2loop}. To summarize, we use the computation of
Broadhurst and Grozin~\cite{Broadhurst:1991fz,Grozin:2000cm} for the divergent
and finite terms of the two-loop heavy quark propagator (the first graph in
Fig.~\ref{fig:2loop}), and compute the remaining Feynman diagrams directly. We
treat the quarks other than top as massless, with $n_f$ such flavors, and thus
do not include effects due to the $b$-quark mass in vacuum polarization
diagrams. We have also confirmed that the resulting $\bar Z_j^{(k)}$ satisfy the
counterterm consistency conditions in Eq.~(\ref{eq:consistency}).
The final result for the renormalized two-loop matrix element is
\begin{align} \label{cB2ren}
m\, \cB_2(\sh,\delta m,\mu) &=
C_F^2 \bigg [  \frac{1}{2} L^4 + L^3  + \Big( \frac{3}{2} +\frac{13\pi^2}{24} \Big) L^2 
 + \Big( 1+\frac{13\pi^2}{24} - 4\zeta_3 \Big) L^1 
 + \Big(  \half  + \frac{7\pi^2}{24} + \frac{53\pi^4}{640} - 2\zeta_3 \Big) L^0  \bigg ] 
  \nn \\[2pt] 
 &  + C_F C_A \bigg [  \Big(  \frac{1}{3} - \frac{\pi^2}{12} \Big) L^2 
   + \Big(  \frac{5}{18}-\frac{\pi^2}{12} -\frac{5\zeta_3}{4}  \Big) L^1 
   + \Big(\! -\frac{11}{54} + \frac{5\pi^2}{48} -\frac{19\pi^4}{960}
   -\frac{5\zeta_3}{8} \Big) L^0
  \bigg] 
\nn \\[2pt]
 &  + C_F \beta_0 \bigg [ \frac{1}{6} L^3  + \frac{2}{3}L^2 
   + \Big(\frac{47}{36} +\frac{\pi^2}{12}  \Big) L^1  
   + \Big(\frac{281}{216} +\frac{23\pi^2}{192}-\frac{17\zeta_3}{48} \Big) L^0
  \bigg ]
 \nn\\[4pt]
 & - 2 \delta m_2 (L^0)^\prime + 2 (\delta m_1)^2 (L^0)^{\prime\prime}
   - 2 \delta m_1 \,C_F \bigg[
   L^2+L^1+\Big(1+\fr{5\pi^2}{24}\Big)L^0 \bigg]^\prime
  \,. 
\end{align}
One can pass from the function $\cB(\hat s,\delta m,\mu)$ to the distribution
$B(\hat s,\delta m,\mu)$ using Eq.~(\ref{eq:BcB}) and the formulas in
Eq.~(\ref{Impart}). This gives
\begin{align} \label{B2NNLO}
 m\, B(\sh,\delta m, \mu) &= \delta(\hat s) 
  + \frac{C_F\alpha_s(\mu)}{\pi} \bigg\{ 2 {\cal L}^1 - {\cal L}^0 
     + \Big(1-\frac{\pi^2}{8}\Big)\delta(\hat s) \bigg\} 
  - \frac{2\, \alpha_s(\mu)}{\pi} \, \delta m_1(\mu)\,
     \delta^\prime(\hat s) 
   \nn \\
&+ \frac{\alpha_s^2(\mu)}{\pi^2} \Bigg \{ 
C_F^2 \bigg [ 2 {\cL}^3 -3 \cL^2  + \Big( 3 - \frac{11\pi^2}{12} \Big) \cL^1 
 + \Big( \! - 1 + \frac{11\pi^2}{24}  +  4\zeta_3 \Big) \cL^0 
 + \Big(  \half  - \frac{5\pi^2}{24} + \frac{13\pi^4}{5760} - 2\zeta_3 \Big)
 \delta(\hat s) \bigg ] 
  \nn \\[2pt] 
& \qquad + C_F C_A \bigg [  \Big(  \frac{2}{3} - \frac{\pi^2}{6} \Big) \cL^1 
   + \Big( \! -\frac{5}{18}+\frac{\pi^2}{12} +\frac{5\zeta_3}{4}  \Big) \cL^0 
   + \Big( \! -\frac{11}{54} -\frac{\pi^2}{144} +\frac{23\pi^4}{2880}
   -\frac{5\zeta_3}{8} \Big) \delta(\hat s)
  \bigg] 
\nn \\[2pt]
& \qquad + C_F \beta_0 \bigg [ \! -\frac{1}{2} \cL^2  + \frac{4}{3} \cL^1 
   + \Big(\!-\frac{47}{36} +\frac{\pi^2}{12}  \Big) \cL^0  
   + \Big(\frac{281}{216} -\frac{59\pi^2}{576}-\frac{17\zeta_3}{48} \Big)
   \delta(\hat s)
  \bigg ] ~~\Bigg \} 
 \nn\\[2pt]
& -  \frac{2\,\alpha_s^2(\mu)}{\pi^2} \Bigg\{ 
   \delta m_2\: \delta^\prime(\hat s) - (\delta m_1)^2\: \delta^{\prime\prime}(\hat s)
   + \delta m_1\, C_F \bigg[
   2 (\cL^1)^\prime -(\cL^0)^\prime
  +\Big(1-\fr{\pi^2}{8}\Big) \delta^\prime(\hat s) \bigg] \ \Bigg\}
\,,
\end{align}
where for the log plus-functions we use the notation
\begin{align} \label{cLk}
 {\cal L}^k = \frac{1}{\mu}\left [ \frac{
      \theta(\sh)\ln^k(\sh/\mu)}{\sh/\mu} \right ]_+ 
   \equiv  \lim_{\xi \to 0}\ \frac{1}{\mu} \bigg[
  \frac{\theta(x\minus \xi)\ln^k x}{x} + 
   \delta(x\minus \xi) \: \frac{\ln^{k+1}\!\xi}{k+1} \bigg]_{x=\hat s/\mu} \, .
\end{align}
The results in Eqs.~(\ref{cB2ren}) and (\ref{B2NNLO}) are presented in an
arbitrary mass scheme, which is specified by the choice for the coefficients
$\delta m_1$ and $\delta m_2$. An appropriate mass-scheme for top-jet cross
sections is described below in section~\ref{sect:jetmass}.  In order to obtain
the distribution $B(\hat s,\delta m,\Gamma_t,\mu)$ with $\Ga_t\neq 0$ one can
input Eq.~(\ref{B2NNLO}) into the integral with the Breit-Wigner in
Eq.~(\ref{eq:dr}). However the simpler method, which we use below in
section~\ref{sect:results}, is to shift $\sh\rightarrow\sh+i\Ga_t$ in
Eqs.~(\ref{cB01ren},\ref{cB2ren}) and then take the imaginary part as in
Eq.~(\ref{eq:bbbb}).
 
From the renormalization constants $Z_1^{(1)}$ and $Z_2^{(1)}$ given in
Appendix~\ref{app:2loop} we also obtain the anomalous dimension terms in
Eq.~(\ref{Gammas}). The cusp anomalous dimension is known up to three-loop order
$\Gamma_{0,1,2}^{\rm c}$~\cite{Korchemsky:1987wg,Moch:2004pa}, and we have
confirmed that our two-loop analysis reproduces the expected result for the
two-loop cusp coefficient, $\Gamma_1^{\rm c}$. For ${\cal B}$ the one-loop
anomalous dimension $\gamma_0$ has been calculated in
Ref.~\cite{Fleming:2007xt}. The two-loop anomalous dimension $\gamma_1$ is
obtained from our calculation of $\bar Z_2^{(1)}$ in
Eq.~(\ref{Zt2}).\footnote{It turns out that the piece of $\ga_1$ proportional to
  $C_F n_f$ is the analog of a contribution in the analysis of a scalar field
  theory made in Ref.~\cite{Beneke:2003xh,Beneke:2004km}. Suitably translated to
  the QCD case their computation agrees with the $C_Fn_f$ term in our $\ga_1$.
  The non-abelian $C_F C_A$ term of $\ga_1$ is original to our work.}  We list
here all the pieces needed for our analysis,
\begin{align}
  \Ga^{\rm c}_0 &=4 C_F \,, \qquad 
 \Ga_1^{\rm c} = \Big( \fr{268}{9} -\fr{4\pi^2}{3} \Big)C_F C_A  - 
   \fr{40}{9} C_F n_f ,
  \nn\\
  \Ga^{\rm c}_2 &= \!
   \Big[\frac{490}{3}\minus \frac{536 \pi ^2}{27}\plus \frac{44 \pi ^4}{45}
      \plus \frac{88 \zeta_3}{3}\Big] C_F C_A^2
   +\!\Big[\frac{80 \pi ^2\minus 836}{27}-\frac{112 \zeta_3}{3}\Big] C_F n_f C_A
   +\!\Big[32 \zeta_3 \minus \frac{110}{3}\Big]C_F^2 n_f
   \!-\! \frac{16 C_F n_f^2}{27},
  \nn\\
\ga_0 &= 4 C_F\ ,\qquad 
\ga_1 =  \Big[\fr{1396}{27}-\fr{23\pi^2}{9} -20\zeta_3\Big]  C_F C_A
   +   \Big[\fr{2\pi^2}{9}-\fr{232}{27}\Big] C_F  n_f
  \,. 
\end{align}
To resum the large logs in the jet function to NNLL order we need to use these
coefficients in the NNLL results for the evolution functions $\omega(\mu,\mu_0)$
and $K(\mu,\mu_0)$. To NNLL order solving Eq.~(\ref{wLfull}) gives
\begin{align} \label{wKNNLL}
   \omega(\mu,\mu_0) &= 
     -\frac{ \Gamma_0^{\rm c} }{\beta_0} \Bigg\{ \ln(r)  
     + \bigg(\frac{\Gamma_1^{\rm c}}{\Gamma_0^{\rm c}}
     -\frac{\beta_1}{\beta_0}\bigg) \frac{\alpha_s(\mu_0)}{4\pi}\, (r-1) 
    + \bigg( \frac{\Gamma_2^{\rm c}}{\Gamma_0^{\rm c}} - \frac{\beta_1
      \Gamma_1^{\rm c}}{\beta_0 \Gamma_0^{\rm c}}
     -\frac{\beta_2}{\beta_0} +\frac{\beta_1^2}{\beta_0^2}
     \bigg) \frac{\alpha_s^2(\mu_0)}{32\pi^2} (r^2-1)
    \Bigg\}
     \,, \\[4pt]
  K(\mu,\mu_0) &=\! \frac{-2
         \pi\Gamma_0^{\rm c}}{\beta_0^2}
   \Bigg\{ \frac{\big( r\minus 1
       \minus r\ln r \big)}{ r\, \alpha_s(\mu_0)}  \! +
     \frac{\gamma_0\,\beta_0}{4\pi\Gamma_0^{\rm c}} \ln r
     \plus \bigg(\frac{\Gamma_1^{\rm c}}{\Gamma_0^{\rm c}}
     -\frac{\beta_1}{\beta_0}\bigg) \frac{(1\minus r\plus \ln r)}{4\pi}
     \plus \frac{\beta_1}{8\pi\beta_0} \ln^2 r 
   +\frac{\alpha_s(\mu_0)}{16\pi^2}\, \Bigg[
   \frac{(\beta_0\gamma_1\minus \beta_1\gamma_0)}{\Gamma_0^{\rm c}} (r\minus 1) 
   \nn\\
  & +
   \bigg( \frac{\beta_1\Gamma_1^{\rm c}}{\beta_0 \Gamma_0^{\rm c}}-\frac{\beta_2}{\beta_0}\bigg) 
   (1\minus r\plus r\ln r) +
   \bigg(\frac{\beta_2}{\beta_0}-\frac{\beta_1^2}{\beta_0^2}\bigg) (r\minus 1)\ln r
   -\bigg(\frac{\Gamma_2^{\rm c}}{\Gamma_0^{\rm c}}-\frac{\beta_1\Gamma_1^{\rm
       c}}{\beta_0\Gamma_0^{\rm
       c}}-\frac{\beta_2}{\beta_0}+\frac{\beta_1^2}{\beta_0^2}\bigg)
   \frac{(1\minus r)^2}{2}
   \Bigg] \Bigg\}\,,\nn
\end{align}
where $r=\alpha_s(\mu)/\alpha_s(\mu_0)$. Eq.~(\ref{wKNNLL}) determines the
evolution functions in terms of coefficients of the anomalous dimensions and
$\beta$-function. It agrees with the NNLL result given in the appendix of
Ref.~\cite{Neubert:2004dd}, which was used to sum large logs in the $B$-meson
shape function for $B\to X_s\gamma$. 

We postpone presenting our final resummed NNLL result for the jet function until
section~\ref{sect:results}.

\section{Wilson Loop Representations, and Comparison with the Heavy Quark Shape
  Function and Fragmentation Function}  \label{sect:Wloop}

\begin{figure*}[t!]
  \centerline{
   \mbox{\epsfxsize=12truecm \hbox{\epsfbox{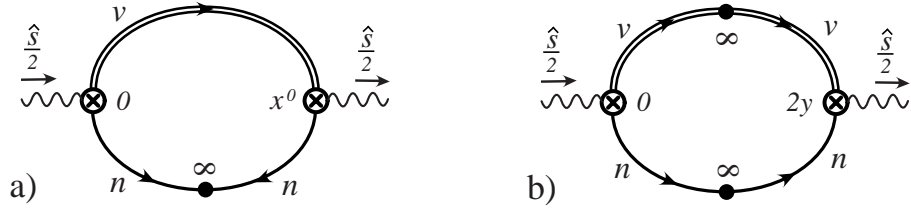}} } \quad
  }
  \vskip-0.2cm
  \caption{Representations of the Wilson line matrix elements for the heavy
    quark jet function,
    The Wilson lines include $W_v$, $W_v^\dagger$ (double lines), and $W_n$,
    $W_n^\dagger$ (single
    lines).  In a)
    we display the result in Eq.~(\ref{cBwloop}) that gives ${\cal B}(\hat
    s,\mu)$.
    In b) we display the result in Eq.~(\ref{Bwloop}) that gives $B(\hat
    s,\mu)$. }
   \label{fig:wilson}
\end{figure*}

It is well known that the leading order coupling of gluons to heavy-quark fields
$h_v$ in HQET can be represented by Wilson lines along the path of the
heavy-quark~\cite{Korchemsky:1991zp}. We define 
\begin{align} \label{WvWv}
  W_v(x) &=
     \overline P\exp\Big(-ig\! \int_0^\infty\!\!\! ds\  v\mcdot A(vs\plus x)\Big)
  \,,
  & W_v^\dagger (x) &=
      P\exp\Big(ig \int_0^\infty\!\!\! ds\  v\mcdot A(vs \plus x)\Big) 
 \,.
\end{align} 
To see how the HQET action reduces to a Wilson line one can make a field
redefinition, $h_v= W_v h_v^{(0)}$, from which we find that $h_v^{(0)}$ is a
free field with Lagrangian ${\cal L}_h = \bar h_v^{(0)} iv\cdot\partial
h_v^{(0)}$, see~\cite{Bauer:2001yt}. Thus, the vacuum matrix element for the
heavy-quark jet function in Eq.~(\ref{eq:CB}) can be written as a matrix element
of Wilson lines
\begin{align} \label{cBwloop}
\cB(2 v\cdot r,\mu)
 &= \fr{-i}{4\pi N_c m}
 \int\! d^4x\, e^{i r\cdot x}\big\langle\,
   0\big|  T\: \bar{h}^{(0)}_{v}(0) W_v^\dagger(0) W_{n}(0)
   W_{n}^\dagger(x) W_v(x) h_{v}^{(0)}(x) \big|0\big\rangle
  \nn\\
 &= \fr{i}{2\pi N_c m}
 \int\! dx^0\, e^{i v\cdot r\, x^0}\: \theta(x^0)\, \left\langle\,
   0\left| {\rm tr}\:  T\: W_v^\dagger(0) W_{n}(0)
   W_{n}^\dagger(x^0) W_v(x^0) \right|0\right\rangle 
 \nn\\
 &= \fr{i}{2\pi N_c m}
 \int\! dx^0\, e^{i v\cdot r\, x^0}\: \theta(x^0)\, \left\langle\,
   0\left| {\rm tr}\:  T\: W_v(x^0,0) W_{n}(0,\infty,x^0) \right|0\right\rangle 
 \,,
\end{align}
where $2v\cdot r=\hat s$ and we use the shorthand $x^0=v\cdot x$, the trace
${\rm tr}$ is over color indices. Here $W_v(x^0,0)=W_v(x^0)W_v^\dagger(0)$ is the
straight Wilson line from $0$ to $x^0$, while $W_n(0,\infty,x^0)\equiv
W_n(0)W_n^\dagger(x^0)$ has a path from $x^0$ to $\infty$ to $0$ that uses two
light-like line-segments.  To obtain the second line of Eq.~(\ref{cBwloop}) we
used the heavy-quark propagator, $\langle 0| T \bar h_v^{(0)a}(0) h_v^{(0)b}(x)
| 0 \rangle= -2 \delta^{ab} \delta^3(\vec x)\theta(x^0)$ where $a$ and $b$ are
color indices. In Fig.~\ref{fig:wilson}a we give a graphical representation for
the Wilson line definition in the last line of Eq.~(\ref{cBwloop}). The arrows
denote the time-ordering.

We can also write the jet function $B(\hat s,\mu)$ as a matrix element of Wilson
lines.  To derive this result we note that
\begin{align} \label{Bwilson}
  B(2v\cdot r,\mu) 
  &= \frac{1}{8\pi N_c m} \int\!\! d^4x \: e^{ir\cdot x}\: \sum_X \, {\rm Tr} \,
    \big\langle 0 \big| \overline T\, W_n^\dagger(x) h_v(x) \big| X \big\rangle 
    \big\langle X \big| T\, \bar h_v(0) W_n(0) \big| 0 \big\rangle 
  \nn\\
  &= \frac{1}{8\pi N_c m} \int\!\! d^4x \: e^{ir\cdot x} \: {\rm Tr} \,
    \big\langle 0 \big| \big[\, \overline T\, W_n^\dagger(x) h_v(x) \big] 
    \big[\,  T\, \bar h_v(0) W_n(0) \big] \big| 0 \big\rangle 
  \nn\\
    &= \frac{1}{8\pi N_c m} \int\!\! d^4x \: e^{ir\cdot x} \: {\rm Tr} \,
    \big\langle 0 \big| \big[\, \overline T\, W_n^\dagger(x) W_v(x) h_v^{(0)}(x) \big] 
    \big[\,  T\, \bar h_v^{(0)}(0) W_v^\dagger(0) W_n(0) \big] \big| 0 \big\rangle 
  \,,
\end{align}
where $T$ is time-ordering, $\overline T$ is anti-time-ordering, and the trace
${\rm Tr}$ is over spin and color indices. Next we use $\langle 0 |
h_{vi}^{(0)a}(x) \bar h_{vi}^{(0)b}(0)|0\rangle = 2 \delta^{ab}\delta^3(\vec x)$,
where here there is no time-ordering and hence no $\theta(x^0)$, and the spin
indices $i$ are contracted. Thus
\begin{align} \label{Bwloop}
  B(\hat s,\mu) &= \frac{1}{2\pi} \int\! dy \: e^{i \hat s\,y} \: \tilde
  B(y,\mu) \,,
  & \tilde B(y,\mu) &=    \frac{1}{ m\, N_c}\:
   \big\langle 0 \big| {\rm tr} \,\big[\, \overline T\, W_n^\dagger(2y) W_v(2y) \big] 
    \big[\,  T\, W_v^\dagger(0) W_n(0) \big] \big| 0
    \big\rangle 
  \,.
\end{align}
Here we took $x^0 = 2y$ in order to agree with the notation for the position
space jet function ${\tilde B}(y,\mu)$ above in Eq.~(\ref{By}).  In
Fig.~\ref{fig:wilson}b we give a graphical representation for the Wilson line
matrix element for $B(\hat s,\mu)$ in Eq.~(\ref{Bwloop}), where the arrows
denote the time-ordering. Comparing to the Wilson loop for ${\cal B}(\hat
s,\mu)$ in Fig.~\ref{fig:wilson}a we note the importance of the $\infty$-points
to determine which fields are time-ordered and which are antitime-ordered.

It is instructive to compare the Wilson line definition of the heavy quark jet
function with the corresponding definitions for the heavy quark shape function
that appears in $B$-decays~\cite{Neubert:1993ch,Bigi:1994ex}, and with the heavy
quark fragmentation function~\cite{Jaffe:1993ie}. Using a variable $\ell^+\ge
0$, the $B$-meson shape function is given by
\begin{align} \label{fwilson}
  f^{B_v}(\ell^+,\mu) 
  &= \frac{1}{8\pi} \int\!\! dx^- \: e^{-i(\ell^+-\bar\Lambda) x^-/2}\: \sum_X \, 
    \big\langle \bar B_v \big| \overline T\, \bar h_v(0) W_n(0) \big| X \big\rangle 
    \big\langle X \big| T\, W_n^\dagger(\tilde x) h_v(\tilde x) \big| \bar B_v \big\rangle 
  \nn\\
  &= \frac{1}{8\pi} \int\!\! dx^- \: e^{-i(\ell^+-\bar\Lambda) x^-/2}\:
    \big\langle \bar B_v \big|
    \big[\,  \overline T\, \bar h_v^{(0)}(0) \tilde W_v^\dagger(0) W_n(0) \big]
    \big[\,  T\, W_n^\dagger(\tilde x) \tilde W_v(\tilde x) h_v^{(0)}(\tilde x) \big]  \big|
    \bar B_v \big\rangle 
  \nn\\
    &= \frac{1}{8\pi} \int\!\! dx^- \: e^{-i(\ell^+-\bar\Lambda) x^-/2}\:
    \big\langle \bar B_v \big|
     \bar h_v^{(0)}(0) \tilde W_v^\dagger(0)  W_n(0,\tilde x) 
     \tilde W_v(\tilde x)h_v^{(0)}(\tilde x)   \big|
    \bar B_v \big\rangle 
  \,,
\end{align}
where we use the shorthand $x^- = \bn\cdot x$ and $\tilde x^\mu = x^- n^\mu/2$,
and set $\bn\cdot v=1$.  To obtain the second line we made the same field
redefinition as above in Eq.~(\ref{WvWv}), but now on both the heavy-quark field
and on the interpolating field for the
$B$-meson~\cite{Chay:2004zn,Arnesen:2005nk}.  Combining the lines from these
sources yielded the Wilson lines
\begin{align} \label{tWv}
  \tilde W_v(x) &=
     P\exp\Big(ig\! \int_{-\infty}^0\!\!\! ds\  v\mcdot A(vs\plus x)\Big)
  \,,
  & \tilde W_v^\dagger (x) &=
      \overline P \exp\Big(-ig \int_{-\infty}^0\!\!\! ds\  v\mcdot A(vs \plus x)\Big) 
 \,.
\end{align} 
To obtain the third line of Eq.~(\ref{fwilson}) we noted that $T W_n^\dagger
=W_n^\dagger$, $\overline T W_n = W_n$, $T \tilde W_v = \tilde W_v$, and
$\overline T \tilde W_v^\dagger = \tilde W_v^\dagger$, and that the gluons in
the Wilson lines which sit next to each other, $W_n^\dagger(\tilde x)\tilde
W_v(\tilde x)$ and $\tilde W_v^\dagger(0)W_n(0)$, are already time-ordered and
anti-time-ordered respectively.

For the B-meson fragmentation function in HQET with variable $\ell^+\ge 0$, the
field redefinition gives
\begin{align} \label{Dwilson} 
  D^{B_v/b}(\ell^+,\mu) 
  &= \frac{1}{16\pi} \int\!\! dx^{-} \: e^{i(\ell^++\bar\Lambda) x^-/2}\: \sum_X \, 
    \big\langle 0 \big| \overline T\, \tilde W_n^\dagger(\tilde x) h_v(\tilde x) \big| 
    \bar B_v X \big\rangle 
    \big\langle \bar B_v X \big| T\,  \bar h_v(0) \tilde W_n(0)  \big| 0 \big\rangle 
  \nn\\
  &= \frac{1}{16\pi} \int\!\! dx^{-} \: e^{i(\ell^++\bar\Lambda) x^-/2}\: \sum_X \, 
    \big\langle 0 \big| \overline T\, \tilde W_n^\dagger(\tilde x) W_{v}(\tilde x)
    h_v^{(0)}(\tilde x) \big| 
    \bar B_v X \big\rangle 
    \big\langle \bar B_v X \big| T\,  \bar h_v^{(0)}(0) W_{v}^\dagger(0) \tilde W_n(0)  
    \big| 0 \big\rangle 
  \,.
\end{align}
Here $\tilde W_n$ and $\tilde W_n^\dagger$ are defined as in Eq.~(\ref{tWv}) but
with $v\to \bn$.  The shape and fragmentation function results in
Eq.~(\ref{fwilson}) and (\ref{Dwilson}) are similar to the heavy-quark
jet function in that all three are defined by matrix elements with heavy-quark
fields and Wilson lines.  They differ because they are non-perturbative
distributions involving a $B$-meson state in contrast to the perturbatively
computable jet function. The shape and fragmentation functions also have a
light-cone separation rather than the time-like separation that we have for the
jet function.

In certain contexts it is also useful to consider the partonic shape function
$f^{b_v}$ and the partonic fragmentation function $D^{b_v/b}$ where the
$B_v$-meson state is replace by a $b_v$-quark with residual momentum $k^\mu$,
where $\bar\Lambda=0$.
In this case we can perform the contraction $h_v^{(0)}(x) |\bar b_v\rangle =
e^{-ik\cdot x} u_v |0\rangle/\sqrt{N_c}$ and write
\begin{align} \label{fDpartonic}
  f^{b_v}_{k^+}(\ell^+,\mu)    &= \frac{1}{4\pi N_c} \int\!\! dx^- \: e^{-i(\ell^++k^+) x^-/2}\:
    \big\langle 0 \big|{\rm tr}\: \tilde W_{v}^\dagger(0)  W_n(0,\tilde x) 
     \tilde W_{v}(\tilde x)   \big| 0 \big\rangle  \,,
  \\[3pt]
  D^{b_v/b}_{k^+}(\ell^+,\mu)  &= 
  \frac{1}{4\pi N_c}\, \int\!\! dx^- \: e^{i(\ell^+-k^+) x^-/2}\:  
    \Big\langle 0 \Big|{\rm tr} \,  \tilde W_n^\dagger(\tilde x) W_{v}(\tilde x)
    W_v^\dagger(0) \tilde W_n(0) 
    \Big| 0 \Big\rangle 
   \nn \\
  &= 
  \frac{1}{4\pi N_c}\, \int\!\! dx^- \: e^{i(\ell^+-k^+) x^-/2}\:
    \big\langle 0 \big| {\rm tr} \, \tilde W_n^\dagger(\tilde x)  \tilde W_{v}(\tilde x)
    \tilde W_v^\dagger(0) \tilde W_n(0)  
    \big| 0 \big\rangle 
  \nn \\
 &= 
  \frac{1}{4\pi N_c}\, \int\!\! dx^- \: e^{i(\ell^+-k^+) x^-/2}\:
    \big\langle 0 \big| {\rm tr} \, \tilde W_v^\dagger(0)  W_{n}(0,\tilde x)
    \tilde W_v(\tilde x)  
    \big| 0 \big\rangle  \nn\\[3pt]
 &  = f_{k^+}^{b_v}(-\ell^+,\mu) 
   \,,\nn 
\end{align}
where we used $\tilde W_n(0) \tilde W_n^\dagger(\tilde x) = W_n(0)
W_n^\dagger(\tilde x) = W_n(0,\tilde x)$.  Eq.~(\ref{fDpartonic}) states that
the partonic shape function and fragmentation function are identical, but with
complementary ranges of support.  This was observed in Ref.~\cite{Gardi:2005yi}
for logs at NNLL accuracy, and was derived to all orders in perturbation theory
in Ref.~\cite{Neubert:2007je} as we outlined above.  Thus, the partonic shape
function in position space, $\tilde f^{b_v}(y,\mu)$ is also given by a vacuum
matrix of Wilson lines.  It differs from $\tilde B(y,\mu)$ in
Eq.~(\ref{cBwloop}) both due to the light-like rather than time-like separation $y =x^-/2$, 
and due to the path.

\section{Non-Abelian Exponentiation}  \label{sect:NAExp}

In the previous section in Eq.~(\ref{Bwloop}) we showed that the position space
heavy-quark jet function $\tilde B(y,\mu)$ is determined by a vacuum matrix
element of Wilson lines.  Due to the non-abelian exponentiation
theorem~\cite{Gatheral:1983cz,Frenkel:1984pz} for matrix elements of Wilson
lines with symmetric restrictions on the phase space of real gluons, $\tilde
B(y,\mu)$ exponentiates.  This is also true of the partonic heavy-quark shape
function in position space.

Taking the Fourier transform of the two-loop jet function result in
Eq.~(\ref{B2NNLO}) using Eq.~(\ref{FTyshat}) we obtain
\begin{align} \label{Babelianexpn}
  m \tilde B(y,\mu) 
  & = 1 
   +\frac{C_F\alpha_s(\mu)}{\pi} \bigg(  \tilde L^2 + \tilde L+\frac{\pi
       ^2}{24}+1\bigg)
  + \frac{\alpha_s^2(\mu)}{\pi^2} \Bigg\{ 
  C_F \beta_0  \bigg[ \frac{1}{6} \tilde L^3 +\frac{2}{3}\tilde L^2
   +\frac{47}{36}\tilde L -\frac{\zeta_3}{48}+\frac{5 \pi
   ^2}{576}+\frac{281}{216}\bigg]
  \nn\\
 & \qquad\qquad
 + C_F C_A \bigg[ \Big(\frac{1}{3}-\frac{\pi ^2}{12}\Big) \tilde L^2
  + \bigg(\frac{5}{18}-\frac{\pi ^2}{12}-\frac{5 \zeta_3}{4}\bigg) \tilde L
  -\frac{5\zeta_3}{8}-\frac{17 \pi ^4}{2880}+\frac{7 \pi ^2}{144}-\frac{11}{54}\bigg]
 \nn\\
 &\qquad\qquad
   + C_F^2 \bigg[ \frac12 \tilde L^4 + \tilde L^3
 + \Big(\frac{3}{2}+\frac{\pi ^2}{24}\Big) \tilde L^2
  + \Big(1+\frac{\pi ^2}{24}\Big) \tilde L 
   +\frac{\pi ^4}{1152}+\frac{\pi ^2}{24}+\frac{1}{2}\bigg] 
  \Bigg\} \,,
\end{align}
where $\tilde L^k = \big(\tilde L\big)^k$ and
\begin{align} \label{tL}
  \tilde L \, \equiv \, \ln\big( i e^{\gamma_E} y\,\mu \big) \,. 
\end{align}
It is evident in Eq.~(\ref{Babelianexpn}) that the two-loop, $C_F^2\alpha_s^2$
term satisfies the exponentiation theorem, being determined by one-half the
square of the one-loop $C_F\alpha_s$ term.  Thus we can write
\begin{align} \label{Bnonabelian}
 m \tilde B(y,\mu)  & = \exp\Bigg\{ 
   \frac{C_F\alpha_s(\mu)}{\pi} \bigg(  \tilde L^2 + \tilde L+\frac{\pi
       ^2}{24}+1\bigg)
 +  \frac{\alpha_s^2(\mu) C_F \beta_0}{\pi^2} 
  \bigg[ \frac{1}{6} \tilde L^3 +\frac{2}{3}\tilde L^2
   +\frac{47}{36}\tilde L -\frac{\zeta_3}{48}+\frac{5 \pi
   ^2}{576}+\frac{281}{216}\bigg] \nn\\
&\qquad\qquad   
 + \frac{\alpha_s^2(\mu) C_F C_A}{\pi^2}
   \bigg[ \Big(\frac{1}{3}-\frac{\pi ^2}{12}\Big) \tilde L^2
  + \bigg(\frac{5}{18}-\frac{\pi ^2}{12}-\frac{5 \zeta_3}{4}\bigg) \tilde L
  -\frac{5\zeta_3}{8}-\frac{17 \pi ^4}{2880}+\frac{7 \pi ^2}{144}-\frac{11}{54}\bigg]
  \Bigg\} \,.
\end{align}
The non-abelian exponentiation theorem guarantees that corrections to this
result are ${\cal O}(\alpha_s^3)$ in the exponent, and that these corrections
will vanish if we take the abelian limit $C_A\to 0$ and $n_f\to 0$.

In the abelian limit with zero $\beta$-function, the exponentiation theorem
implies that $\ln [ m \tilde B(y,\mu)]$ is one-loop exact. Thus taking $C_A=
n_f= 0$, and a charge $C_F$ we have to all orders in perturbation theory
\begin{align} \label{Babelian}
  m \tilde B(y,\mu)^{\rm abelian}\: &=\:  \exp\bigg[ 
   \frac{\alpha_s}{4\pi}\, \Big(\Gamma_0^{\rm c} \tilde L^2
    + \gamma_0 \tilde L
    + T_0 \Big) \bigg] \,,
\end{align}
where the constants are $\Gamma_0^c=\gamma_0=4C_F$ and $T_0=4 C_F(1+\pi^2/24)$.
The exact result in Eq.~(\ref{Babelian}) provides a simple way of testing the
properties of different possible jet-mass definitions at higher orders in
perturbation theory, as discussed in section~\ref{sect:jetmass}.

We can also consider the implications of the non-abelian exponentiation theorem
for the solution of the renormalization group equation~(\ref{Byrun}).  Following
Ref.~\cite{Hoang:2007vb} we first use the evolution kernel $K(\mu,\mu_0)$ to
solve for $\tilde {B}(y,\mu)$ by taking $\mu_0 = \mu_y\equiv -ie^{-\gamma_E}/y$.
This makes all the logs in $\tilde {B}(y,\mu_y)$ vanish since $\tilde L(\mu_y)=\ln(i
e^{\gamma_E} y\mu_y)= 0$. Thus
\begin{align} \label{RGEexp}
  m \tilde {B}(y,\mu) = e^{K(\mu,\mu_y)}\: m \tilde {B}(y,\mu_y) =
  e^{K(\mu,\mu_y)+T[\alpha_s(\mu_y)]} \: \,.
\end{align}
Here the boundary condition for the RGE, denoted $m \tilde {B}(y,\mu_y)$, is just
a perturbative series in $\alpha_s(\mu_y)$. Due to the non-abelian
exponentiation theorem this series must exponentiate to give $\exp(
T[\alpha_s(\mu_y)])$, and the coefficients in the perturbative series for
$T[\alpha_s]$ have color factors that satisfy the exponentiation theorem
constraints. It is a straightforward exercise to verify that expanding the
result for $K(\mu,\mu_0)$ in Eq.~(\ref{wKNNLL}) to ${\cal O}(\alpha_s^2)$ gives
a result from Eq.~(\ref{RGEexp}) that is consistent with
Eq.~(\ref{Bnonabelian}).

The Fourier transformed partonic b-quark shape function is also given by a
vacuum matrix element of Wilson lines via Eq.~(\ref{fDpartonic}). Thus, it too
satisfies the criteria of the non-abelian exponentiation
theorem~\cite{Korchemsky:1992xv}. Taking the Fourier transform of the two-loop
computation of $f^{b_v}(\ell^+,\mu)$ in Ref.~\cite{Becher:2005pd} we have verified that
the $C_F^2\alpha_s^2$ terms satisfy the non-abelian exponentiation theorem. This
calculation gives
\begin{align} \label{fpos}
  \tilde f^{b_v}(y,\mu) &= \exp\Bigg\{ 
   \frac{-C_F\alpha_s(\mu)}{\pi} \bigg( \tbreve L^2 - \tbreve L +\frac{5\pi
       ^2}{24}\bigg)\!
 +  \frac{\alpha_s^2(\mu) C_F \beta_0}{\pi^2} 
  \bigg[\! -\frac{1}{6} \tbreve L^3 -\frac{1}{6}\tbreve L^2
   + \frac{(1\minus 3\pi^2)}{36}\tbreve L -\frac{11\zeta_3}{48}
   -\frac{7 \pi^2}{192}\plus \frac{1}{216}\bigg] \nn\\
&\qquad  
 + \frac{\alpha_s^2(\mu) C_F C_A}{\pi^2}
   \bigg[\! -\Big(\frac{1}{3}-\frac{\pi ^2}{12}\Big) \tbreve L^2
  + \bigg(\! -\frac{11}{18}-\frac{\pi ^2}{12}+\frac{9 \zeta_3}{4}\bigg) \tbreve L
  -\frac{9\zeta_3}{8}+\frac{107 \pi ^4}{2880}-\frac{13 \pi ^2}{48}-\frac{29}{108}\bigg]
  \Bigg\} \,,
\end{align}
where $\tilde L$ is defined in Eq.~(\ref{tL}), but now $y$ is the conjugate
variable to $\ell^+$, $y=x^-/2$.
Corrections to this result are again ${\cal O}(\alpha_s^3)$ in the exponent, and
vanish when $C_A=n_f=0$.  Comparing Eq.~(\ref{Bnonabelian}) and (\ref{fpos}) we
explicitly observe the difference between the heavy quark jet function and the
partonic shape function.  Up to a sign the highest powers of $\tilde L$ agree at
each order in $\alpha_s$, because of the relation between their cusp anomalous
dimension terms. The subleading logs and constant terms differ.

\section{A Transitive Jet-Mass Scheme} \label{sect:jetmass}

The last remaining ingredient needed for the NNLO and NNLL computations of the
heavy-quark jet function is the specification of the mass scheme counterterm
$\delta m$ at two-loop order. Since the jet function will be used to describe
momenta $\hat s\sim \Gamma$, where $\Gamma$ is the width of the physical
invariant mass distribution, we must have $\delta m\sim \Gamma$ or smaller to
not upset the power counting in the HQET Lagrangian, Eq.~(\ref{eq:LHQET}).  In
the $\overline {\rm MS}$ scheme $\delta \overline m\sim m (\alpha_s + \alpha_s^2
+\ldots)$, and since $m\alpha_s \gg \Gamma$ this scheme does not satisfy the
power counting criteria.  In the pole-mass scheme $\delta m=0$ to all orders,
however this scheme has instabilities related to its infrared sensitivity. In
particular the pole-mass has an infrared renormalon that leads to an asymptotic
ambiguity $\delta m^{\rm pole}\sim \Lambda_{\rm QCD}$, and hence is not a useful
scheme for precision computations.  Schemes that satisfy $\delta m\sim \Gamma$
and do not suffer from infrared renormalons were called top ``jet-mass''
schemes in Ref.~\cite{Fleming:2007qr}. We refer to them more generally as ``top
resonance mass-schemes'' here and reserve the name jet-mass for a specific
example of this type of scheme. These mass-schemes are suitable for use in the
factorization theorem for the top-invariant mass distribution in
Eq.~(\ref{Fthm}) and related observables. We start by defining a jet-mass scheme
with nice renormalization properties in section~\ref{subsect:massA}, and then
relate this jet-mass to the pole, $\overline {\rm MS}$, and 1S mass schemes in
section~\ref{subsect:massB}.

\subsection{Potential Jet-Mass Definitions and Anomalous Dimensions} \label{subsect:massA}

In this section we explore three resonance mass-schemes for $m$. With the
notation for $\delta m$ in Eq.~(\ref{dm}) they are defined by
\begin{align} \label{mscheme}
 & \text{a) } \qquad\qquad  
     \frac{d}{d\hat s}\, B(\hat s,\delta m^{\rm peak},\Gamma_t,\mu) 
    \bigg|_{\hat s=0}  = 0 
  \,,
   \\[3pt]
 & \text{b) } \qquad\qquad
  \int_{-\infty}^{R} \!\! d\hat s \ \hat s\  B(\hat s,\delta m^{\rm mom},\mu)  =0 
  \,,
  \nn\\[3pt]
 & \text{c) } \qquad\qquad
  \delta m_J = \frac{-i}{2\,  \tilde {B}(y,\mu)}\: 
   \frac{d}{dy}\, \tilde {B}(y,\mu) \bigg|_{y = -ie^{-\gamma_E}/R}
   = e^{\gamma_E}\: \frac{R}{2}\, \frac{d}{d\ln(i y)} \ln \tilde {B}(y,\mu) 
  \bigg|_{iye^{\gamma_E} =1/R}
    \,. \nn
\end{align}
We refer to a), b), c) as the peak-mass, moment-mass, and position-mass
respectively.  The peak-mass definition uses the jet function with a non-zero
width and satisfies the $\delta m\sim\Gamma_t$ power counting
criteria~\cite{Fleming:2007qr}. In b) and c) the schemes depend on a parameter
$R$, and we must take $R\sim \Gamma_t$ in order to satisfy the power counting
criteria.  Different choices for $R$ specify different schemes, and are
analogous to the difference between the ${\rm MS}$ and $\overline {\rm MS}$
mass-schemes.  All three schemes in Eq.~(\ref{mscheme}) are free from leading
renormalon ambiguities~\cite{FHMSmass:inprep}. In the following we will argue
that only the definition in c) is a reasonable scheme for higher order
computations. Thus we will only use the name jet-mass for this position-scheme
mass definition.

The definitions in Eq.~(\ref{mscheme}) are all perturbative mass-schemes which
stabilize the peak position of the jet function $B(\hat s,\delta
m,\Gamma_t,\mu)$. In scheme a) the peak position is fixed to all orders in
perturbation theory by definition. In scheme b) we instead fix the first moment,
which provides a more local observable that is still sensitive to the peak
location.  However, scheme b) still has non-locality induced by the cutoff $R$
on the momentum space moment. A finite $R$ is necessary due to ultraviolet
divergences that occur for $R\to\infty$. This type of moment divergence is a
general property of functions that have a cusp anomalous dimension (see for
example Refs.~\cite{Bauer:2003pi,Lange:2003ff}). If it was not for the UV
divergences then the schemes b) and c) would be equivalent in the limit $R\to
\infty$.  In the situation at hand, c) provides an independent mass scheme
definition. A jet-mass definition from c) is explicitly local since it just
involves the position space jet function at a particular position $y$.

An additional criteria for a reasonable jet-mass scheme is to have a
renormalization group evolution that is transitive, as discussed in
Ref.~\cite{Fleming:2007xt}.  Transitivity is a well-known feature of the
$\overline {\rm MS}$ mass, and implies that we will obtain the same result if we
evolve directly from $\mu_0\to \mu_2$, or if we first evolve from $\mu_0\to
\mu_1$ and then from $\mu_1\to \mu_2$. Transitivity is guaranteed by any
mass-scheme with a consistent anomalous dimension and renormalization group
equation.  Since in HQET the scale independent $m^{\rm pole}= m(\mu) +\delta m(\mu)$,
the general form for the RGE equation for the mass is
\begin{align} \label{mad}
  \mu\frac{d}{d\mu} m(\mu) &= \gamma_m[R, m(\mu),\alpha_s(\mu)] \,,
\qquad\qquad
  \gamma_m =  - \mu\frac{d}{d\mu} \delta m(\mu) \,,
\end{align}
where $R$ is a mass dimension-1 scheme parameter.  Transitivity of $m(\mu)$ is
guaranteed by this anomalous dimension equation, as long as $\gamma_m$ is
proportional to $[m(\mu)]^k R^{1-k}$ for some $k$ (and thus, for example, is not
a sum of two types of terms with different powers of $k$). In the $\overline
{\rm MS}$ scheme $k=1$ and the anomalous dimension is proportional to $m(\mu)$,
while in all three schemes in Eq.~(\ref{mscheme}) we have $k=0$.  However, it
turns out that the peak-scheme and moment-scheme do not have consistent
anomalous dimension equations of the form in Eq.~(\ref{mad}), because there
$\gamma_m$'s depend on explicit powers $\ln^j(\mu/\Gamma_t)$ and $\ln^j(\mu/R)$
with higher and higher powers of $j\ge 1$ occurring for higher orders in
$\alpha_s$. These logs render the moment scheme anomalous dimension equation
inconsistent at NLO order, and the peak scheme does not even have an anomalous
dimension equation of the form in (\ref{mad}) at LO order.

In order to illustrate the difference between the three schemes in
Eq.~(\ref{mscheme}) we first consider the simplified case of the jet function in
the abelian limit, $C_A\to 0$ and $n_f\to 0$. The all-order result for $\tilde
B(y,\mu)$ is given in Eq.~(\ref{Babelian}), and can be directly used to
determine $\delta m$ in the position-mass scheme. The derivative of the
exponential gives back an exponential which cancels against the $1/\tilde
B(y,\mu)$ in $\delta m_J$. Thus the abelian result in the position-mass scheme
is one-loop exact,
\begin{align} \label{dmJabelian}
 \delta m_J^{\rm abelian} &= e^{\gamma_E} R  \frac{C_F\alpha_s}{\pi} \Big[ \ln\frac{\mu}{R}
 +\frac{1}{2} \Big] \,.
\end{align}
Since for the abelian limit $d\alpha_s/d\mu=0$, the abelian anomalous dimension
computed from Eq.~(\ref{dmJabelian}) is $(\gamma_m^{J})^{\rm abelian}= - Re^{\gamma_E} C_F
\alpha_s/\pi$ to all orders. Thus this position-scheme anomalous dimension has
the desired form in Eq.~(\ref{mad}).  To compute results for the peak and moment
mass-schemes we need the abelian jet function in momentum space, $B(\hat
s,\mu)$. Tree, one-loop, and two-loop terms are given by the abelian terms in
Eq.~(\ref{B2NNLO}).  To determine three-loop and higher order results we can
simply expand Eq.~(\ref{Babelian}) in $\alpha_s$ and take the Fourier transform.
For the three-loop term in the abelian jet function this gives
\begin{align}
  &m B_3(\hat s,\mu)^{\rm abelian} = C_F^3 \bigg[ {\cal L}^5 -\frac{5}{2} {\cal
    L}^4 + \Big( 4 \minus \frac{19\pi^2}{12}\Big) {\cal L}^3
   + \Big(\minus \frac{7}{2} \plus \frac{19\pi^2}{8}\plus 20\zeta_3\Big) {\cal L}^2
   + \Big(2\minus \frac{15\pi^2}{8} \plus \frac{25\pi^4}{576}\minus
   20\zeta_3\Big){\cal L}^1
  \\
  &\ + \Big(\minus \frac{1}{2}\plus \frac{13\pi^2}{24}\minus \frac{25\pi^4}{1152} 
    \plus 8\zeta_3 \minus\frac{19\pi^2\zeta_3}{6} \plus 24\zeta_5\Big) {\cal L}^0
   + \Big( \frac{1}{6} \minus \frac{7 \pi ^2}{48} \plus \frac{41 \pi ^4}{5760}
      \minus \frac{13777 \pi^6}{2903040} \minus \frac{7 \zeta_3}{3} 
      \plus \frac{19\pi^2 \zeta_3}{12} \plus \frac{20 \zeta_3^2}{3}\minus 12\zeta_5
    \Big)\delta(\hat s) \bigg]
  .\nn 
\end{align}
Using $B_{0,1,2,3}(\hat s,\mu)^{\rm abelian}$ we find that up to three-loop
order
\begin{align} \label{dmababelian}
  \delta m^{\rm peak}_{\rm abelian} &= \frac{\pi\Gamma_t}{4} \bigg\{ 
  \frac{C_F\alpha_s}{\pi} \bigg[  \ln\frac{\mu}{\Gamma_t} +\frac{3}{2} \bigg] 
  + \frac{C_F^2\alpha_s^2}{\pi^2} \bigg[ \minus \ln^2\!\frac{\mu}{\Gamma_t} \plus
   \Big(\frac{\pi^2}{3}\minus 5\Big)\ln\frac{\mu}{\Gamma_t} \minus \frac{13}{4}\plus
   \frac{\pi^2}{2} \minus  2\zeta_3 \bigg]
  + \frac{C_F^3\alpha_s^3}{\pi^3} 
  \bigg[ \Big( 1\plus \frac{\pi^2}{12}\Big) \ln^3\!\frac{\mu}{\Gamma_t} 
  \nn \\
  & 
  \ + \Big(\frac{25}{2}\minus \frac{5\pi^2}{12}\minus 4\zeta_3\Big)
   \ln^2\!\frac{\mu}{\Gamma_t} + \Big(\frac{75}{4}\minus \frac{151\pi^2}{48}
   \plus \frac{11\pi^4}{45} \minus 8\zeta_3 \Big)\ln\frac{\mu}{\Gamma_t}
  \plus \frac{59}{8} \minus \frac{5 \pi ^2}{2}\plus \frac{11 \pi ^4}{30}
  \plus 5 \zeta_3 \minus \pi^2 \zeta_3 \minus 12 \zeta_5
  \bigg] \bigg\} \,,
  \nn\\
  \delta m^{\rm mom}_{\rm abelian} &= R\, \bigg\{ 
  \frac{C_F\alpha_s}{\pi} \bigg[  \ln\frac{\mu}{R}  +\frac{3}{2} \bigg] 
  +\frac{C_F^2\alpha_s^2}{\pi^2} \bigg[ \Big(4\minus \frac{\pi^2}{3}\Big)\ln\frac{\mu}{R} 
    \plus 8 \minus \frac{\pi^2}{2} \minus 2\zeta_3 \bigg]
    + \frac{C_F^3\alpha_s^3}{\pi^3} 
  \bigg[ (6\minus 4\zeta_3) \ln^2\!\frac{\mu}{R} \nn\\
 &\  +\Big( 46\minus \frac{8 \pi
    ^2}{3} \minus \frac{\pi ^4}{45} \minus 12 \zeta_3\Big) \ln\frac{\mu}{R} 
   + \frac{159}{2} \minus \frac{16 \pi ^2}{3} \minus \frac{\pi^4}{30}
   \minus 21 \zeta_3 \plus \frac{4 \pi^2 \zeta_3}{3} \minus 12 \zeta_5
  \bigg] \bigg\} \,. 
\end{align}
At one-loop order the $\delta m$ factors in the three schemes a), b), c) are
quite similar.\footnote{In the position-scheme it might appear that there is a
  freedom in the overall normalization of $\delta m$ in Eq.~(\ref{mscheme})c,
  and in the choice of $R$. In fact to obtain a renormalon free jet-mass
  scheme there is no freedom in the overall normalization, there is only freedom
  in the choice of $R$~\cite{FHMSmass:inprep}.}  However the three schemes are
quite different at two-loop order.  Computing $\gamma_m$ from these counterterms
we see that a $\ln(\mu/\Gamma_t)$ appears in $\gamma_{m,{\rm abelian}}^{\rm
  peak}$ at two-loop order, and that a $\ln(\mu/R)$ appears in $\gamma_{m,{\rm
    abelian}}^{\rm mom}$ at three-loop order. At higher orders in $\alpha_s$,
higher and higher powers of these logarithms, $\ln^j(\mu/R)$, appear in
$\gamma_m$ in the peak and moment schemes. In particular we see from
Eq.~(\ref{dmababelian}) that at three-loops for the peak-scheme there is a
$C_F^3\ln^3(\mu/\Gamma_t)$ term that generates a $\ln^2(\mu/\Gamma_t)$ in the
computation of $\gamma_m$. For the moment scheme we have extended the abelian
computation to four-loops, and find a term
\begin{align}
  \big(\delta m_{\rm abelian}^{\rm mom}\big)^{\rm 4loop}
  \, &=\,  R \frac{C_F^4\alpha_s^4}{\pi^4} 
    \bigg[ \Big(\frac{32}{3}-\frac{4\pi^4}{45}\Big)\ln^3\!\frac{\mu}{R} 
     + \ldots \bigg] \,.
\end{align}  
This gives a $\ln^2(\mu/R)$ in the moment scheme $\gamma_m$ at four-loop
order.\footnote{The presence of these higher logs in $\gamma_m$ for the peak and
  moment schemes implies that these masses also do not fall into the
  cusp-anomalous dimension category, which requires an anomalous dimension of
  the form $\gamma_m = R\, \gamma_1[\alpha_s] + R \ln(\mu/R)\,
  \gamma_2[\alpha_s]$, i.e.  with just a single $\ln(\mu/R)$ to all orders in
  $\alpha_s$.} The absence of $\alpha_s^k \ln^k(\mu/R)$ terms in $\delta m^{\rm
  mom}_{\rm abelian}$ is a reflection of the fact that the moment-mass has a
consistent anomalous dimension at LO.  Neither the peak-scheme nor the
moment-scheme have consistent anomalous dimension equations in general.  This
inconsistency arises because of the non-locality inherent in their definitions.
This is problematic because we would like to be able to evolve our mass as a
function of $\mu$, for instance to run it up to large mass scales and connect it
to the $\overline {\rm MS}$ scheme at a scale $\mu=m_t$.  On the other hand the
position-scheme is entirely local, and so far we have demonstrated that in the
abelian limit it has a consistent mass anomalous dimension. We extend this proof
to the non-abelian case below.

In table~\ref{table:mass} we present non-abelian results for the two-loop
computation of $\delta m$ for all three schemes in Eq.~(\ref{mscheme}). The
position-scheme jet-mass is no longer one-loop exact, and has corrections at
each order in $\alpha_s(\mu)$.  Since now the coupling $\alpha_s(\mu)$ evolves,
higher powers of $\ln^j(\mu/R)$ are unavoidable. In order for the scheme to
yield an anomalous dimension of the form in Eq.~(\ref{mad}) these higher powers
must appear along with $\beta_i$ coefficients in just the right way to ensure
that the $\ln(\mu/R)$ terms do not appear in $\gamma_m$. This is precisely what
happens for the position-scheme (jet-mass scheme) at two-loop order.  Note that
the main difference between the moment scheme and the jet-mass scheme is the
presence of $C_F^2$ terms in $\delta m^{\rm mom}$, but that there are also
differences in the subleading log and constant $C_F \beta_0$ terms at two-loops.
In Ref.~\cite{Fleming:2007xt} it was proven by an explicit construction at LL
order that the moment-mass has a consistent anomalous dimension which sums the
leading logs.  Thus the one-loop analysis in Ref.~\cite{Fleming:2007xt} (which
used the moment-mass) is fully consistent.  However, beyond one-loop order it is
mandatory to use the position-scheme definition of the jet-mass given in
Eq.~(\ref{mscheme})c.
\begin{table}[t!]
\begin{center}
\begin{tabular}{l|l|l|l}
\hline
  order 
  & \hspace{1.5cm}   peak scheme  
  & \hspace{1.3cm}  moment scheme
   & \hspace{1.3cm}  jet-mass scheme
   \\ 
 & \hspace{1.8cm} $\frac{4}{\pi\Gamma_t}\: \delta m^{\rm peak} =$ 
    & \hspace{1.6cm}  $\frac{1}{R}\, \delta m^{\rm mom}=$
    & \hspace{1.4cm}  $e^{-\gamma_E} \frac{1}{R}\, \delta m_J =$
  \\[3pt]  \hline 
  & & &  \\[-8pt]
 $\alpha_s/\pi$ 
   &  $  C_F  \big[ \ln \frac{\mu}{\Gamma_t} +\frac{3}{2} \big]$
   &  $  C_F  \big[ \ln\frac{\mu}{R} +\frac{3}{2} \big]$
   & $  C_F  \big[ \ln\frac{\mu}{R} +\frac{1}{2} \big]$ 
   \\[14pt]
 $\alpha_s^2/\pi^2$ 
 &  $ C_F^2\big[\minus \ln^2\!\frac{\mu}{\Gamma_t} \plus
   (\frac{\pi^2}{3}\minus 5)\ln\frac{\mu}{\Gamma_t} \minus \frac{13}{4}\plus
   \frac{\pi^2}{2} \minus  2\zeta_3 \big]$
   &\  $ C_F^2\big[ (4\minus \frac{\pi^2}{3})\ln\frac{\mu}{R} 
    \plus 8 \minus \frac{\pi^2}{2} \minus 2\zeta_3 \big]$
   &\  0 \\[3pt]
 &   $+ C_F\beta_0\big[ \frac{1}{4}\ln^2\!\frac{\mu}{\Gamma_t} \plus
   \frac{7}{6}\ln\frac{\mu}{\Gamma_t} \plus \frac{95}{72} \plus \frac{\pi^2}{48}  \big]$
   & $+ C_F\beta_0\big[ \frac{1}{4}\ln^2\!\frac{\mu}{R} \plus
   \frac{7}{6}\ln\frac{\mu}{R} \plus \frac{131}{72} \minus \frac{\pi^2}{24} \big]$
   & $+ C_F\beta_0\big[ \frac{1}{4}\ln^2\!\frac{\mu}{R} \plus
   \frac{2}{3}\ln\frac{\mu}{R} \plus \frac{47}{72} \big]$
   \\[3pt]
 &   $+ C_F C_A\big[ (\frac{1}{3}\minus \frac{\pi^2}{12})\ln\frac{\mu}{\Gamma_t} 
   \plus \frac{17}{36} \minus \frac{\pi^2}{8} \minus \frac{5}{8}\zeta_3 \big]$
   & $+ C_F C_A\big[ (\frac{1}{3}\minus \frac{\pi^2}{12})\ln\frac{\mu}{R} 
   \plus \frac{17}{36} \minus \frac{\pi^2}{8} \minus \frac{5}{8}\zeta_3 \big]$
   & $+ C_F C_A\big[ (\frac{1}{3}\minus \frac{\pi^2}{12})\ln\frac{\mu}{R} 
   \plus \frac{5}{36} \minus \frac{\pi^2}{24} \minus \frac{5}{8}\zeta_3 \big]$
 \\[8pt]
\hline
\end{tabular}
\end{center}
\caption{Coefficients of the HQET counterterm $\delta m$  for different mass
  schemes at one and two-loop order.
  \label{table:mass}}
\end{table}

Lets extend the proof of consistency of the anomalous dimension in the
position-scheme (jet-mass scheme) to the full non-abelian case. At the same time
we will derive the very nice result that $\gamma_m$ for the jet-mass scheme is
entirely determined by the cusp-anomalous dimension.  
To all orders in perturbation theory, using Eq.~(\ref{mscheme}), the
jet-mass anomalous dimension is
\begin{align}
  \gamma_m^J &= - \frac{d\delta m(\mu)}{d\ln\mu} = - e^{\gamma_E} \frac{R}{2}\,
  \frac{d}{d\ln\mu}\, \frac{d}{d\ln(i y)}\,  \ln \tilde B(y,\mu)
  \bigg|_{i ye^{\gamma_E}=1/R} \,.
\end{align}
Using Eq.~(\ref{RGEexp}) and then Eq.~(\ref{wLfull}) this gives [$\mu_y =
e^{-\gamma_E}/(i y)$]
\begin{align}
  \gamma_m^J &
  = -e^{\gamma_E}\frac{R}{2} \, \frac{d}{d\ln\mu}\,\frac{d}{d\ln(iy)}\,K(\mu,\mu_y)
  = e^{\gamma_E}\frac{R}{2} \,\frac{d}{d\ln\mu}\, \frac{d}{d\ln\mu_y}\, K(\mu,\mu_y)
  \\
  & = e^{\gamma_E} R\: \beta[\alpha_s(\mu)] \: \beta[\alpha_s(\mu_y)] \:
  \frac{d^2 }{d\alpha_s(\mu_y)d\alpha_s(\mu)}
  \int_{\alpha_s(\mu_y)}^{\alpha_s(\mu)}
  \frac{d\alpha}{\beta[\alpha]}\, \Gamma^{\rm c}[\alpha]
  \int_{\alpha_s(\mu_y)}^\alpha \frac{d\alpha'}{\beta[\alpha']}
   \nn \,,
\end{align} 
where we should evaluate the final result at $\mu_y=R$. Performing the
derivatives with respect to the couplings we see that at any order in
perturbation theory the anomalous dimension for $m_J(\mu)$ is actually
independent of $\mu_y$.  Furthermore the result is given by the cusp-anomalous
dimension, $\gamma_m= - e^{\gamma_E}\,R\, \Gamma^c[\alpha_s(\mu)]$. Thus, to all
orders in perturbation theory the jet-mass scheme, defined by c) in
Eq.~(\ref{mscheme}), has a consistent anomalous dimension as in Eq.~(\ref{mad}),
and yields a transitive running mass, $m_J(\mu)$. The final anomalous dimension
equation for the jet-mass is
\begin{align} \label{dmdlnmu}
  \frac{dm_J(\mu)}{d\ln\mu} &= - e^{\gamma_E} R\ \Gamma^{\rm c}[\alpha_s(\mu)] \,,
\end{align}
and is fully determined by the cusp-anomalous dimension. The all-orders solution
of this equation is
\begin{align}
 m_J(\mu) = m_J(\mu_0) - \frac{e^{\gamma_E} R}{2}\: \omega(\mu,\mu_0) \,. 
\end{align}
Since $\Gamma^{\rm c}$ is known to three-loop order we can
use Eq.~(\ref{wKNNLL}) to obtain the running jet-mass at NNLL
\begin{align} \label{mjetrun}
  m_J(\mu) &= m_J(\mu_0) + e^{\gamma_E} R\: \frac{2 C_F }{\beta_0} \ln\bigg[
  \frac{\alpha_s(\mu)}{\alpha_s(\mu_0)}\bigg]
  + e^{\gamma_E} R \Big( \frac{\Gamma_1^{\rm c}}{\beta_0} - \frac{\beta_1 \Gamma_0^{\rm
       c}}{\beta_0^2}\Big)  
   \bigg[\frac{\alpha_s(\mu)-\alpha_s(\mu_0)}{8\pi}\bigg] 
  \nn\\
  &\quad  + e^{\gamma_E} R \Big( \frac{\Gamma_2^{\rm c}}{\beta_0} - \frac{\Gamma_1^{\rm
      c}\beta_1}{\beta_0^2} +\frac{\Gamma_0^{\rm c}\beta_1^2}{\beta_0^3}
   -\frac{\Gamma_0^{\rm c}\beta_2}{\beta_0^2}\Big) \bigg[
   \frac{\alpha_s^2(\mu)-\alpha_s^2(\mu_0)}{64\pi^2}\bigg]
  \,.
\end{align}
Note that the form of the anomalous dimension in $\mu d/d\mu\,[ m_J(\mu)/R]$ has
the same structure as that in $\mu d/d\mu\,[ \ln \overline{m}(\mu)]$, where
$\overline{m}(\mu)$ is the $\overline {\rm MS}$ mass. In the remaining sections
we will use the position mass-scheme and refer to it exclusively as the jet-mass.

\subsection{Relating the Jet-Mass to other Mass Schemes} \label{subsect:massB}

Having obtained a suitable mass definition for measurements of the top-mass from
jets, we now turn to perturbatively connecting it to other schemes. Using the
result for $\delta m_J$ from Table~\ref{table:mass} we obtain the two-loop
relation between the jet-mass and pole-mass 
\begin{align}
\label{mjetpole}
  m_{\rm pole}&=   m_J(\mu)  + e^{\gamma_E} R \frac{\alpha_s(\mu)\,
  C_F}{\pi} \Big[
 \ln\frac{\mu}{R} + \frac{1}{2} 
  \,\Big] \\
 &\quad + e^{\gamma_E} R  \frac{\alpha_s^2(\mu)}{\pi^2} \bigg\{ C_F\beta_0 \Big[
  \frac{1}{4}\ln^2\!\frac{\mu}{R} \plus
   \frac{2}{3}\ln\frac{\mu}{R} \plus \frac{47}{72} \Big] 
  + C_F C_A \Big[  \Big(\frac{1}{3}\minus \frac{\pi^2}{12}\Big)\ln\frac{\mu}{R} 
   \plus \frac{5}{36} \minus \frac{\pi^2}{24} \minus \frac{5}{8}\zeta_3  \Big] 
  \bigg\} \,. \nn
\end{align}
This relation can be compared to other well known two-loop mass relations, such
as i) between the pole-mass and $\overline {\rm MS}$-mass~\cite{Gray:1990yh,Fleischer:1998dw},
\begin{align} \label{MSbar}
  m_{\rm pole}&=  {\overline m}(\mu) \bigg\{ 1 
   + \frac{C_F\alpha_s(\mu)}{\pi} \bigg[1 + \frac{3}{2} \ln\!\frac{\mu}{ {\overline
     m}}\bigg]
   + \frac{\alpha_s^2(\mu)}{\pi^2} \bigg[ 
   C_F \beta_0^{(n_f+1)} \bigg( \frac{3}{8} \ln^2\!\frac{\mu }{\overline m_t} 
    + \frac{13}{16} \ln\frac{\mu }{\overline m_t} + \frac{71}{128} 
  + \frac{\pi^2}{16} \bigg)
  + C_F \Big( \frac{\pi^2-3}{8}\Big)
  \nn\\
 & \hspace{-0.5cm}
  + C_F C_A\bigg(  \frac{7}{8} \ln\frac{\mu }{\overline m}
   + \frac{55}{64} -\frac{5\pi^2}{16} \plus \frac{\pi^2 \ln 2}{4}- \frac{3\zeta_3}{8} 
  \bigg) 
+ C_F^2\bigg(  \frac{9}{8}\ln^2\!\frac{\mu }{\overline m}
  -\frac{9}{16} \ln\frac{\mu }{\overline m}
   - \frac{71}{128} \plus\frac{5\pi^2}{16} -\frac{\pi^2 \ln 2}{2}\plus \frac{3\zeta_3}{4} 
  \bigg) 
  \bigg] \bigg\} \,,
\end{align}
where $\overline m = \overline m(\mu)$, $\beta_0^{(n_f+1)}=\beta_0-2/3$, and
$\alpha_s(\mu)$ here is in the $(n_f+1)$-flavor theory, and
ii) between the pole-mass and 1S-mass \cite{Hoang:1998hm},
\begin{align} \label{M1S}
  m_{\rm pole} &=  m^{1S} \bigg\{ 1 + \frac{C_F^2\alpha_s^2(\mu) }{8} 
   + \frac{C_F^2 \alpha_s^3(\mu)}{8\pi} \bigg[ \beta_0 \ln\Big(\frac{\mu}{C_F
     \alpha_s(\mu) m^{1S} }\Big) + \frac{11\beta_0}{6} - \frac{4 C_A}{3}
   \bigg] \bigg\} \,.
\end{align}
Here $\alpha_s(\mu)$ is in the $n_f$-flavor theory. Note that in
Eqs.~(\ref{MSbar}) and (\ref{M1S}) we have $n_f=5$ light massless flavors, and
thus did not write for example vacuum polarization terms depending on the
b-quark mass.

Lets imagine that the jet-mass $m_J(\mu_J)$ is determined from a fit to massive
event shapes using a scale $\mu_J\sim\Gamma$.  In order to connect this
$m_J(\mu_J)$ to the high-energy $\overline {\rm MS}$-mass, we proceed as
follows. First because the renormalization group evolution of the $\overline {\rm MS}$-mass does
not make sense below the mass itself, we evolve the jet-mass $m_J(\mu_J)$ up to
the scale $\overline m_t=\overline m(\overline m_t)$ to obtain $m_J(\overline
m_t)$ using the NNLL running result in Eq.~(\ref{mjetrun}).  At this scale we
then connect the jet and $\overline {\rm MS}$-masses by eliminating the pole
mass from Eqs.~(\ref{mjetpole},\ref{MSbar}). Thus the two-loop relation between
the jet-mass and $\overline {\rm MS}$-mass is
\begin{align} \label{Mbartojet}
    \overline {m}(\overline {m}_t) &=  m_J(\overline{m}_t) 
 + \bigg\{ e^{\gamma_E} R \frac{\alpha_s(\overline{m}_t)C_F}{\pi}  \Big[
 \ln\frac{\overline{m}_t}{R} 
  + \frac{1}{2} \Big]- m_J(\overline{m}_t)
 \frac{\alpha_s(\overline{m}_t)C_F}{\pi}  \bigg\} 
  +\bigg\{
  - m_J(\overline{m}_t)  \frac{\alpha_s^2(\overline{m}_t)}{\pi^2} \bigg[
    C_F \bigg( \frac{\pi^2}{12} \minus \frac{143}{192}\bigg)
  \nn\\
  & \hspace{-1cm} 
  + C_F \beta_0 \bigg(  \frac{71}{128} \plus  \frac{\pi^2}{16} \bigg)
  \plus C_F C_A \bigg( \frac{55}{64} \minus \frac{5\pi^2}{16} \plus \frac{\pi^2 \ln 2}{4}
  \minus  \frac{3\zeta_3}{8} \bigg) 
  \plus C_F^2 \bigg( \frac{5\pi^2}{16} \minus \frac{199}{128} \minus \frac{\pi^2\ln 2}{2}   
   +\frac{3\zeta_3}{4} \bigg)
  \bigg]  
  \nn\\
 & \hspace{-1cm} 
    + e^{\gamma_E} R  \frac{\alpha_s^2(\overline{m}_t)}{\pi^2} \bigg[ C_F\beta_0 \Big[
  \frac{1}{4}\ln^2\!\frac{\overline{m}_t}{R}\, \plus
   \frac{2}{3}\ln\!\frac{\overline{m}_t}{R} \,\plus \frac{47}{72} \Big] 
  + C_F C_A \Big[ \Big(\frac{1}{3}\minus \frac{\pi^2}{12}\Big)\ln\frac{\overline{m}_t}{R} 
   \,\plus \frac{5}{36} \minus \frac{\pi^2}{24} \minus \frac{5}{8}\zeta_3  \Big] 
  - C_F^2 \Big[ \ln\frac{\overline{m}_t}{R} + \frac{1}{2} \Big] \bigg] \bigg\} \,.
\end{align}
Since we take $\mu = \overline m_t$ there is no threshold correction at the order we are working, and $\alpha_s(\overline m_t)$ in Eq.~(\ref{Mbartojet}) is the same in the $n_f$ and $(n_f+1)$-flavor theories. Together with Eq.~(\ref{mjetrun}) this formula inputs a jet-mass determined from
production of tops far above threshold, and outputs an $\overline {\rm MS}$ mass
that can be used in other processes, such as the analysis of precision
electroweak data. Since a high precision result for the top-mass in the
1S mass-scheme can be determined from a threshold cross-section analysis, we
also quote the two-loop conversion between the jet-mass and 1S-mass schemes.
An extra power of $\alpha_s$ is kept in the 1S-scheme terms to properly
ensure a renormalon free series~\cite{Hoang:1998hm}. At a scale $\mu$ the
conversion at second order is
\begin{align}\label{1Stojet}
     {m}_t^{1S} &=  m_J(\mu) 
 + \bigg\{ e^{\gamma_E} R \frac{\alpha_s(\mu)C_F}{\pi}  \Big[
 \ln\frac{\mu}{R} + \frac{1}{2} \Big]
 - m_J(\mu) \frac{ \alpha_s^2(\mu) C_F^2}{8} \bigg\} 
  \nn\\
  &   +\bigg\{ - e^{\gamma_E} R  \frac{\alpha_s^3(\mu)C_F^3}{8\pi}\Big[
 \ln\frac{\mu}{R} + \frac{1}{2} \Big] 
  - m_J(\mu)  \frac{ \alpha_s^3(\mu)C_F^2}{8\pi} \bigg[
   \beta_0 \ln\Big(\frac{\mu}{C_F\alpha_s(\mu)m_J(\mu)}\Big) +\frac{11\beta_0}{6}
  -\frac{4 C_A}{3}  \bigg] 
  \nn\\
 & + e^{\gamma_E} R  \frac{\alpha_s^2(\mu)}{\pi^2} \bigg[ C_F\beta_0 \Big[
  \frac{1}{4}\ln^2\!\frac{\mu}{R}\, \plus
   \frac{2}{3}\ln\!\frac{\mu}{R} \,\plus \frac{47}{72} \Big] 
  + C_F C_A \Big[ \Big(\frac{1}{3}\minus \frac{\pi^2}{12}\Big)
   \ln\frac{\mu}{R}
   \,\plus \frac{5}{36} \minus \frac{\pi^2}{24} \minus \frac{5}{8}\zeta_3  \Big]
   \bigg]
  \bigg\} 
  \,.
\end{align}

\section{Results for the NNLL Jet Function } \label{sect:results}

In this section we present the final result for the heavy quark jet function $B(\hat
s,\delta m,\Gamma_t,\mu)$, with NNLO perturbative corrections and a NNLL
resummation of large logs. We study the numerical effect of these two-loop
corrections as well as of the log-resummation, including the perturbative
convergence and $\mu$-dependence of $B$ as a function of $\hat s$, and in
particular the stability of its peak position which is important for a top-mass
measurement. At tree-level $B(\hat s,\delta m,\mu)=\delta(\hat s)$ and we
see from Eq.~(\ref{eq:dr}) that $B(\hat s,\delta m,\Gamma_t,\mu)$ is simply a
Breit-Wigner centered at $\hat s=0$ with a width $\Gamma_t$. Beyond tree-level
the jet function becomes dependent on $\mu$ and on the choice of mass-scheme
through $\delta m$.

For the cross-section $d^2\sigma/dM_t^2 dM_{\bar t}^2$ in Eq.~(\ref{Fthm}) it
has been proven that at any order in perturbation theory, the only large logs
that effect the shape of the invariant mass distribution are those due to the
resummation in the heavy-quark jet function~\cite{Fleming:2007xt}.\footnote{In
  principle both the logs in the jet function and in the soft-function can
  modify the invariant mass distribution. However due to the consistency
  conditions discussed in Ref.~\cite{Fleming:2007xt} it is always possible to
  exchange a summation of large logs in the soft function in favor of large logs
  in the jet function and in the hard function normalization factors. }
Furthermore these large logs only exist between scales $\mu_\Gamma\sim
\Gamma\equiv \Gamma_t+ Q\Lambda_{\rm QCD}/m$ and $\mu_\Lambda \gtrsim
\Lambda_{\rm QCD}+ m\Gamma_t/Q$.  The remaining large logs only modify the
cross-sections normalization. The expression which resums all logs between the
scales $\mu_Q\simeq Q\gg \mu_m\simeq m\gg \mu_\Gamma \simeq \Gamma\gg
\mu_\Lambda\gtrsim \Lambda_{\rm QCD}$ is
\begin{align} \label{Fthm2}
  \frac{d^2\sigma}{dM_t dM_{\bar t} } &=
    4 \sigma_0 M_t M_{\bar t}\, H_Q(Q,\mu_Q) U_{H_Q}(Q,\mu_Q,\mu_m) H_m(m_J,\mu_m)
  U_{H_m}(Q/m_J,\mu_m,\mu_\Lambda) 
   \\
 & \hspace{-1cm}
\times \int_{-\infty}^{+\infty}\!\!\! d\ell^+\,d\ell^- 
  B_+\Big(\hat s_t -\frac{Q\ell^+}{m_J} ,\delta m_J,\Gamma_t,\mu_\Lambda,\mu_\Gamma\Big) 
  B_-\Big(\hat s_{\bar t} -\frac{Q\ell^-}{m_J},\delta m_J,\Gamma_t,\mu_\Lambda,\mu_\Gamma\Big) 
  S\big(\ell^+,\ell^-,\mu_\Lambda,\delta,\bar\Delta(\mu_\Lambda)\big) 
  \,,\nn
\end{align}
where we have defined the resummed jet function as
\begin{align} \label{eq:bmumu}
  B(\hat s , \delta m_J,\Gamma_t, \mu_\Lambda,\mu_\Gamma) &\equiv 
   \int\!\! d\hat s'\  U_B(\hat s-\hat s',\mu_\Lambda,\mu_\Gamma)
  \  B(\hat s',\delta m_J,\Gamma_t,\mu_\Gamma) 
  \nn\\
  &= 
   \int\! d\hat s'\, d\hat s''\  U_B(\hat s-\hat s',\mu_\Lambda,\mu_\Gamma)
  \  B(\hat s'-\hat s'',\delta m_J,\mu_\Gamma) \ \frac{\Gamma_t}{\pi(\hat
    s^{\prime\prime\,2}+\Gamma_t^2)} \,.
\end{align}
In Eqs.~(\ref{Fthm2},\ref{eq:bmumu}) large logs are resummed by the evolution
factors $U_{H_Q}$, $U_{H_m}$, and $U_B$, and of these, the first two only affect
the overall normalization. Since the scales $\mu_\Gamma$ and $\mu_\Lambda$
differ by a factor of $Q/m$ it is necessary to sum the large logs between these
scales. Recall that Eq.~(\ref{Fthm2}) is valid for $Q\gg m$, which is mandatory
for the top quark and antitop quarks to decay to well separated jets.  The
numerical importance of this particular resummation has already been
demonstrated at NLL order in Ref.~\cite{Fleming:2007xt}.

In the following we study the resummed jet function $B(\hat s , \delta
m,\Gamma_t, \mu_\Lambda,\mu_\Gamma)$ and its dependence on $\hat s$ and
$\mu_\Gamma$. In particular the $\mu_\Gamma$ dependence cancels out
order-by-order in renormalization group improved perturbation theory, and thus
the residual $\mu_\Gamma$ dependence provides a method for estimating the effect
of higher order corrections to the jet function. This $\mu_\Gamma$ dependence
cancels order-by-order between the evolutor $U_B(\hat s-\hat
s',\mu_\Lambda,\mu_\Gamma)$ and the fixed-order jet function matrix element that
gives $B(\hat s'-\hat s'',\delta m,\mu_\Gamma)$ in Eq.~(\ref{eq:bmumu}). On the
other hand, the dependence of the resummed jet function on $\mu_\Lambda$ cancels
out only in the complete cross-section, where there is additional dependence on
$\mu_\Lambda$ in both the evolution function $U_{H_m}$ and the soft-function
$S$. The analysis of the invariant mass dependence of the full NNLL
cross-section requires constructing a consistent model for the soft-function at
two-loop order, since $S$ contains both perturbative and non-perturbative
pieces. The procedure in Ref.~\cite{Hoang:2007vb} can be used to carry out this
analysis, but we leave the study of the full cross-section to a future
publication. Here we focus on the resummed jet function.

Following the strategy in appendix~E of Ref.~\cite{Fleming:2007xt} we can obtain analytic
results for the NNLL jet function even in the presence of the width.
At NNLL order we find
\begin{align} \label{BNNLL}
 &m  B(\hat s , \delta m,\Gamma_t, \mu_\Lambda,\mu_\Gamma) = 
    G_0 + \frac{C_F\alpha_s(\mu_\Gamma)}{\pi} \Big[
    G_2-G_1 + \Big(1+\frac{5\pi^2}{24}\Big) G_0 \Big] 
  - \frac{2\alpha_s(\mu_\Gamma)}{\pi}\, \delta m_1(\mu_\Gamma)\, (G_0)^{\prime} 
  \nn\\
 &\quad
 + \frac{\alpha_s^2(\mu_\Gamma)}{\pi^2} \bigg \{ 
C_F^2 \bigg [  \frac{1}{2} G_4 - G_3  + \Big( \frac{3}{2} +\frac{13\pi^2}{24} \Big) G_2 
 - \Big( 1+\frac{13\pi^2}{24} - 4\zeta_3 \Big) G_1 
 + \Big(  \half  + \frac{7\pi^2}{24} + \frac{53\pi^4}{640} - 2\zeta_3 \Big) G_0  \bigg ] 
  \nn \\[2pt] 
&\quad\qquad\qquad
   + C_F C_A \bigg [  \Big(  \frac{1}{3} - \frac{\pi^2}{12} \Big) G_2 
   - \Big(  \frac{5}{18}-\frac{\pi^2}{12} -\frac{5\zeta_3}{4}  \Big) G_1 
   + \Big(\! -\frac{11}{54} + \frac{5\pi^2}{48} -\frac{19\pi^4}{960}
   -\frac{5\zeta_3}{8} \Big) G_0
  \bigg] 
\nn \\[2pt]
&\quad\qquad\qquad
 + C_F \beta_0 \bigg [ -\frac{1}{6} G_3  + \frac{2}{3} G_2 
   - \Big(\frac{47}{36} +\frac{\pi^2}{12}  \Big) G_1  
   + \Big(\frac{281}{216} +\frac{23\pi^2}{192}-\frac{17\zeta_3}{48} \Big) G_0
  \bigg ] ~\Bigg \} 
 \nn\\[2pt]
&\quad
 - \frac{2 \alpha_s^2(\mu_\Gamma)}{\pi^2} \Bigg\{
  \delta m_2 \, (G_0)^\prime  - (\delta m_1)^2 \, (G_0)^{\prime\prime}
   + \delta m_1\, {C_F} \bigg[
   (G_2)^\prime - (G_1)^\prime +\Big(1+\fr{5\pi^2}{24}\Big)(G_0)^\prime \bigg]~ \Bigg\}
\, . 
\end{align}
The result is expressed in terms of the functions $G_n=G_n(\hat
s,\Gamma_t,\mu_\Lambda,\mu_\Gamma)$ and their $\hat s$ derivatives, with 
\begin{align}
  G_n = \frac{1}{\pi}\: {\rm Im}\: \Bigg[ \frac{e^{K} (\mu_\Gamma e^{\gamma_E})^\omega\,
    \Gamma(1+\omega)}{(-\hat s-i\Gamma_t)^{1+\omega} } \ I_n\bigg(\frac{\hat
    s+i\Gamma_t}{\mu_\Gamma}, \omega\bigg) \Bigg]\,.
\end{align} 
Here $\omega=\omega(\mu_\Lambda,\mu_\Gamma)$ and $K=K(\mu_\Lambda,\mu_\Gamma)$
are given in Eq.~(\ref{wKNNLL}) and
\begin{align}
 I_0(x,\omega) & = 1\,, 
   \\[2mm]
 I_1(x,\omega)\, & =  \ln(-x\minus i0)-H(\omega) \,,
   \nn\\[2mm]
 I_2(x,\omega) & =
 \big[\ln(-x\minus i0)-H(\omega)\big]^2 + \psi^{(1)}(1\plus \omega)-\zeta_2
   \,,
   \nn\\[2mm]
 I_3(x,\omega) & =
 \big[\ln(-x\minus i0)-H(\omega)\big]^3
  +3 \big[\psi^{(1)}(1\plus \omega)-\zeta_2 \big]\big[\ln(-x\minus i0)-H(\omega)\big]
   +\psi^{(2)}(1) - \psi^{(2)}(1\plus \omega)
   \,,
   \nn\\[2mm]
 I_4(x,\omega) & = 
   \big[\ln(-x\minus i0)-H(\omega)\big]^4 
   +6 \big[\psi^{(1)}(1\plus \omega)-\zeta_2 \big]\big[\ln(-x\minus
   i0)-H(\omega)\big]^2 
  \nn\\
 &  -4 \big[\psi^{(2)}(1\plus \omega)-\psi^{(2)}(1) \big]\big[\ln(-x\minus i0)-H(\omega)\big]
   + \psi^{(3)}(1\plus \omega)-\psi^{(3)}(1) 
   +3\big[\psi^{(1)}(1\plus \omega) -\zeta_2\big]^2
  \,, \nn
\end{align}
with $H(\omega)$ the harmonic-number function, and $\psi^{(k)}(x)$ the $k$'th
derivative of the digamma function or equivalently the $(k+1)$'th derivative of
the log of the gamma function.

We focus our numerical analysis on two mass schemes for the jet function. In the
pole scheme we take $\delta m_1=\delta m_2=0$ in Eq.~(\ref{BNNLL}) and use a
fixed pole mass $m=m^{\rm pole}=172\,{\rm GeV}$ in the formula for $\hat s$ in
Eq.~(\ref{shat}). In the jet-mass scheme we use $\delta m_1$ and $\delta m_2$
from the last column of Table~\ref{table:mass} with a scheme parameter
$R=0.8\,{\rm GeV}$ that corresponds to a scale $e^{\gamma_E} R\simeq 1.4\,{\rm
  GeV}$.  Here $m=m_J(\mu_\Gamma)$ is the mass in the jet-scheme, and
\begin{align}
 \hat s = \frac{M_t^2-m_J^2(\mu_\Gamma)}{m_J(\mu_\Gamma)} \,.
\end{align}
The value of $m_J(\mu_\Gamma)$ to be used here is obtained using the evolution
equation in Eq.~(\ref{mjetrun}), running up from an input scale $\mu_0$. For
this scheme it is the parameter $m_J(\mu_0)$ that one will extract with future
linear-collider data.  In our analysis we take $\mu_0=2\,{\rm GeV}$ and simply
fix $m_J(\mu_0)=172\,{\rm GeV}$.  We use the three-loop result for the running
coupling everywhere,
\begin{align} \label{as3}
  \frac{1}{\alpha_s(\mu)}  = \frac{X}{\alpha_s(\mu_0)}
  +\frac{\beta_1}{4\pi\beta_0}  \ln X + \frac{\alpha_s(\mu_0)}{16\pi^2} 
  \bigg[ \frac{\beta_1^2}{\beta_0^2} \Big( \frac{\ln X}{X} +\frac{1}{X} -1\Big)
  + \frac{\beta_2}{\beta_0} \Big(1-\frac{1}{X}\Big) \bigg] 
  \,,
\end{align}
where $X\equiv 1+\alpha_s(\mu_0)\beta_0 \ln(\mu/\mu_0)/(2\pi)$ and we evolve to
lower scales using the reference value $\alpha_s(\mu_0=m_Z)=0.118$ with $n_f=5$.
Since we have systematically treated the $b$-quark as massless we also ignore
the $b$-quark threshold in our coupling evolution. We also fix
$\Gamma_t=1.43\,{\rm GeV}$, and $\mu_\Lambda=1\,{\rm GeV}$. For $\mu_\Gamma$ we
take a central value of $\mu_\Gamma=5\,{\rm GeV}$ and consider variations about
this scale in the range $3.3\; {\rm GeV}<\mu_\Gamma< 7.5\; {\rm GeV}$.  Even
though it may slightly underestimate higher-order uncertainties, we have chosen
not to make the canonical choice of varying $\mu_\Gamma$ up and down by a factor
of two because of the importance of retaining the hierarchy
$\mu_\Gamma/\mu_\Lambda \simeq Q/m_J$ as emphasized in
Ref.~\cite{Fleming:2007xt}.

\begin{figure}[t!]
  \centerline{ 
   \includegraphics[width=8.5cm]{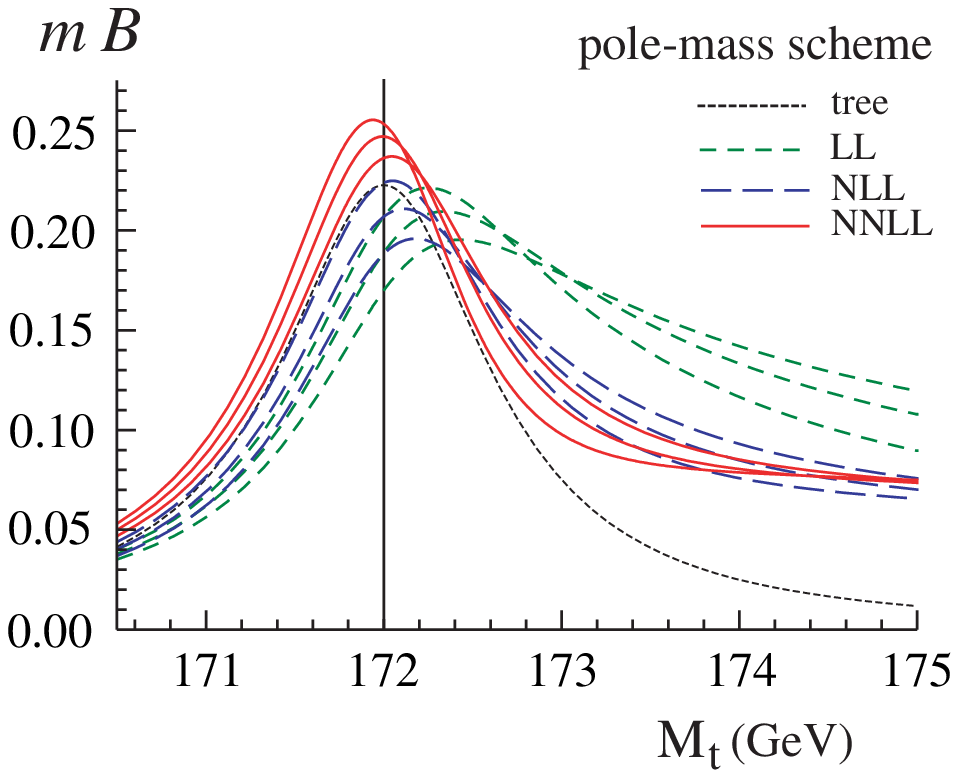} \qquad
   \includegraphics[width=8.5cm]{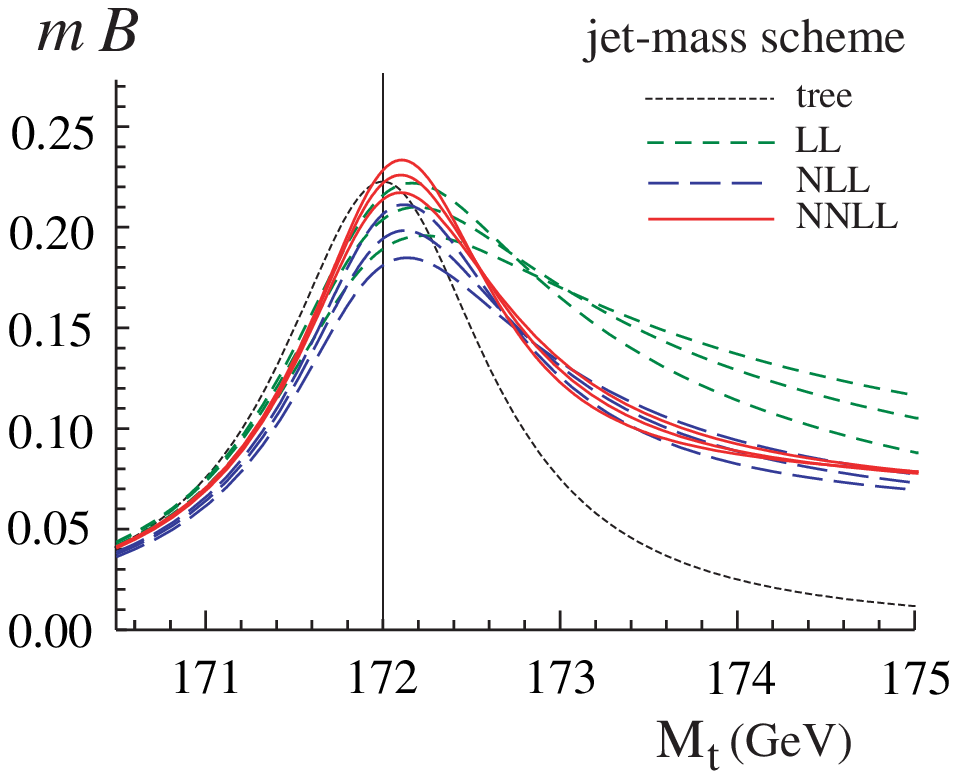}
  }
\vskip-0.2cm
\caption{
  The jet function, $m B(\hat s,\delta m,\Gamma_t,\mu)$ versus $M_t$, where
  $\hat s=(M_t^2-m^2)/m$ and $\Gamma_t=1.43\,{\rm GeV}$. The left panel shows
  results in the pole-mass scheme and the right panel shows results in the
  jet-mass scheme. The black dotted curve is the tree-level Breit-Wigner, the
  green short-dashed curves are LL results, blue long-dashed curves are NLL, and
  the solid red curves are at NNLL order. For each of the LL, NLL, and NNLL results we
  show three curves with $\mu_\Gamma=3.3,5.0,7.5\,{\rm GeV}$ respectively. Other
  parameters are discussed in the text.  }
\label{fig:BSrun}
\end{figure}
In Fig.~\ref{fig:BSrun} we plot the resummed heavy-quark jet function $B(\hat s
, \delta m,\Gamma_t, \mu_\Lambda,\mu_\Gamma)$ versus $M_t$.  In the left panel
we show results for the pole scheme, and in the right panel we show results for
the jet-scheme. In each panel we plot tree level results (black dotted-line), LL
results (green short-dashed lines), NLL results (blue long-dashed lines), and
NNLL results (solid red lines). The tree results are the pure Breit-Wigner, the
LL results use the tree-level $B(\hat s,\delta m,\mu)$ with the LL result for
$U_B$ in Eq.~(\ref{eq:bmumu}), and thus correspond to just the first $G_0$ term
in Eq.~(\ref{BNNLL}). From the LL results we see that beyond tree-level the jet
function grows a perturbative tail above the peak. The NLL results use the
one-loop $B$ with the NLL result for $U_B$ and thus include the ${\cal
  O}(\alpha_s(\mu_\Gamma))$ terms in Eq.~(\ref{BNNLL}), and the NNLL curves use
the two-loop $B$ with the NNLL result for $U_B$ and thus all of the terms in
Eq.~(\ref{BNNLL}).  At each of the LL, NLL, and NNLL orders we show three curves
with $\mu_\Gamma=3.3, 5.0, 7.5\,{\rm GeV}$, which are the curves from top to
bottom near the peak respectively.  Recall that in the jet scheme we fix
$m_J(\mu_0\!  =\!2\,{\rm GeV})=172\,{\rm GeV}$ and use the solution of the mass
renormalization group equation in Eq.~(\ref{mjetrun}). Thus the conversion from $M_t$ to $\hat s$ depends
on the value of $\mu_\Gamma$ and order-by-order compensates for the $\mu_\Gamma$
dependence of the residual mass terms $\delta m_{1,2}(\mu_\Gamma)$.

Examining the LL, NLL, and NNLL results for the jet function in
Fig.~\ref{fig:BSrun} we observe that the jet-scheme results in the second panel
exhibit better perturbative convergence than the pole-scheme results in the
first panel.  This is true of all features, including the slope to the left of
the peak, the perturbative tail to right of the peak, the peak location, and the
peak height.  Comparing the spread of the curves we see that at both NLL and
NNLL order the residual $\mu_\Gamma$ dependence is smaller in the jet-scheme
than in the pole-scheme. The numerical size of the residual $\mu_\Gamma$ scale
dependence varies region by region. In the pole-mass scheme the scale dependence
in the slope before the peak is $\sim 17\%$ at NLL and $\sim 14\%$ at NNLL,
while the maximum variation near the peak is $23\%$ at NLL and $17\%$ at NNLL,
and then in the tail region well above the peak it is $\sim 19\%$ at NLL and
$\sim 13\%$ at NNLL. Hence, in the pole scheme including the NNLL results does
not significantly decrease the $\mu_\Gamma$ dependence. In the jet-mass scheme
the scale dependence in the slope before the peak is $\sim 6\%$ at NLL and $\sim
2\%$ at NNLL, while the maximum variation near the peak is $14\%$ at NLL and
$7\%$ at NNLL, and then in the tail above the peak it is $\sim 12\%$ at NLL and
$\sim 5\%$ at NNLL. Thus, in the jet-mass scheme the $\mu_\Gamma$ dependence is
reduced by a factor of two or more. The same level of improvement is observed for
different values of the scheme parameter $R$ than the value used in our
analysis.

\begin{figure}[t!]
  \centerline{ 
   \includegraphics[width=12cm]{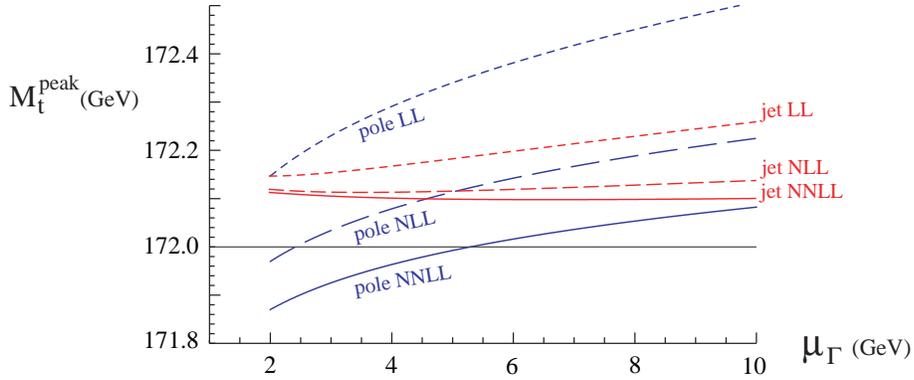} 
   } 
\vskip-0.2cm
\caption{ 
  Peak position $M_t^{\rm peak}$ of the jet function versus $\mu_\Gamma$.
  Short-dashed results are at LL order, long-dashed are at NLL order, and solid
  are at NNLL order. Results are labeled for the pole mass-scheme (blue) and
  jet mass-scheme (red). }
\label{fig:peak}
\end{figure}
In Fig.~\ref{fig:peak} we plot the peak position $M_t^{\rm peak}$ of the jet
function curves, versus $\mu_\Gamma$. This figure displays the convergence and
$\mu_\Gamma$ dependence of the jet function peak position in more detail than
Fig.~\ref{fig:BSrun}.  The stability of the jet function peak has a direct
influence on the peak of the cross-section, and both are very sensitive to the
value of the short-distance top-mass. Hence the peak-position is important to
gauge the effect of perturbative corrections for the mass measurement.  We use a
wider range for $\mu_\Gamma$ than that of the curves in Fig.~\ref{fig:BSrun},
but note that results for $\mu_\Gamma\le 3\,{\rm GeV}$ upset the hierarchy
$\mu_\Gamma/\mu_\Lambda\simeq 5$ and hence can be safely ignored. In the
pole-mass scheme we observe that there is limited sign of convergence for the
peak position, although the shifts with $\mu_\Gamma=5\,{\rm GeV}$ at each order
are still relatively small being $\simeq 230\,{\rm MeV}$ from LL to NLL order
and $\simeq 120\,{\rm MeV}$ from NLL to NNLL order.  The lack of convergence of
the peak-position in the pole-scheme is a reflection of the infrared renormalon
in the pole-mass, which destabilizes perturbative predictions. We also observe
from Fig.~\ref{fig:peak} that the $\mu_\Gamma$ dependence of the peak-position
is not reduced in going from LL, to NLL, to NNLL in the pole-scheme. However, in the
jet-mass scheme the peak location converges nicely from LL to NLL to NNLL, with
a numerical value of $M_t^{\rm peak}=172.099\,{\rm GeV}$ at NNLL order. At
$\mu_\Gamma=5\,{\rm GeV}$ the shifts are $\simeq 67\,{\rm MeV}$ from LL to NLL
and $\simeq 17\,{\rm MeV}$ from NLL to NNLL. Also in the jet-scheme,
Fig.~\ref{fig:peak} shows that the $\mu_\Gamma$ dependence of the peak-position
decreases from LL, to NLL, to NNLL order, with the curves becoming flatter as
the order increases. The residual $\mu_\Gamma$ scale dependence of the peak
position in the jet-scheme is $\delta M_t^{\rm peak}=0.005\,{\rm GeV}$, where we
quote the difference in $M_t^{\rm peak}$ from $\mu_\Gamma=3\,{\rm GeV}$ to
$\mu_\Gamma=10\,{\rm GeV}$.

Utilizing the two-loop computation with NNLL renormalization group improvement,
and a jet-mass scheme with good renormalization group behavior, we have achieved
stable results for the heavy quark jet function.  However, we caution that the
final result for the jet function is dependent on the choice of $\mu_\Lambda$,
and so a more detailed phenomenological analysis must be made only after
combining the results reported here with perturbative corrections in the soft
function to yield a $\mu_\Lambda$ independent prediction for the cross-section.
In particular the size of the perturbative tail above the peak in the
cross-section is affected by both perturbative corrections to the jet function
and soft-function, and is strongly $\mu_\Lambda$ dependent in each of these
functions individually.

\section{Conclusion}  \label{sect:conc}

Effective field theories are an important tool for making high precision
predictions for jet observables, and facilitate a measurement of the top mass
with theoretical uncertainty less than ${\cal O}(\Lambda_{\rm QCD})$. Starting
from the expression for the cross section for double top production at an
$e^+e^-$ collider, Eq.~(\ref{Fthm}) derived in Ref.~\cite{Fleming:2007qr}, we
have studied the properties of the heavy-quark jet function $B$ at higher loop
orders.  The function $B$ can be calculated in HQET, and was studied at one-loop
and NLL order in Ref.~\cite{Fleming:2007xt}. Here we have performed the 2-loop
computation of $B$ to obtain a NNLO result. Our analysis also yielded the
two-loop anomalous dimension of $B$, which when combined with the three-loop
cusp anomalous dimension from Ref.~\cite{Moch:2004pa} was used to obtain a
renormalization group improved heavy quark jet function at NNLL order.  Using
the formulation in terms of vacuum matrix elements of Wilson lines we have
explained precisely how the jet function is different from the heavy-quark shape
function and fragmentation function that have been considered previously in the
literature at two-loop
order~\cite{Korchemsky:1992xv,Becher:2005pd,Gardi:2005yi,Neubert:2007je}.

The two-loop computation also allowed us to study the higher loop behavior of
$B$ and arrive at a suitable definition of a short distance top mass-scheme, for
higher order analysis. In particular we gave a definition for a jet-mass scheme
with nice renormalization group properties, and demonstrated that in this scheme
the mass anomalous dimension is determined by the cusp anomalous dimension to
all orders in perturbation theory. To study the properties of different
mass-scheme definitions we exploited the fact that the heavy-quark jet function
in position space fulfills the requirements to obey the non-abelian
exponentiation theorem of Refs.~\cite{Gatheral:1983cz,Frenkel:1984pz}. In
particular in the abelian limit ($C_A,\; n_f\rightarrow 0$) the all orders
result for $B$ is simply the exponentiated one-loop result.  We considered
differences between a peak-position mass definition, a first moment mass
definition, and a mass definition based on the position space jet function.  We
have checked that among these three possibilities, the peak definition and
moment definition do not yield consistent mass renormalization group equations
at LO and NLO order respectively. Thus only the position space definition
provides a reasonable way of defining the jet-mass beyond LL order.  This
definition is given in Eq.~(\ref{mscheme})c.  Relations between the jet-mass and
the $\overline {\rm MS}$, 1S, and pole masses were also given at two-loop order. The
proof that the jet-mass is a short-distance mass, free from leading order
renormalon ambiguities, is given in Ref.~\cite{FHMSmass:inprep}.

Our final result for the heavy-quark jet function $B$ uses the jet-mass scheme
with NNLO fixed order results and a NNLL resummation of large logarithms, which
we refer to as the NNLL order result. The logs in this summation are a well
defined set for a physical observable, being the only large logs that effect the
shape of the top-invariant mass cross-section $d^2\sigma/dM_t^2dM_{\bar t}^2$.  We
have studied the numerical stability of $B$, both in terms of perturbative
convergence from LL, to NLL, to NNLL order, and with respect to its scale
dependence. In the jet-mass scheme the convergence of $B$ improves by a factor
of two or better in going from NLL to NNLL order. Very stable results were also
obtained for the peak position of the heavy-quark jet function, with residual
perturbative uncertainties estimated to be at the $5\,{\rm MeV}$ level. This
level of precision and stability for the jet function and its peak-position were
not observed in the pole-mass scheme.

Future applications of our work include the extension to complete NNLL results
for the cross-section $d^2\sigma/dM_t^2dM_{\bar t}^2$ by including the
convolution with the soft-function and its perturbative and non-perturbative
components. The use of position space as a convenient way of defining a top
mass-scheme, could also be extended to the b-quark where currently a moment mass
scheme, called the shape-function scheme~\cite{Bosch:2004th}, is often employed.
Based on our analysis we expect that this shape function mass also does not have
a consistent anomalous dimension beyond LL order, but that this can be
rectified by using a modified definition of the scheme in position space.
Finally, the same position space technique can be applied to the definition of
the renormalon free gap parameter~\cite{Hoang:2007vb}, a parameter which is
important for combining perturbative and non-perturbative results for the
soft-function for jet production.
 
\section*{Acknowledgments}
We would like to thank A.~Hoang for helpful comments.  This work was
supported in part by the Department of Energy Office of Nuclear Science under
the grant DE-FG02-94ER40818.  I.~Scimemi was also supported in part by a Marie
Curie International Fellowship from the European Union, grant number 021379
(BDECMIT), and thanks the Fundaci\'o Bosch i Gimpera and J.~Soto of the
University of Barcelona (Spain) for support.  I.W.~Stewart was also supported in
part by the DOE Outstanding Junior Investigator program and Sloan Foundation.


\appendix

\section{Two-Loop Graphs and Renormalization} \label{app:2loop}

In this section we briefly summarize results for the two-loop jet function
graphs and their renormalization factors. We use Feynman gauge and dimensional
regularization with $d=4-2\epsilon$. Numbering the two-loop graphs in
Fig.~\ref{fig:2loop} from left-to-right and top-to-bottom we have
\begin{align} \label{B2bareSum}
 \iota^{2\epsilon}\mu^{4\epsilon} \: {\cal B}_2^{\rm bare}(\hat s)
  &= \sum_{i=1}^{16} G_i \,,
  & G_i & = \frac{i\alpha_s^2(\mu)}{4\pi^2 a} \ \hat G_i \,,
\end{align}
where we have defined $\hat G_i$ by pulling out a common prefactor, and we let
$a=\hat s+i 0$ and $\iota = \exp(\gamma_E)/(4\pi)$. The sum of terms for ${\cal
  B}_2^{\rm bare}$ is gauge invariant. In Feynman gauge the results for the $\hat
G_i$'s in terms of master integrals are
\begin{align} \label{Gi}
 \hat G_1 & =  4 C_F T_F n_f h_{1\epsilon} + 4 C_F^2 h_{2\epsilon}  - C_F
  C_A h_{3\epsilon}   \, ,  
 & \hat G_2 &=  8 C_F^2 \,  F_0(1,1,1) F_0(1,1,0)
  \\
 \hat G_3 &= 2  C_F^2 \,  \big[ F_0(1,1,1) \big]^2 \,,
 & \hat G_4 &=  - 8  C_F^2  \,  F(101,210,100) \,, 
  \nn \\
 \hat G_5 &=  - 8 C_F \Big( C_F - \half  C_A\Big)  F(101,111,100) \, ,
 & \hat G_6 &=  - 2 C_F\Big( C_F-\half C_A \Big) F(101,111,101)  \,,
  \nn \\
 \hat G_7 &= 
  C_F C_A \Big[ F(111,100,010) -2\,   F(111,100,100) + 4\, F(111,110,10 \,
  -\!\!1)   +2 \, F(111,110,000) +    F(111,110,100) \Big] ,
  \hspace{-9cm} & & \hspace{9cm}
 \nn \\
 \hat G_8 &=   C_F C_A  F (111,110,100)  \,,
 & \hat G_9 &=  -4 C_F^2\, F(101,110,101) +2 C_F C_A\, F(101,110,011) \, ,
  \nn \\
 \hat G_{10} &= -
 C_F C_A   \big[ 2\, F(111,010,010) - F(111,010,100) \big] \, , 
   \hspace{-2cm} & & \hspace{2cm}
  \nn \\
 \hat G_{11} & = 
 (4 C_F T_F  n_f f_{1\epsilon} - 2 C_F C_A f_{2\epsilon})  \, 
  \big[ 2 F_0(1+\epsilon,1,1) 
  +F_0(2+\epsilon,1,0) \big]
  \, , 
  \hspace{-8cm} & & \hspace{8cm}
  \nn \\
 \hat G_{12} & = 
    (-2 C_F T_F  n_f f_{1\epsilon} +  C_F C_A f_{2\epsilon})
     \,  F_0(2+\epsilon,1,0) \,,
  \hspace{-8cm} & \hat G_{13} &= \hat G_{14} = \hat G_{15} = \hat G_{16} =0 \,. 
  \hspace{8cm}
 \nn
\end{align}
Here $T_F=1/2$, $C_A=3$, $C_F=4/3$. The $\hat G_1$ and the $h_{i\epsilon}$ are
determined by the two-loop computation of the heavy-quark self-energy in
Ref.~\cite{Broadhurst:1991fz,Grozin:2000cm}, while the $f_{i\epsilon}$ are
determined from the standard sum of one-loop quark, ghost, and gluon vacuum
polarization graphs.  We have
\begin{align}
f_{1\epsilon} &= 
   \left ( \frac{\mu}{-a} \right )^{2\epsilon} \,
\frac{  (1\minus \epsilon)\, e^{\epsilon\gamma_E} (-1)^{1+\epsilon} }
 {\epsilon(3-2\epsilon)(1-2\epsilon)}\,
 \frac{\Gamma^2(1\minus \epsilon)\Gamma(1\plus \epsilon)}{\Gamma(1\minus
   2\epsilon)}
  \,,
 & f_{2\epsilon} &=  \left ( \frac{\mu}{-a} \right )^{2\epsilon} \,
\frac{  (5\minus 3\epsilon)\, e^{\epsilon\gamma_E} (-1)^{1+\epsilon} 
    }{2\epsilon(3\minus 2\epsilon)(1\minus 2\epsilon)} \,
\frac{\Gamma^2(1\minus \epsilon)\Gamma(1\plus \epsilon)}{\Gamma(1\minus 2\epsilon)} 
 \,  ,
  \nn\\
h_{1\epsilon} &= 
   \Big( \frac{\mu}{-a} \Big)^{4\epsilon} \,
\frac{(1\minus \epsilon) \Gamma^2(1\minus \epsilon)\Gamma(1\plus 4\epsilon)
    e^{2\epsilon\gamma_E}} {2\epsilon^2(1\minus 2\epsilon)(-2\minus 2\epsilon)(1\minus
    4\epsilon)} 
  \,,
& h_{2\epsilon} &= 
 \left ( \frac{\mu}{-a} \right )^{4\epsilon} \,
 \frac{\Gamma^2 (1-\epsilon ) \Gamma (1+4 \epsilon )  e^{2\epsilon\gamma_E}}
 {4 \epsilon ^2 (1-2\epsilon )^2} 
  \,, 
 \nn \\
h_{3\epsilon} &= -
   \Big( \frac{\mu}{-a} \Big)^{4\epsilon} \,
  \frac{\Gamma^2 (1\minus \epsilon ) e^{2\epsilon\gamma_E}}{4\epsilon ^2(1\plus
    \epsilon) (1\minus 4\epsilon)(1\minus 2\epsilon )^2} 
   \Big[ \left(10\epsilon ^2\minus 9\epsilon \plus 5\right) \Gamma
  (1\plus 4\epsilon )+4(1\plus \epsilon )(1\minus 4\epsilon )\Gamma^2 (1\plus 2\epsilon ) \Big]
\,  . \hspace{-7cm} && \hspace{+7cm}
\end{align}
The results in Eq.~(\ref{Gi}) are given in terms of the one-loop master integral
\begin{align} \label{F0}
    F_0(\lambda_1\lambda_2\lambda_3)
    &\equiv -i N_d \int\!\! \frac{d^{d} \ell}{(2\pi)^{d}}\,
   \frac{a^{-4 +2\lambda_1+\lambda_2+\lambda_3}(\bar{n}\mcdot v)^{\lambda_3} }
   { [\ell^2]^{\lambda_1}[2 v\mcdot \ell\plus a]^{\lambda_2}
   [\bar n\mcdot \ell\,]^{\lambda_3} } 
 = (-1)^{4-\lambda_1} \Big(\frac{\mu}{-a}\Big)^{\! 2\epsilon}\: \frac{
    e^{\epsilon\gamma_E}\: \Gamma(2\lambda_1\plus\lambda_2\plus \lambda_3-d)\,
    \Gamma(d/2 \minus \lambda_1\minus \lambda_3)}{\Gamma(\lambda_1)\Gamma(\lambda_2)}
\end{align}
where $N_d=(4\pi)^{d/2}\mu^{2\epsilon} e^{\ga_E \epsilon}$, and the two-loop master integral
\begin{align} \label{F}
 & F(\lambda_1\lambda_2\lambda_3,\lambda_4\lambda_5\lambda_6,
  \lambda_7\lambda_8\lambda_9)
 \\
 &\quad\quad
\equiv  \int\frac{d^{d} \ell\, d^{d} k}{(2\pi)^{2d}}\,
 \frac{N_d^2\: 
  a^{(2\lambda_1+ 2\lambda_2+ 2\lambda_3+ \lambda_4+
    \lambda_5+ \lambda_6+ \lambda_7 + \lambda_8+ \lambda_9- 8)} \:
  [\bar n\cdot v]^{(\lambda_7+\lambda_8+\lambda_9)}
  }
 {[\ell^2]^{\lambda_1}
  [(\ell+k)^2]^{\lambda_2}[k^2]^{\lambda_3} 
 [2 v\mcdot \ell\plus a]^{\lambda_4}[2 v\mcdot (\ell\plus k) \plus a]^{\lambda_5}
 [2 v\mcdot k \plus a]^{\lambda_6} [\bar n\mcdot \ell]^{\lambda_7}
  [\bar n\mcdot \ell\plus \bar n\mcdot k]^{\lambda_8}[\bar n\mcdot k]^{\lambda_9}
 }
 \nn \,.
\end{align}
All denominator factors in square brackets in Eq.~(\ref{F0}) and (\ref{F}) have
$+i0$. As written, in the light-like propagators $[\bn\cdot k]$, $[\bn\cdot l +
\bn\cdot k]$, $[\bn\cdot k]$ this prescription for $+i0$'s does not precisely
match the $i0$'s from the definition in Eq.~(\ref{cBwloop}). However we have
checked that the difference results in scaleless integrals which are cancelled
by 0-bin subtraction terms~\cite{Manohar:2006nz}, which are part of the
definition of propagators of the collinear fields in the jet
function~\cite{Fleming:2007qr}.  All 0-bin subtraction terms are also scaleless
and therefore have not been shown explicitly. For the heavy quark jet function
the result of adding these contributions is simply that the $1/\epsilon$
divergences from the integrals in Eq.~(\ref{Gi}) are all UV.  The general result
for $F(\lambda_1\lambda_2\lambda_3,\lambda_4\lambda_5\lambda_6,
\lambda_7\lambda_8\lambda_9)$ is not known. Several of these master integrals
could be obtained from the calculation in Ref.~\cite{Becher:2005pd} by shifts of
variable that move $a$ into the light-like denominators, but these are not the
integrals needed for our analysis.  We have therefore computed the master
integral for the cases appearing in Eq.~(\ref{Gi}). Two cases are iterations of
the one-loop master integral, $F(101,110,101)=-F_0(1,1+2\epsilon,1) F_0(1,1,1)$
and $F(101,210,100)=-F_0(1,1+2\epsilon,1) F_0(1,1,0)$.  In order to evaluate
some of the remaining cases we have used the integration by parts
technique~\cite{Tkachov:1981wb,Chetyrkin:1981qh}.  For simplicity we quote the
results as series in $\epsilon$,
\begin{align}
   F(101,110,011) &=
   \Big(\frac{\mu}{-a}\Big)^{\!4\epsilon}\: \Big(
   - \frac{1}{8 \epsilon^4}  -\frac{11 \pi^2}{48 \epsilon^2}
   +\frac{17 \zeta_3}{6 \epsilon}-\frac{907 \pi ^4}{2880}
     \Big) \,,
  \nonumber \\
  F(101,111,100) &= 
  \Big(\frac{\mu}{-a}\Big)^{\!4\epsilon}\: \Big(
  \frac{1}{8 \epsilon^3} +\frac{1}{4 \epsilon^2}
  +\frac{1}{2 \epsilon} +\frac{\pi ^2}{48 \epsilon} 
  +\frac{17 \zeta_3}{12}+\frac{\pi^2}{24}+1
    \Big)\,, 
 \nonumber \\
   F(101,111,101) &= 
   \Big(\frac{\mu}{-a}\Big)^{\!4\epsilon}\: \Big(
   -\frac{\pi ^2}{6
   \epsilon^2} +\frac{4
   \zeta_3}{\epsilon} -\frac{7 \pi ^4}{20} \Big) \,, 
\nonumber \\
   F(111,100,010) &= 
   \Big(\frac{\mu}{-a}\Big)^{\!4\epsilon}\: \Big(
   \frac{1}{8 \epsilon^4}  +\frac{3 \pi^2}{16 \epsilon^2}
    -\frac{31\zeta_3}{12 \epsilon} + \frac{221 \pi ^4}{960}\Big) \,, 
  \nonumber \\
  F(111,100,100) &= 
    \Big(\frac{\mu}{-a}\Big)^{\!4\epsilon}\: \Big(
   -\frac{1}{8 \epsilon^3} -\frac{1}{4 \epsilon^2} 
   -\frac{3 \pi^2}{16 \epsilon}-\frac{1}{2 \epsilon}
  + \frac{31 \zeta_3}{12}-\frac{3\pi ^2}{8}-1\Big) \,, 
  \nonumber  \\
  F(111,110,000) &= 
   \Big(\frac{\mu}{-a}\Big)^{\!4\epsilon}\: \Big(
  \frac{\pi ^2}{6 \epsilon} -4 \zeta_3+\frac{\pi^2}{3} \Big) \,, 
 \nonumber \\
  F(111,110,100) &=
    \Big(\frac{\mu}{-a}\Big)^{\!4\epsilon}\: \Big(
   \frac{\pi ^2}{12 \epsilon^2} -\frac{7 \zeta_3}{2\epsilon}+\frac{13 \pi
   ^4}{72} \Big) \,, 
\nonumber \\
  F(111,110,10-\!1) &= 
    \Big(\frac{\mu}{-a}\Big)^{\!4\epsilon}\: \Big(
  -\frac{1}{8 \epsilon^3} -\frac{1}{4 \epsilon^2}
  -\frac{7 \pi ^2}{48 \epsilon}-\frac{1}{2 \epsilon}
     -\frac{2 \zeta_3}{3}+\frac{\pi^2}{24}-1   \Big) \,.
\end{align}

To determine the counterterms $\bar Z_i^{(j)}$ we use the analog of
Eq.~(\ref{twoloopren}) where we set $\delta m=0$ and take the imaginary part of
both sides. Since ${\rm Im}[m {\cal B}_0(\hat s,\mu)] =\delta(\hat s)$ this
gives a simpler set of equations for the terms in the renormalized jet function.
At one-loop we have ${mB}_{1}(\sh,\mu) = \bar Z_1(\sh,\mu) + \iota^\epsilon
\mu^{2\epsilon}\, {m B}^{\rm{bare}}_{1}(\sh)$, and at two-loops
\begin{align} \label{twolooprenB} 
{mB}_{2}(\sh,\mu) &= 
   \bar Z_2(\sh,\mu)
  + \iota^{2\epsilon} \mu^{4\epsilon}\, {mB}^{\rm{bare}}_{2}(\sh)
   + 2 z_{g1}\, \iota^\epsilon \mu^{2\epsilon} {mB}^{\rm bare}_1(\sh) 
   + \int\!\! d\sh' \, \bar Z_1(\sh\minus \sh',\mu)\ 
      \iota^\epsilon \mu^{2\epsilon} {mB}^{\rm{bare}}_{1}(\sh')  
   \,. 
\end{align}
Here $z_{g1}= -\beta_0/8$ enters from coupling constant renormalization.  To
evaluate the convolution integral term in Eq.~(\ref{twolooprenB}) we need ${
  B}^{\text{bare}}_1$ up to ${\cal O}(\epsilon^2)$, so the required ingredients
from one-loop graphs are
\begin{align} \label{oneloopresult}
   \iota^\epsilon \mu^{2\epsilon}\,{mB}^{\text{bare}}_{1} (\sh) & = 
  C_F \bigg[   
 \frac{4\epsilon^2}{3}  {\cal L}^3
 \minus \big( {2 \epsilon }\plus {2 \epsilon^2} \big) {\cal L}^2 
 \plus \Big(  2 \plus 2\epsilon  \plus 4 \epsilon^2  
   \minus \frac{ \pi ^2 }{2}\epsilon^2 \Big)  {\cal L}^1 
 \minus \Big(\frac{1}{\epsilon } \plus 1 
   \plus   2 \epsilon \minus \frac{\pi^2}{4} \epsilon
   \minus  \frac{ 7\zeta_3 }{3}\epsilon ^2 \minus\frac{\pi ^2}{4} \epsilon^2
  \plus 4 \epsilon ^2
   \Big) {\cal L}^0  
 \nn \\ 
& \qquad  \plus \Big( 
  \frac{1}{2 \epsilon ^2}  +\frac{1}{2 \epsilon } 
  + 1  - \frac{\pi^2}{8}  
  +2 \epsilon -\frac{\pi^2}{8} \epsilon -\frac{7\zeta_3}{6}  \epsilon 
  -\frac{7\zeta_3}{6}  \epsilon ^2 -\frac{\pi^4}{192} \epsilon^2
  -\frac{\pi ^2 }{4} \epsilon^2 +4 \epsilon^2
   \Big)
  \delta(\hat s)  
   \bigg] ~ + ~ {\cal O}(\epsilon^3)
  \nn\,, \\ 
\bar Z_1(\sh,\mu) & = 
 C_F \bigg[
  \frac{1}{\epsilon} {\cal L}^0 - \left (
  \frac{1}{2\epsilon^2}+\frac{1}{2\epsilon}\right ) \delta(\sh)  \bigg] 
  \, ,
\end{align}
where the distribution ${\cal L}^k$ was defined in Eq.~(\ref{cLk}).  From
Eq.~(\ref{B2bareSum}) the sum of the two-loop graphs gives
\begin{align}
\label{B2bare} \nn
\iota^{2\epsilon} \mu^{4\epsilon}\,{mB}_{2}^{\text{bare}}(\sh)
&=
 C_F C_A \bigg \{ 
  \Big( \frac{2}{3}-\frac{\pi^2}{6} \Big) {\cal L}^1 
  -\Big( \frac{1}{6 \epsilon }-\frac{\pi ^2}{24 \epsilon }
    +\frac{5}{18}-\frac{\pi^2}{12} -\frac{5 \zeta_3}{4}
    \Big) {\cal L}^0
\nn \\
& \hspace{-.4cm} 
   + \Big( \frac{1}{24 \epsilon ^2} -\frac{\pi ^2}{96 \epsilon ^2}
    +\frac{5}{72 \epsilon}-\frac{\pi^2}{48\epsilon}-\frac{5\zeta_3}{16\epsilon}
    -\frac{11}{54}-\frac{\pi^2}{144}+\frac{23\pi^4}{2880}-\frac{5\zeta_3}{8}
   \Big)\, \delta(\hat s) \bigg\}
\nn \\
&\hspace{-.8cm} 
+C_F^2 \bigg\{ 
  \frac{16}{3} {\cal L}^3
  -\Big( 8+\frac{4}{\epsilon } \Big) {\cal L}^2 
  + \Big( \frac{2}{\epsilon ^2} +\frac{4}{\epsilon } + 10
    - \frac{7 \pi ^2}{3} \Big) {\cal L}^1 
  - \Big(\frac{1}{2 \epsilon^3}  +\frac{1}{\epsilon ^2} +\frac{5}{2 \epsilon }
    -\frac{7 \pi ^2}{12\epsilon } -\frac{31 \zeta_3}{3} - \frac{7 \pi ^2}{6} + 6
    \Big) {\cal L}^0
\nn \\
&\hspace{-.4cm} + \Big(
   \frac{1}{8 \epsilon^4} +\frac{1}{4 \epsilon^3} 
   +\frac{5}{8\epsilon^2} - \frac{7 \pi^2}{48 \epsilon ^2}
    +\frac{3}{2\epsilon}-\frac{7 \pi^2}{24\epsilon}-\frac{31 \zeta_3}{12\epsilon}
   +\frac{7}{2}-\frac{35 \pi^2}{48} -\frac{\pi^4}{320}-\frac{31 \zeta_3}{6} 
   \Big)\,
  \delta(\hat s) \bigg\}
\nn \\
& \hspace{-.8cm} 
  + C_F \beta_0 \bigg\{ 
   - {\cal L}^2 
   + \Big(\frac{1}{2\epsilon}+\frac{11}{6}\Big) {\cal L}^1
   - \Big(  \frac{1}{8\epsilon^2} +\frac{11}{24\epsilon} 
      +\frac{65}{36} - \frac{7\pi^2}{48} \Big) {\cal L}^0
\nn \\
&\hspace{-0.4cm}
 + \Big(
   \frac{1}{32 \epsilon^3} +\frac{11}{96 \epsilon^2} 
   +\frac{65}{144\epsilon } -\frac{7\pi^2}{192 \epsilon} 
   +\frac{389}{216}-\frac{77\pi^2}{576}-\frac{31 \zeta_3}{48}
    \Big)\,
  \delta(\hat s) 
  \bigg \}
+ {\cal O}(\epsilon) \,.
\end{align}
The convolution integral required in Eq.~(\ref{twolooprenB}) is given by
\begin{align}  \label{conv11}
 & \int d\sh' \, \bar Z_1(\sh-\sh',\mu) \, 
  \iota^\epsilon \mu^{2\epsilon} {mB}^{\text{bare}}_{1}(\sh')  
 \nn \\ 
 & = 
  C_F^2\, \bigg\{
   - \frac{10}{3} {\cal L}^3
   + \Big(\frac{4}{\epsilon }+  5  \Big) {\cal L}^2 
   - \Big(\frac{3}{\epsilon ^2}+\frac{4}{\epsilon }+7- \frac{17 \pi ^2}{12} \Big) {\cal L}^1  
   + \Big( \frac{1}{\epsilon ^3} +\frac{3}{2 \epsilon ^2} +\frac{5}{2 \epsilon }
    -\frac{7 \pi ^2}{12 \epsilon }
    +5 -\frac{17 \pi ^2}{24} -\frac{19 \zeta_3}{3}
      \Big) {\cal L}^0    
   \nn \\ 
& \quad  
   - \Big( \frac{1}{4 \epsilon ^4} + \frac{1}{2 \epsilon^3}
    + \frac{3}{4 \epsilon^2}  -\frac{11 \pi ^2}{48 \epsilon ^2}
    +\frac{3}{2 \epsilon } -\frac{7 \pi ^2}{24 \epsilon }
       -\frac{31 \zeta_3}{12 \epsilon }
    +3 -\frac{25 \pi ^2}{48} -\frac{31 \pi ^4}{5760} -\frac{19 \zeta_3}{6}
    \Big)\, \delta(\sh) 
  \bigg\} \, .
\end{align}
In order to obtain this result we have used Eq.~(\ref{oneloopresult}) along with
Eq.~(\ref{convolution}) from Appendix~\ref{appB}. Combining the last three terms
in Eq.~(\ref{twolooprenB}) the remaining $1/\epsilon^k$ terms must be canceled
by $\bar Z_2$, hence uniquely fixing it. This gives
\begin{align}\label{Zt2} 
  \nn
 \tilde Z_2(\sh) & =
C_F^2 \bigg\{ 
   \frac{1}{\epsilon^2 }{\cal L}^1
  - \Big(\frac{1}{2 \epsilon ^3}   + \frac{1}{2 \epsilon ^2}  \Big)
  {\cal L}^0
  +\Big(\frac{1}{8 \epsilon^4} +\frac{1}{4 \epsilon ^3}
    +\frac{1}{8 \epsilon ^2} -\frac{\pi ^2}{12 \epsilon ^2}
   \Big) \delta(\sh) 
  \bigg \}
   \\ \nn
&  \quad + C_F C_A \bigg \{ 
  \Big(  \frac{1}{6 \epsilon }-\frac{\pi ^2}{24 \epsilon }  \Big) {\cal L}^0
  + \Big(  -\frac{1}{24 \epsilon^2}  +\frac{\pi ^2}{96 \epsilon^2}
  -\frac{5}{72\epsilon }+\frac{\pi^2}{48\epsilon} + \frac{5\zeta_3}{16\epsilon } 
  \Big) \delta(\sh)  
  \bigg \} 
  \\ 
& \quad + C_F \beta_0 \, \bigg\{ 
   \Big( -\frac{1}{8 \epsilon ^2}  + \frac{5}{24 \epsilon }\Big) 
   {\cal L}^0
  + \Big( \frac{3}{32\epsilon^3} + \frac{1}{96 \epsilon^2}
    -\frac{29}{144 \epsilon } + \frac{\pi ^2}{192 \epsilon } 
   \Big) \delta(\sh) 
  \bigg\} \, 
\, .
\end{align}
Using the notation in Eq.~(\ref{ZZ12}) the counterterm consistency equations
that follow from Eq.~(\ref{eq:consistency}) are
\begin{align}
  \mu\frac{\partial}{\partial\mu}\, \bar Z_1^{(1)} &= 2 \bar Z_1^{(2)} 
   \,,
 \qquad\qquad
 \mu\frac{\partial}{\partial\mu}\, \bar Z_1^{(2)} = 0
   \,,
 \qquad\qquad
  \mu\frac{\partial}{\partial\mu}\, \bar Z_2^{(3)} = 4 \bar Z_2^{(4)} 
   \,,
   \nn\\
  4 \bar Z_2^{(3)} &=  \mu\frac{\partial}{\partial\mu}\, \bar Z_2^{(2)}
    - \frac{\beta_0}{2} \bar Z_1^{(2)} + 2 \int\! d\hat s' \bar Z_1^{(2)}(\hat
    s\minus \hat s',\mu) \bar Z_1^{(1)}(\hat s') 
   \,,
  \nn\\
  4 \bar Z_2^{(2)} &=  \mu\frac{\partial}{\partial\mu}\, \bar Z_2^{(1)}
    - \frac{\beta_0}{2} \bar Z_1^{(1)} + 2 \int\! d\hat s' \bar Z_1^{(1)}(\hat
    s\minus \hat s',\mu) \bar Z_1^{(1)}(\hat s') 
   \,.
\end{align}
Reading off the coefficients of various powers of $1/\epsilon^k$ from the
results in Eqs.~(\ref{oneloopresult}) and (\ref{Zt2}) we can verify that they
are all satisfied.

\section{Relations for plus-distributions}\label{appB}

In the text in several places we converted between momentum space, position
space, and plus-distributions arising from imaginary parts. Useful conversion
formulas include
\begin{align} \label{FTyshat}
  {\rm FT}\big[ \ln^k(i\, y\, \mu e^{\gamma_E} )\big] = 
  \frac{d^k}{d\epsilon^k} \: \frac{e^{\epsilon\gamma_E}}{\Gamma(1\minus\epsilon)}
  \bigg\{ \delta(\hat s)  -
    \frac{\epsilon}{\mu}\,
    \left[\frac{\theta(\hat s) e^{-\epsilon \ln(\hat s/\mu)}}{\hat s/\mu}\right]_+
  \bigg\} \bigg|_{\epsilon=0}
  \,.
\end{align}
and
\begin{align} \label{Imformula}
  {\rm Im}\bigg[ \frac{\ln^n(- x\minus i0)}{\pi(- x\minus i0)}\bigg]
   \! = \cos^2\!\Big(\frac{n\pi}{2}\Big) \frac{(-\pi^2)^{n/2}}{n+1} \delta(x) 
    +
    \sum_{j=0}^{\left[\left[\frac{n-1}{2}\right]\right]}
   \frac{(-1)^j\, n!\, \pi^{2j}}{(2j\plus 1)!
      (n\minus 2j\minus 1)!} \bigg[ \frac{ \theta(x) \ln^{n-2j-1}(x)}{x}\bigg]_+
   \! ,
\end{align}
where $[[p]]$ on the sum is the greatest integer not exceeding $p$. From
Eq.~(\ref{Imformula}) the cases we used include
\begin{align}\label{Impart}
  {\rm Im}[ L^0] &= \delta(\hat s) \,,
 & {\rm Im}[ L^1 ] &= -{\cal L}^0   \,,
 & {\rm Im}[ L^2 ] &=  2 {\cal L}^1 - \, \frac{\pi^2}{3} \delta(\hat s)
    \,,\nn \\ 
 {\rm Im}[ L^3 ] &= - 3 {\cal L}^2  + \, \pi^2 {\cal L}^0 \,, 
 & {\rm Im}[L^4 ] &=
  4 {\cal L}^3  - \, 4 \pi^2 {\cal L}^1 + \, \frac{\pi^4}{5} \delta(\hat s)  
  \,,
 & {\rm Im}[L^5 ] &=
  -5 {\cal L}^4  + \, 10 \pi^2 {\cal L}^2 - \, \pi^4 {\cal L}^0
   \,,\nn \\ 
 {\rm Im}[L^6 ] &= 6 {\cal L}^5 - \, 20\pi^2 {\cal L}^3 + \,6 \pi^4 {\cal L}^1
  - \, \frac{\pi^6}{7} \delta(\hat s) \, .  \hspace{-2.5cm}
\end{align}
Here $L^k$ is defined by Eq.~(\ref{Lk}) and the distribution ${\cal L}^k$ is defined
in Eq.~(\ref{cLk}).  The following rescaling identity is also useful
\begin{equation}\frac{1}{\lambda}\left [  \frac{\lambda \,
      \theta(x)\log^p(x/\lambda)}{x}  \right ]_+ = \sum_{k=0}^{p}\frac{p\,
    !}{(p-k)\,\! ! \, k \, !} \log^{p-k}\left (  \frac{1}{\lambda}  \right )
  \left [  \frac{\theta(x)\log^k x}{x}  \right ]_+  +
  \frac{\delta(x)}{(p+1)}\log^{p+1}\left (  \frac{1}{\lambda}  \right ) \, .
\end{equation}
For $k\ge 0$ this result readily gives
\begin{equation} \label{rgecL}
  \mu \frac{d}{d\mu} {\cal L}^k =
  - k {\cal L}^{k-1} - \delta_{k,0} \, \delta(\hat s)  \,.
\end{equation}
Eq.~(\ref{rgecL}) can be used to verify the expected $\mu$-dependence at various
stages.  Finally, we need the convolution of two plus distributions. The general
formula is
\begin{align}
  \int \! d\hat s'\ {\cal L}^{j}(\hat s-\hat s')\, {\cal L}^{k}(\hat s')
   &= \delta(\hat s) (-1)^{k+j} \frac{d^k}{dw^{\prime k}} \frac{d^j}{dw^{j}} 
  \bigg\{ \frac{\Gamma(-w)\Gamma(-w')}{\Gamma(1-w-w')} 
  - \frac{1}{w w'}\bigg\}  \bigg|_{{w=w'=0}}
  \\[5pt]
 &\hspace{-3.2cm}
  + \frac{1}{\mu} \bigg[ (-1)^{k+j} \frac{d^k}{dw^{\prime k}}
   \frac{d^j}{dw^{j}} \ \Big(\frac{\hat s}{\mu}\Big)^{-1-w-w'} \bigg\{ \frac{1}{w}+\frac{1}{w'}
  + \frac{\Gamma(-w)\Gamma(-w')}{\Gamma(-w-w')} \bigg\}
  \bigg|_{{w=w'=0}}
  + \Big(\frac{1}{k+1}+\frac{1}{j+1}\Big) \frac{(\ln \hat s/\mu)^{k+j+1}}{\hat s/\mu} \bigg]_+ 
  .\nn
\end{align}
The following cases were used in
our analysis
\begin{align}\label{convolution}
 \int \! d\hat s'\ {\cal L}^{0}(\hat s-\hat s')\, {\cal L}^{0}(\hat s')
  & = 2 {\cal L}^1(\hat s) - \frac{\pi^2}{6}\, \delta(\hat s)  \,, 
  \\ 
\int \! d\hat s'\ {\cal L}^{0}(\hat s-\hat s')\, {\cal L}^{1}(\hat s')
  & = \frac{3}{2} {\cal L}^2(\hat s) - \frac{\pi^2}{6}\, {\cal L}^0 + \zeta_3\,
  \delta(\hat s)  \,, 
   \nn\\
\int \! d\hat s'\ {\cal L}^{0}(\hat s-\hat s')\, {\cal L}^{2}(\hat s')
 &=
\frac{4}{3} {\cal L}^3(\hat s) - \frac{\pi^2}{3} {\cal L}^1(\hat s) +
 2\zeta_3\, {\cal L}^0 - \frac{\pi^4}{45}\, \delta(\hat s) 
 \, .  \nn
\end{align}

\bibliographystyle{iain}

\bibliography{topjet}

\begin{thebibliography}{71}
\expandafter\ifx\csname natexlab\endcsname\relax\def\natexlab#1{#1}\fi
\expandafter\ifx\csname bibnamefont\endcsname\relax
  \def\bibnamefont#1{#1}\fi
\expandafter\ifx\csname bibfnamefont\endcsname\relax
  \def\bibfnamefont#1{#1}\fi
\expandafter\ifx\csname citenamefont\endcsname\relax
  \def\citenamefont#1{#1}\fi
\expandafter\ifx\csname url\endcsname\relax
  \def\url#1{\texttt{#1}}\fi
\expandafter\ifx\csname urlprefix\endcsname\relax\def\urlprefix{URL }\fi
\providecommand{\bibinfo}[2]{#2}
\providecommand{\eprint}[2][]{\url{#2}}

\bibitem[{unk(2007)}]{unknown:2007bxa}
\bibinfo{journal}{Tevatron Electroweak Working Group}  (\bibinfo{year}{2007}),
  \eprint{hep-ex/0703034}.

\bibitem[{\citenamefont{Gruenewald}(2007)}]{private1}
\bibinfo{author}{\bibfnamefont{M.}~\bibnamefont{Gruenewald}},
  \bibinfo{journal}{EPS-HEP, Manchester,}  (\bibinfo{year}{2007}),
  \eprint{\href{http://arXiv.org/abs/0709.3744}{arXiv:0709.3744}}.

\bibitem[{\citenamefont{Yao et~al.}(2006)}]{Yao:2006px}
\bibinfo{author}{\bibfnamefont{W.~M.} \bibnamefont{Yao}} \bibnamefont{et~al.}
  (\bibinfo{collaboration}{Particle Data Group}), \bibinfo{journal}{J. Phys.}
  \textbf{\bibinfo{volume}{G33}}, \bibinfo{pages}{1} (\bibinfo{year}{2006}).

\bibitem[{HFA(2007)}]{HFAG}
\bibinfo{journal}{HFAG (Heavy Flavor Averaging Group)}  (\bibinfo{year}{2007}).

\bibitem[{\citenamefont{Bigi et~al.}(1997)\citenamefont{Bigi, Shifman,
  Uraltsev, and Vainshtein}}]{Bigi:1996si}
\bibinfo{author}{\bibfnamefont{I.~I.~Y.} \bibnamefont{Bigi}},
  \bibinfo{author}{\bibfnamefont{M.~A.} \bibnamefont{Shifman}},
  \bibinfo{author}{\bibfnamefont{N.}~\bibnamefont{Uraltsev}}, \bibnamefont{and}
  \bibinfo{author}{\bibfnamefont{A.~I.} \bibnamefont{Vainshtein}},
  \bibinfo{journal}{Phys. Rev.} \textbf{\bibinfo{volume}{D56}},
  \bibinfo{pages}{4017} (\bibinfo{year}{1997}),
  \eprint{\href{http://arXiv.org/abs/hep-ph/9704245}{hep-ph/9704245}}.

\bibitem[{\citenamefont{Hoang et~al.}(1999{\natexlab{a}})\citenamefont{Hoang,
  Ligeti, and Manohar}}]{Hoang:1998hm}
\bibinfo{author}{\bibfnamefont{A.~H.} \bibnamefont{Hoang}},
  \bibinfo{author}{\bibfnamefont{Z.}~\bibnamefont{Ligeti}}, \bibnamefont{and}
  \bibinfo{author}{\bibfnamefont{A.~V.} \bibnamefont{Manohar}},
  \bibinfo{journal}{Phys. Rev.} \textbf{\bibinfo{volume}{D59}},
  \bibinfo{pages}{074017} (\bibinfo{year}{1999}{\natexlab{a}}),
  \eprint{\href{http://arXiv.org/abs/hep-ph/9811239}{hep-ph/9811239}}.

\bibitem[{\citenamefont{Hoang et~al.}(1999{\natexlab{b}})\citenamefont{Hoang,
  Ligeti, and Manohar}}]{Hoang:1998ng}
\bibinfo{author}{\bibfnamefont{A.~H.} \bibnamefont{Hoang}},
  \bibinfo{author}{\bibfnamefont{Z.}~\bibnamefont{Ligeti}}, \bibnamefont{and}
  \bibinfo{author}{\bibfnamefont{A.~V.} \bibnamefont{Manohar}},
  \bibinfo{journal}{Phys. Rev. Lett.} \textbf{\bibinfo{volume}{82}},
  \bibinfo{pages}{277} (\bibinfo{year}{1999}{\natexlab{b}}),
  \eprint{\href{http://arXiv.org/abs/hep-ph/9809423}{hep-ph/9809423}}.

\bibitem[{\citenamefont{Hoang and Teubner}(1999)}]{Hoang:1999zc}
\bibinfo{author}{\bibfnamefont{A.~H.} \bibnamefont{Hoang}} \bibnamefont{and}
  \bibinfo{author}{\bibfnamefont{T.}~\bibnamefont{Teubner}},
  \bibinfo{journal}{Phys. Rev.} \textbf{\bibinfo{volume}{D60}},
  \bibinfo{pages}{114027} (\bibinfo{year}{1999}),
  \eprint{\href{http://arXiv.org/abs/hep-ph/9904468}{hep-ph/9904468}}.

\bibitem[{\citenamefont{Bosch et~al.}(2004)\citenamefont{Bosch, Lange, Neubert,
  and Paz}}]{Bosch:2004th}
\bibinfo{author}{\bibfnamefont{S.~W.} \bibnamefont{Bosch}},
  \bibinfo{author}{\bibfnamefont{B.~O.} \bibnamefont{Lange}},
  \bibinfo{author}{\bibfnamefont{M.}~\bibnamefont{Neubert}}, \bibnamefont{and}
  \bibinfo{author}{\bibfnamefont{G.}~\bibnamefont{Paz}},
  \bibinfo{journal}{Nucl. Phys.} \textbf{\bibinfo{volume}{B699}},
  \bibinfo{pages}{335} (\bibinfo{year}{2004}),
  \eprint{\href{http://arXiv.org/abs/hep-ph/0402094}{hep-ph/0402094}}.

\bibitem[{\citenamefont{Beneke}(1999)}]{Beneke:1998ui}
\bibinfo{author}{\bibfnamefont{M.}~\bibnamefont{Beneke}},
  \bibinfo{journal}{Phys. Rept.} \textbf{\bibinfo{volume}{317}},
  \bibinfo{pages}{1} (\bibinfo{year}{1999}), \eprint{hep-ph/9807443}.

\bibitem[{\citenamefont{Smith and Willenbrock}(1997)}]{Smith:1996xz}
\bibinfo{author}{\bibfnamefont{M.~C.} \bibnamefont{Smith}} \bibnamefont{and}
  \bibinfo{author}{\bibfnamefont{S.~S.} \bibnamefont{Willenbrock}},
  \bibinfo{journal}{Phys. Rev. Lett.} \textbf{\bibinfo{volume}{79}},
  \bibinfo{pages}{3825} (\bibinfo{year}{1997}), \eprint{hep-ph/9612329}.

\bibitem[{\citenamefont{Fleming
  et~al.}(2008{\natexlab{a}})\citenamefont{Fleming, Hoang, Mantry, and
  Stewart}}]{Fleming:2007qr}
\bibinfo{author}{\bibfnamefont{S.}~\bibnamefont{Fleming}},
  \bibinfo{author}{\bibfnamefont{A.~H.} \bibnamefont{Hoang}},
  \bibinfo{author}{\bibfnamefont{S.}~\bibnamefont{Mantry}}, \bibnamefont{and}
  \bibinfo{author}{\bibfnamefont{I.~W.} \bibnamefont{Stewart}},
  \bibinfo{journal}{Phys. Rev.} \textbf{\bibinfo{volume}{D77}},
  \bibinfo{pages}{074010} (\bibinfo{year}{2008}{\natexlab{a}}),
  \eprint{\href{http://arXiv.org/abs/hep-ph/0703207}{hep-ph/0703207}}.

\bibitem[{\citenamefont{Hoang et~al.}(2000)}]{Hoang:2000yr}
\bibinfo{author}{\bibfnamefont{A.~H.} \bibnamefont{Hoang}}
  \bibnamefont{et~al.}, \bibinfo{journal}{Eur. Phys. J. direct}
  \textbf{\bibinfo{volume}{C2}}, \bibinfo{pages}{1} (\bibinfo{year}{2000}),
  \eprint{hep-ph/0001286}.

\bibitem[{\citenamefont{Hoang et~al.}(2002)\citenamefont{Hoang, Manohar,
  Stewart, and Teubner}}]{Hoang:2001mm}
\bibinfo{author}{\bibfnamefont{A.~H.} \bibnamefont{Hoang}},
  \bibinfo{author}{\bibfnamefont{A.~V.} \bibnamefont{Manohar}},
  \bibinfo{author}{\bibfnamefont{I.~W.} \bibnamefont{Stewart}},
  \bibnamefont{and} \bibinfo{author}{\bibfnamefont{T.}~\bibnamefont{Teubner}},
  \bibinfo{journal}{Phys. Rev.} \textbf{\bibinfo{volume}{D65}},
  \bibinfo{pages}{014014} (\bibinfo{year}{2002}),
  \eprint{\href{http://arXiv.org/abs/hep-ph/0107144}{hep-ph/0107144}}.

\bibitem[{\citenamefont{Pineda and Signer}(2007)}]{Pineda:2006ri}
\bibinfo{author}{\bibfnamefont{A.}~\bibnamefont{Pineda}} \bibnamefont{and}
  \bibinfo{author}{\bibfnamefont{A.}~\bibnamefont{Signer}},
  \bibinfo{journal}{Nucl. Phys.} \textbf{\bibinfo{volume}{B762}},
  \bibinfo{pages}{67} (\bibinfo{year}{2007}), \eprint{hep-ph/0607239}.

\bibitem[{\citenamefont{Martinez and Miquel}(2003)}]{Martinez:2002st}
\bibinfo{author}{\bibfnamefont{M.}~\bibnamefont{Martinez}} \bibnamefont{and}
  \bibinfo{author}{\bibfnamefont{R.}~\bibnamefont{Miquel}},
  \bibinfo{journal}{Eur. Phys. J.} \textbf{\bibinfo{volume}{C27}},
  \bibinfo{pages}{49} (\bibinfo{year}{2003}), \eprint{hep-ph/0207315}.

\bibitem[{\citenamefont{Juste et~al.}(2005)}]{Juste:2006sv}
\bibinfo{author}{\bibfnamefont{A.}~\bibnamefont{Juste}} \bibnamefont{et~al.},
  \bibinfo{journal}{ECONF C0508141, PLEN0043}  (\bibinfo{year}{2005}),
  \eprint{\href{http://arXiv.org/abs/hep-ph/0601112}{hep-ph/0601112}}.

\bibitem[{\citenamefont{Hoang and Reisser}(2006)}]{Hoang:2006pd}
\bibinfo{author}{\bibfnamefont{A.~H.} \bibnamefont{Hoang}} \bibnamefont{and}
  \bibinfo{author}{\bibfnamefont{C.~J.} \bibnamefont{Reisser}},
  \bibinfo{journal}{Phys. Rev.} \textbf{\bibinfo{volume}{D74}},
  \bibinfo{pages}{034002} (\bibinfo{year}{2006}), \eprint{hep-ph/0604104}.

\bibitem[{\citenamefont{Fleming et~al.}(2007)\citenamefont{Fleming, Hoang,
  Mantry, and Stewart}}]{Fleming:2007xt}
\bibinfo{author}{\bibfnamefont{S.}~\bibnamefont{Fleming}},
  \bibinfo{author}{\bibfnamefont{A.~H.} \bibnamefont{Hoang}},
  \bibinfo{author}{\bibfnamefont{S.}~\bibnamefont{Mantry}}, \bibnamefont{and}
  \bibinfo{author}{\bibfnamefont{I.~W.} \bibnamefont{Stewart}}
  (\bibinfo{year}{2007}), \eprint{\href{http://arXiv.org/abs/0711.2079
  [hep-ph]}{0711.2079 [hep-ph]}}.

\bibitem[{\citenamefont{Manohar and Wise}(2000)}]{Manohar:2000dt}
\bibinfo{author}{\bibfnamefont{A.~V.} \bibnamefont{Manohar}} \bibnamefont{and}
  \bibinfo{author}{\bibfnamefont{M.~B.} \bibnamefont{Wise}},
  \bibinfo{journal}{Camb. Monogr. Part. Phys. Nucl. Phys. Cosmol.}
  \textbf{\bibinfo{volume}{10}}, \bibinfo{pages}{1} (\bibinfo{year}{2000}).

\bibitem[{\citenamefont{Neubert}(1994{\natexlab{a}})}]{Neubert:1993mb}
\bibinfo{author}{\bibfnamefont{M.}~\bibnamefont{Neubert}},
  \bibinfo{journal}{Phys. Rept.} \textbf{\bibinfo{volume}{245}},
  \bibinfo{pages}{259} (\bibinfo{year}{1994}{\natexlab{a}}),
  \eprint{hep-ph/9306320}.

\bibitem[{\citenamefont{Hoang and Stewart}(2008)}]{Hoang:2007vb}
\bibinfo{author}{\bibfnamefont{A.~H.} \bibnamefont{Hoang}} \bibnamefont{and}
  \bibinfo{author}{\bibfnamefont{I.~W.} \bibnamefont{Stewart}},
  \bibinfo{journal}{Phys. Lett.} \textbf{\bibinfo{volume}{B660}},
  \bibinfo{pages}{483} (\bibinfo{year}{2008}),
  \eprint{\href{http://arXiv.org/abs/0709.3519 [hep-ph]}{0709.3519 [hep-ph]}}.

\bibitem[{\citenamefont{Korchemsky and Tafat}(2000)}]{Korchemsky:2000kp}
\bibinfo{author}{\bibfnamefont{G.~P.} \bibnamefont{Korchemsky}}
  \bibnamefont{and} \bibinfo{author}{\bibfnamefont{S.}~\bibnamefont{Tafat}},
  \bibinfo{journal}{JHEP} \textbf{\bibinfo{volume}{10}}, \bibinfo{pages}{010}
  (\bibinfo{year}{2000}),
  \eprint{\href{http://arXiv.org/abs/hep-ph/0007005}{hep-ph/0007005}}.

\bibitem[{\citenamefont{Bauer et~al.}(2000)\citenamefont{Bauer, Fleming, and
  Luke}}]{Bauer:2000ew}
\bibinfo{author}{\bibfnamefont{C.~W.} \bibnamefont{Bauer}},
  \bibinfo{author}{\bibfnamefont{S.}~\bibnamefont{Fleming}}, \bibnamefont{and}
  \bibinfo{author}{\bibfnamefont{M.~E.} \bibnamefont{Luke}},
  \bibinfo{journal}{Phys. Rev.} \textbf{\bibinfo{volume}{D63}},
  \bibinfo{pages}{014006} (\bibinfo{year}{2000}),
  \eprint{\href{http://arXiv.org/abs/hep-ph/0005275}{hep-ph/0005275}}.

\bibitem[{\citenamefont{Bauer et~al.}(2001)\citenamefont{Bauer, Fleming,
  Pirjol, and Stewart}}]{Bauer:2000yr}
\bibinfo{author}{\bibfnamefont{C.~W.} \bibnamefont{Bauer}},
  \bibinfo{author}{\bibfnamefont{S.}~\bibnamefont{Fleming}},
  \bibinfo{author}{\bibfnamefont{D.}~\bibnamefont{Pirjol}}, \bibnamefont{and}
  \bibinfo{author}{\bibfnamefont{I.~W.} \bibnamefont{Stewart}},
  \bibinfo{journal}{Phys. Rev.} \textbf{\bibinfo{volume}{D63}},
  \bibinfo{pages}{114020} (\bibinfo{year}{2001}),
  \eprint{\href{http://arXiv.org/abs/hep-ph/0011336}{hep-ph/0011336}}.

\bibitem[{\citenamefont{Bauer et~al.}(2002)\citenamefont{Bauer, Pirjol, and
  Stewart}}]{Bauer:2001yt}
\bibinfo{author}{\bibfnamefont{C.~W.} \bibnamefont{Bauer}},
  \bibinfo{author}{\bibfnamefont{D.}~\bibnamefont{Pirjol}}, \bibnamefont{and}
  \bibinfo{author}{\bibfnamefont{I.~W.} \bibnamefont{Stewart}},
  \bibinfo{journal}{Phys. Rev.} \textbf{\bibinfo{volume}{D65}},
  \bibinfo{pages}{054022} (\bibinfo{year}{2002}),
  \eprint{\href{http://arXiv.org/abs/hep-ph/0109045}{hep-ph/0109045}}.

\bibitem[{\citenamefont{Bauer and Stewart}(2001)}]{Bauer:2001ct}
\bibinfo{author}{\bibfnamefont{C.~W.} \bibnamefont{Bauer}} \bibnamefont{and}
  \bibinfo{author}{\bibfnamefont{I.~W.} \bibnamefont{Stewart}},
  \bibinfo{journal}{Phys. Lett.} \textbf{\bibinfo{volume}{B516}},
  \bibinfo{pages}{134} (\bibinfo{year}{2001}),
  \eprint{\href{http://arXiv.org/abs/hep-ph/0107001}{hep-ph/0107001}}.

\bibitem[{\citenamefont{Fadin and Khoze}(1987)}]{Fadin:1987wz}
\bibinfo{author}{\bibfnamefont{V.~S.} \bibnamefont{Fadin}} \bibnamefont{and}
  \bibinfo{author}{\bibfnamefont{V.~A.} \bibnamefont{Khoze}},
  \bibinfo{journal}{JETP Lett.} \textbf{\bibinfo{volume}{46}},
  \bibinfo{pages}{525} (\bibinfo{year}{1987}).

\bibitem[{\citenamefont{Beneke et~al.}(2004{\natexlab{a}})\citenamefont{Beneke,
  Chapovsky, Signer, and Zanderighi}}]{Beneke:2003xh}
\bibinfo{author}{\bibfnamefont{M.}~\bibnamefont{Beneke}},
  \bibinfo{author}{\bibfnamefont{A.~P.} \bibnamefont{Chapovsky}},
  \bibinfo{author}{\bibfnamefont{A.}~\bibnamefont{Signer}}, \bibnamefont{and}
  \bibinfo{author}{\bibfnamefont{G.}~\bibnamefont{Zanderighi}},
  \bibinfo{journal}{Phys. Rev. Lett.} \textbf{\bibinfo{volume}{93}},
  \bibinfo{pages}{011602} (\bibinfo{year}{2004}{\natexlab{a}}),
  \eprint{hep-ph/0312331}.

\bibitem[{\citenamefont{Beneke et~al.}(2004{\natexlab{b}})\citenamefont{Beneke,
  Chapovsky, Signer, and Zanderighi}}]{Beneke:2004km}
\bibinfo{author}{\bibfnamefont{M.}~\bibnamefont{Beneke}},
  \bibinfo{author}{\bibfnamefont{A.~P.} \bibnamefont{Chapovsky}},
  \bibinfo{author}{\bibfnamefont{A.}~\bibnamefont{Signer}}, \bibnamefont{and}
  \bibinfo{author}{\bibfnamefont{G.}~\bibnamefont{Zanderighi}},
  \bibinfo{journal}{Nucl. Phys.} \textbf{\bibinfo{volume}{B686}},
  \bibinfo{pages}{205} (\bibinfo{year}{2004}{\natexlab{b}}),
  \eprint{\href{http://arXiv.org/abs/hep-ph/0401002}{hep-ph/0401002}}.

\bibitem[{\citenamefont{Hoang and Reisser}(2005)}]{Hoang:2004tg}
\bibinfo{author}{\bibfnamefont{A.~H.} \bibnamefont{Hoang}} \bibnamefont{and}
  \bibinfo{author}{\bibfnamefont{C.~J.} \bibnamefont{Reisser}},
  \bibinfo{journal}{Phys. Rev.} \textbf{\bibinfo{volume}{D71}},
  \bibinfo{pages}{074022} (\bibinfo{year}{2005}), \eprint{hep-ph/0412258}.

\bibitem[{\citenamefont{Korchemsky and Sterman}(1995)}]{Korchemsky:1994is}
\bibinfo{author}{\bibfnamefont{G.~P.} \bibnamefont{Korchemsky}}
  \bibnamefont{and} \bibinfo{author}{\bibfnamefont{G.}~\bibnamefont{Sterman}},
  \bibinfo{journal}{Nucl. Phys.} \textbf{\bibinfo{volume}{B437}},
  \bibinfo{pages}{415} (\bibinfo{year}{1995}),
  \eprint{\href{http://arXiv.org/abs/hep-ph/9411211}{hep-ph/9411211}}.

\bibitem[{\citenamefont{Korchemsky and Sterman}(1999)}]{Korchemsky:1999kt}
\bibinfo{author}{\bibfnamefont{G.~P.} \bibnamefont{Korchemsky}}
  \bibnamefont{and} \bibinfo{author}{\bibfnamefont{G.}~\bibnamefont{Sterman}},
  \bibinfo{journal}{Nucl. Phys.} \textbf{\bibinfo{volume}{B555}},
  \bibinfo{pages}{335} (\bibinfo{year}{1999}),
  \eprint{\href{http://arXiv.org/abs/hep-ph/9902341}{hep-ph/9902341}}.

\bibitem[{\citenamefont{Bauer et~al.}(2003)\citenamefont{Bauer, Manohar, and
  Wise}}]{Bauer:2002ie}
\bibinfo{author}{\bibfnamefont{C.~W.} \bibnamefont{Bauer}},
  \bibinfo{author}{\bibfnamefont{A.~V.} \bibnamefont{Manohar}},
  \bibnamefont{and} \bibinfo{author}{\bibfnamefont{M.~B.} \bibnamefont{Wise}},
  \bibinfo{journal}{Phys. Rev. Lett.} \textbf{\bibinfo{volume}{91}},
  \bibinfo{pages}{122001} (\bibinfo{year}{2003}),
  \eprint{\href{http://arXiv.org/abs/hep-ph/0212255}{hep-ph/0212255}}.

\bibitem[{\citenamefont{Bauer et~al.}(2004)\citenamefont{Bauer, Lee, Manohar,
  and Wise}}]{Bauer:2003di}
\bibinfo{author}{\bibfnamefont{C.~W.} \bibnamefont{Bauer}},
  \bibinfo{author}{\bibfnamefont{C.}~\bibnamefont{Lee}},
  \bibinfo{author}{\bibfnamefont{A.~V.} \bibnamefont{Manohar}},
  \bibnamefont{and} \bibinfo{author}{\bibfnamefont{M.~B.} \bibnamefont{Wise}},
  \bibinfo{journal}{Phys. Rev.} \textbf{\bibinfo{volume}{D70}},
  \bibinfo{pages}{034014} (\bibinfo{year}{2004}),
  \eprint{\href{http://arXiv.org/abs/hep-ph/0309278}{hep-ph/0309278}}.

\bibitem[{\citenamefont{Gatheral}(1983)}]{Gatheral:1983cz}
\bibinfo{author}{\bibfnamefont{J.~G.~M.} \bibnamefont{Gatheral}},
  \bibinfo{journal}{Phys. Lett.} \textbf{\bibinfo{volume}{B133}},
  \bibinfo{pages}{90} (\bibinfo{year}{1983}).

\bibitem[{\citenamefont{Frenkel and Taylor}(1984)}]{Frenkel:1984pz}
\bibinfo{author}{\bibfnamefont{J.}~\bibnamefont{Frenkel}} \bibnamefont{and}
  \bibinfo{author}{\bibfnamefont{J.~C.} \bibnamefont{Taylor}},
  \bibinfo{journal}{Nucl. Phys.} \textbf{\bibinfo{volume}{B246}},
  \bibinfo{pages}{231} (\bibinfo{year}{1984}).

\bibitem[{\citenamefont{Korchemsky and Marchesini}(1993)}]{Korchemsky:1992xv}
\bibinfo{author}{\bibfnamefont{G.~P.} \bibnamefont{Korchemsky}}
  \bibnamefont{and}
  \bibinfo{author}{\bibfnamefont{G.}~\bibnamefont{Marchesini}},
  \bibinfo{journal}{Nucl. Phys.} \textbf{\bibinfo{volume}{B406}},
  \bibinfo{pages}{225} (\bibinfo{year}{1993}),
  \eprint{\href{http://arXiv.org/abs/hep-ph/9210281}{hep-ph/9210281}}.

\bibitem[{\citenamefont{Gardi}(2005)}]{Gardi:2005yi}
\bibinfo{author}{\bibfnamefont{E.}~\bibnamefont{Gardi}},
  \bibinfo{journal}{JHEP} \textbf{\bibinfo{volume}{02}}, \bibinfo{pages}{053}
  (\bibinfo{year}{2005}),
  \eprint{\href{http://arXiv.org/abs/hep-ph/0501257}{hep-ph/0501257}}.

\bibitem[{\citenamefont{Becher and Neubert}(2006)}]{Becher:2005pd}
\bibinfo{author}{\bibfnamefont{T.}~\bibnamefont{Becher}} \bibnamefont{and}
  \bibinfo{author}{\bibfnamefont{M.}~\bibnamefont{Neubert}},
  \bibinfo{journal}{Phys. Lett.} \textbf{\bibinfo{volume}{B633}},
  \bibinfo{pages}{739} (\bibinfo{year}{2006}), \eprint{hep-ph/0512208}.

\bibitem[{\citenamefont{Neubert}(2007)}]{Neubert:2007je}
\bibinfo{author}{\bibfnamefont{M.}~\bibnamefont{Neubert}}
  (\bibinfo{year}{2007}), \eprint{arXiv:0706.2136 [hep-ph]}.

\bibitem[{\citenamefont{Neubert}(1994{\natexlab{b}})}]{Neubert:1993ch}
\bibinfo{author}{\bibfnamefont{M.}~\bibnamefont{Neubert}},
  \bibinfo{journal}{Phys. Rev.} \textbf{\bibinfo{volume}{D49}},
  \bibinfo{pages}{3392} (\bibinfo{year}{1994}{\natexlab{b}}),
  \eprint{hep-ph/9311325}.

\bibitem[{\citenamefont{Bigi et~al.}(1994)\citenamefont{Bigi, Shifman,
  Uraltsev, and Vainshtein}}]{Bigi:1994ex}
\bibinfo{author}{\bibfnamefont{I.~I.~Y.} \bibnamefont{Bigi}},
  \bibinfo{author}{\bibfnamefont{M.~A.} \bibnamefont{Shifman}},
  \bibinfo{author}{\bibfnamefont{N.~G.} \bibnamefont{Uraltsev}},
  \bibnamefont{and} \bibinfo{author}{\bibfnamefont{A.~I.}
  \bibnamefont{Vainshtein}}, \bibinfo{journal}{Int. J. Mod. Phys.}
  \textbf{\bibinfo{volume}{A9}}, \bibinfo{pages}{2467} (\bibinfo{year}{1994}),
  \eprint{\href{http://arXiv.org/abs/hep-ph/9312359}{hep-ph/9312359}}.

\bibitem[{\citenamefont{Jaffe and Randall}(1994)}]{Jaffe:1993ie}
\bibinfo{author}{\bibfnamefont{R.~L.} \bibnamefont{Jaffe}} \bibnamefont{and}
  \bibinfo{author}{\bibfnamefont{L.}~\bibnamefont{Randall}},
  \bibinfo{journal}{Nucl. Phys.} \textbf{\bibinfo{volume}{B412}},
  \bibinfo{pages}{79} (\bibinfo{year}{1994}), \eprint{hep-ph/9306201}.

\bibitem[{\citenamefont{Manohar and Stewart}(2007)}]{Manohar:2006nz}
\bibinfo{author}{\bibfnamefont{A.~V.} \bibnamefont{Manohar}} \bibnamefont{and}
  \bibinfo{author}{\bibfnamefont{I.~W.} \bibnamefont{Stewart}},
  \bibinfo{journal}{Phys. Rev.} \textbf{\bibinfo{volume}{D76}},
  \bibinfo{pages}{074002} (\bibinfo{year}{2007}),
  \eprint{\href{http://arXiv.org/abs/hep-ph/0605001}{hep-ph/0605001}}.

\bibitem[{\citenamefont{Falk et~al.}(1992)\citenamefont{Falk, Neubert, and
  Luke}}]{Falk:1992fm}
\bibinfo{author}{\bibfnamefont{A.~F.} \bibnamefont{Falk}},
  \bibinfo{author}{\bibfnamefont{M.}~\bibnamefont{Neubert}}, \bibnamefont{and}
  \bibinfo{author}{\bibfnamefont{M.~E.} \bibnamefont{Luke}},
  \bibinfo{journal}{Nucl. Phys.} \textbf{\bibinfo{volume}{B388}},
  \bibinfo{pages}{363} (\bibinfo{year}{1992}), \eprint{hep-ph/9204229}.

\bibitem[{\citenamefont{Polyakov}(1979)}]{Polyakov:1979gp}
\bibinfo{author}{\bibfnamefont{A.~M.} \bibnamefont{Polyakov}},
  \bibinfo{journal}{Phys. Lett.} \textbf{\bibinfo{volume}{B82}},
  \bibinfo{pages}{247} (\bibinfo{year}{1979}).

\bibitem[{\citenamefont{Dotsenko and Vergeles}(1980)}]{Dotsenko:1979wb}
\bibinfo{author}{\bibfnamefont{V.~S.} \bibnamefont{Dotsenko}} \bibnamefont{and}
  \bibinfo{author}{\bibfnamefont{S.~N.} \bibnamefont{Vergeles}},
  \bibinfo{journal}{Nucl. Phys.} \textbf{\bibinfo{volume}{B169}},
  \bibinfo{pages}{527} (\bibinfo{year}{1980}).

\bibitem[{\citenamefont{Brandt et~al.}(1981)\citenamefont{Brandt, Neri, and
  Sato}}]{Brandt:1981kf}
\bibinfo{author}{\bibfnamefont{R.}~\bibnamefont{Brandt}},
  \bibinfo{author}{\bibfnamefont{F.}~\bibnamefont{Neri}}, \bibnamefont{and}
  \bibinfo{author}{\bibfnamefont{M.}~\bibnamefont{Sato}},
  \bibinfo{journal}{Phys. Rev.} \textbf{\bibinfo{volume}{D24}},
  \bibinfo{pages}{879} (\bibinfo{year}{1981}).

\bibitem[{\citenamefont{Korchemsky and Radyushkin}(1987)}]{Korchemsky:1987wg}
\bibinfo{author}{\bibfnamefont{G.~P.} \bibnamefont{Korchemsky}}
  \bibnamefont{and} \bibinfo{author}{\bibfnamefont{A.~V.}
  \bibnamefont{Radyushkin}}, \bibinfo{journal}{Nucl. Phys.}
  \textbf{\bibinfo{volume}{B283}}, \bibinfo{pages}{342} (\bibinfo{year}{1987}).

\bibitem[{\citenamefont{Balzereit et~al.}(1998)\citenamefont{Balzereit, Mannel,
  and Kilian}}]{Balzereit:1998yf}
\bibinfo{author}{\bibfnamefont{C.}~\bibnamefont{Balzereit}},
  \bibinfo{author}{\bibfnamefont{T.}~\bibnamefont{Mannel}}, \bibnamefont{and}
  \bibinfo{author}{\bibfnamefont{W.}~\bibnamefont{Kilian}},
  \bibinfo{journal}{Phys. Rev.} \textbf{\bibinfo{volume}{D58}},
  \bibinfo{pages}{114029} (\bibinfo{year}{1998}),
  \eprint{\href{http://arXiv.org/abs/hep-ph/9805297}{hep-ph/9805297}}.

\bibitem[{\citenamefont{Neubert}(2005)}]{Neubert:2004dd}
\bibinfo{author}{\bibfnamefont{M.}~\bibnamefont{Neubert}},
  \bibinfo{journal}{Eur. Phys. J. C} \textbf{\bibinfo{volume}{40}},
  \bibinfo{pages}{165} (\bibinfo{year}{2005}),
  \eprint{\href{http://arXiv.org/abs/hep-ph/0408179}{hep-ph/0408179}}.

\bibitem[{\citenamefont{Gross and Wilczek}(1973)}]{Gross:1973id}
\bibinfo{author}{\bibfnamefont{D.~J.} \bibnamefont{Gross}} \bibnamefont{and}
  \bibinfo{author}{\bibfnamefont{F.}~\bibnamefont{Wilczek}},
  \bibinfo{journal}{Phys. Rev. Lett.} \textbf{\bibinfo{volume}{30}},
  \bibinfo{pages}{1343} (\bibinfo{year}{1973}).

\bibitem[{\citenamefont{Politzer}(1973)}]{Politzer:1973fx}
\bibinfo{author}{\bibfnamefont{H.~D.} \bibnamefont{Politzer}},
  \bibinfo{journal}{Phys. Rev. Lett.} \textbf{\bibinfo{volume}{30}},
  \bibinfo{pages}{1346} (\bibinfo{year}{1973}).

\bibitem[{\citenamefont{Jones}(1974)}]{Jones:1974mm}
\bibinfo{author}{\bibfnamefont{D.~R.~T.} \bibnamefont{Jones}},
  \bibinfo{journal}{Nucl. Phys.} \textbf{\bibinfo{volume}{B75}},
  \bibinfo{pages}{531} (\bibinfo{year}{1974}).

\bibitem[{\citenamefont{Caswell}(1974)}]{Caswell:1974gg}
\bibinfo{author}{\bibfnamefont{W.~E.} \bibnamefont{Caswell}},
  \bibinfo{journal}{Phys. Rev. Lett.} \textbf{\bibinfo{volume}{33}},
  \bibinfo{pages}{244} (\bibinfo{year}{1974}).

\bibitem[{\citenamefont{Tarasov et~al.}(1980)\citenamefont{Tarasov, Vladimirov,
  and Zharkov}}]{Tarasov:1980au}
\bibinfo{author}{\bibfnamefont{O.~V.} \bibnamefont{Tarasov}},
  \bibinfo{author}{\bibfnamefont{A.~A.} \bibnamefont{Vladimirov}},
  \bibnamefont{and} \bibinfo{author}{\bibfnamefont{A.~Y.}
  \bibnamefont{Zharkov}}, \bibinfo{journal}{Phys. Lett.}
  \textbf{\bibinfo{volume}{B93}}, \bibinfo{pages}{429} (\bibinfo{year}{1980}).

\bibitem[{\citenamefont{Larin and Vermaseren}(1993)}]{Larin:1993tp}
\bibinfo{author}{\bibfnamefont{S.~A.} \bibnamefont{Larin}} \bibnamefont{and}
  \bibinfo{author}{\bibfnamefont{J.~A.~M.} \bibnamefont{Vermaseren}},
  \bibinfo{journal}{Phys. Lett.} \textbf{\bibinfo{volume}{B303}},
  \bibinfo{pages}{334} (\bibinfo{year}{1993}), \eprint{hep-ph/9302208}.

\bibitem[{\citenamefont{Broadhurst and Grozin}(1991)}]{Broadhurst:1991fz}
\bibinfo{author}{\bibfnamefont{D.~J.} \bibnamefont{Broadhurst}}
  \bibnamefont{and} \bibinfo{author}{\bibfnamefont{A.~G.}
  \bibnamefont{Grozin}}, \bibinfo{journal}{Phys. Lett.}
  \textbf{\bibinfo{volume}{B267}}, \bibinfo{pages}{105} (\bibinfo{year}{1991}),
  \eprint{hep-ph/9908362}.

\bibitem[{\citenamefont{Grozin}(2000)}]{Grozin:2000cm}
\bibinfo{author}{\bibfnamefont{A.~G.} \bibnamefont{Grozin}}
  (\bibinfo{year}{2000}), \eprint{hep-ph/0008300}.

\bibitem[{\citenamefont{Moch et~al.}(2004)\citenamefont{Moch, Vermaseren, and
  Vogt}}]{Moch:2004pa}
\bibinfo{author}{\bibfnamefont{S.}~\bibnamefont{Moch}},
  \bibinfo{author}{\bibfnamefont{J.~A.~M.} \bibnamefont{Vermaseren}},
  \bibnamefont{and} \bibinfo{author}{\bibfnamefont{A.}~\bibnamefont{Vogt}},
  \bibinfo{journal}{Nucl. Phys.} \textbf{\bibinfo{volume}{B688}},
  \bibinfo{pages}{101} (\bibinfo{year}{2004}), \eprint{hep-ph/0403192}.

\bibitem[{\citenamefont{Korchemsky and Radyushkin}(1992)}]{Korchemsky:1991zp}
\bibinfo{author}{\bibfnamefont{G.~P.} \bibnamefont{Korchemsky}}
  \bibnamefont{and} \bibinfo{author}{\bibfnamefont{A.~V.}
  \bibnamefont{Radyushkin}}, \bibinfo{journal}{Phys. Lett.}
  \textbf{\bibinfo{volume}{B279}}, \bibinfo{pages}{359} (\bibinfo{year}{1992}),
  \eprint{\href{http://arXiv.org/abs/hep-ph/9203222}{hep-ph/9203222}}.

\bibitem[{\citenamefont{Chay et~al.}(2005)\citenamefont{Chay, Kim, Kim, and
  Lee}}]{Chay:2004zn}
\bibinfo{author}{\bibfnamefont{J.}~\bibnamefont{Chay}},
  \bibinfo{author}{\bibfnamefont{C.}~\bibnamefont{Kim}},
  \bibinfo{author}{\bibfnamefont{Y.~G.} \bibnamefont{Kim}}, \bibnamefont{and}
  \bibinfo{author}{\bibfnamefont{J.-P.} \bibnamefont{Lee}},
  \bibinfo{journal}{Phys. Rev.} \textbf{\bibinfo{volume}{D71}},
  \bibinfo{pages}{056001} (\bibinfo{year}{2005}), \eprint{hep-ph/0412110}.

\bibitem[{\citenamefont{Arnesen et~al.}(2005)\citenamefont{Arnesen, Kundu, and
  Stewart}}]{Arnesen:2005nk}
\bibinfo{author}{\bibfnamefont{C.~M.} \bibnamefont{Arnesen}},
  \bibinfo{author}{\bibfnamefont{J.}~\bibnamefont{Kundu}}, \bibnamefont{and}
  \bibinfo{author}{\bibfnamefont{I.~W.} \bibnamefont{Stewart}},
  \bibinfo{journal}{Phys. Rev.} \textbf{\bibinfo{volume}{D72}},
  \bibinfo{pages}{114002} (\bibinfo{year}{2005}),
  \eprint{\href{http://arXiv.org/abs/hep-ph/0508214}{hep-ph/0508214}}.

\bibitem[{\citenamefont{Fleming
  et~al.}(2008{\natexlab{b}})\citenamefont{Fleming, Hoang, Mantry, and
  Stewart}}]{FHMSmass:inprep}
\bibinfo{author}{\bibfnamefont{S.}~\bibnamefont{Fleming}},
  \bibinfo{author}{\bibfnamefont{A.~H.} \bibnamefont{Hoang}},
  \bibinfo{author}{\bibfnamefont{S.}~\bibnamefont{Mantry}}, \bibnamefont{and}
  \bibinfo{author}{\bibfnamefont{I.}~\bibnamefont{Stewart}},
  \bibinfo{journal}{in preparation}  (\bibinfo{year}{2008}{\natexlab{b}}).

\bibitem[{\citenamefont{Bauer and Manohar}(2004)}]{Bauer:2003pi}
\bibinfo{author}{\bibfnamefont{C.~W.} \bibnamefont{Bauer}} \bibnamefont{and}
  \bibinfo{author}{\bibfnamefont{A.~V.} \bibnamefont{Manohar}},
  \bibinfo{journal}{Phys. Rev.} \textbf{\bibinfo{volume}{D70}},
  \bibinfo{pages}{034024} (\bibinfo{year}{2004}), \eprint{hep-ph/0312109}.

\bibitem[{\citenamefont{Lange and Neubert}(2003)}]{Lange:2003ff}
\bibinfo{author}{\bibfnamefont{B.~O.} \bibnamefont{Lange}} \bibnamefont{and}
  \bibinfo{author}{\bibfnamefont{M.}~\bibnamefont{Neubert}},
  \bibinfo{journal}{Phys. Rev. Lett.} \textbf{\bibinfo{volume}{91}},
  \bibinfo{pages}{102001} (\bibinfo{year}{2003}), \eprint{hep-ph/0303082}.

\bibitem[{\citenamefont{Gray et~al.}(1990)\citenamefont{Gray, Broadhurst,
  Grafe, and Schilcher}}]{Gray:1990yh}
\bibinfo{author}{\bibfnamefont{N.}~\bibnamefont{Gray}},
  \bibinfo{author}{\bibfnamefont{D.~J.} \bibnamefont{Broadhurst}},
  \bibinfo{author}{\bibfnamefont{W.}~\bibnamefont{Grafe}}, \bibnamefont{and}
  \bibinfo{author}{\bibfnamefont{K.}~\bibnamefont{Schilcher}},
  \bibinfo{journal}{Z. Phys.} \textbf{\bibinfo{volume}{C48}},
  \bibinfo{pages}{673} (\bibinfo{year}{1990}).

\bibitem[{\citenamefont{Fleischer et~al.}(1999)\citenamefont{Fleischer,
  Jegerlehner, Tarasov, and Veretin}}]{Fleischer:1998dw}
\bibinfo{author}{\bibfnamefont{J.}~\bibnamefont{Fleischer}},
  \bibinfo{author}{\bibfnamefont{F.}~\bibnamefont{Jegerlehner}},
  \bibinfo{author}{\bibfnamefont{O.~V.} \bibnamefont{Tarasov}},
  \bibnamefont{and} \bibinfo{author}{\bibfnamefont{O.~L.}
  \bibnamefont{Veretin}}, \bibinfo{journal}{Nucl. Phys.}
  \textbf{\bibinfo{volume}{B539}}, \bibinfo{pages}{671} (\bibinfo{year}{1999}),
  \eprint{hep-ph/9803493}.

\bibitem[{\citenamefont{Tkachov}(1981)}]{Tkachov:1981wb}
\bibinfo{author}{\bibfnamefont{F.~V.} \bibnamefont{Tkachov}},
  \bibinfo{journal}{Phys. Lett.} \textbf{\bibinfo{volume}{B100}},
  \bibinfo{pages}{65} (\bibinfo{year}{1981}).

\bibitem[{\citenamefont{Chetyrkin and Tkachov}(1981)}]{Chetyrkin:1981qh}
\bibinfo{author}{\bibfnamefont{K.~G.} \bibnamefont{Chetyrkin}}
  \bibnamefont{and} \bibinfo{author}{\bibfnamefont{F.~V.}
  \bibnamefont{Tkachov}}, \bibinfo{journal}{Nucl. Phys.}
  \textbf{\bibinfo{volume}{B192}}, \bibinfo{pages}{159} (\bibinfo{year}{1981}).

\end{thebibliography}

\end{document}